\documentclass[aps,onecolumn,showpacs,showkeys,nofootinbib]{revtex4}
\usepackage{epsfig}
\usepackage{amsmath}
\usepackage{amsfonts}
\usepackage{amssymb}
\usepackage{graphicx}
\usepackage{colordvi}
\begin{document}

\title{The Weak Field Limit of Fourth Order Gravity}

\author{S. Capozziello$^{1,2}$\footnote{capozziello@na.infn.it}, A. Stabile$^3$\footnote{arturo.stabile@gmail.com}}

\affiliation{$^1$Dipartimento di Scienze Fisiche, Universita' di
Napoli \emph{Federico II}, $^2$INFN Sez. di Napoli, Compl. Univ.
di Monte S. Angelo, Edificio G, Via Cinthia, I-80126, Napoli,
Italy
\\
$^3$Dipartimento di Ingegneria, Universita' del Sannio, Palazzo
Dell'Aquila Bosco Lucarelli, Corso Garibaldi, 107 - 82100,
Benevento, Italy}

\begin{abstract}

Fourth Order Theories of Gravity have recently attracted a lot of
interest as  candidates to explain the observed cosmic
acceleration, the flatness of the rotation curves of spiral
galaxies, the large scale structure and other relevant
astrophysical phenomena. This means that  the "Dark Side" issue of
the Universe could be completely reversed considering dark matter
and dark energy as "shortcomings" of General Relativity in its
simplest formulation (a linear theory in the Ricci scalar $R$,
minimally coupled to the standard perfect fluid matter) and
claiming for the "correct" theory of Gravity as that derived by
matching the largest number of observational data, without
imposing any theory a priori. As a working hypothesis,
accelerating behavior of cosmic fluid, large scale structure,
potential of galaxy clusters, rotation curves of spiral galaxies
could be reproduced by means of extending  General Relativity to
generic actions containing higher order and non-minimally coupled
terms in curvature invariants. In other words, gravity could act
in different ways at different scales and the above "shortcomings"
could be due to the incorrect extrapolations of the Einstein
theory, actually tested at short scales and low energy regimes.
Very likely, what we call "dark matter and "dark energy" could be
nothing else but signals of the breakdown of General Relativity at
large scales. Then, it is a crucial point testing these  Extended
Theories in the  weak field limit. In this sense, comparing these
theories to General Relativity could be a fundamental step to
retain or rule out them. In this review paper, after a survey of
what is intended for Extended Theories of Gravity in the so called
\emph{metric approach}, we extensively discuss their Newtonian and
the post-Newtonian limits pointing out, in details, their
resemblances and differences with respect to General Relativity.
Particular emphasis is placed on the exact solutions and methods
used to obtain them. Finally,it is clearly shown that General
Relativity  results, in the Solar System context, are easily
recovered since Einstein theory is a particular case of this
extended approach. This is a crucial point against several wrong
results in literature stating that these theories (e.g.
$f(R)$-gravity) are not viable at local scales.

\end{abstract}
\pacs{04.25.Nx; 04.50.Kd; 04.40.Nr} \keywords{Alternative theories
of gravity; Newtonian and post-Newtonian limit; weak field limit.}
\date{\today}
\maketitle

\section{Introduction}

In recent years, the effort to give a physical explanation to the
today observed cosmic acceleration \cite{cosmic_acceleration} has
attracted a good amount of interest in Fourth Order Gravity (FOG)
considered as a viable mechanism to explain the cosmic
acceleration by extending the geometric sector of field equations
without the introduction of dark matter and dark energy. At
fundamental level, several efforts have been aimed towards the
unification of gravity with the other interactions of physics,
like Electromagnetism, assuming GR as the only fundamental theory
capable of explaining the gravitational interaction. The failure
of such attempts led to the common belief that GR had to be
revised in the ultraviolet limit in order to address issues like
quantization and renormalization. These are only some aspects of
the several physical and mathematical motivations to enlarge GR to
more general approaches. For comprehensive reviews of the problem,
see
\cite{theo_aspect_1,theo_aspect_2,theo_aspect_3,theo_aspect_4,theo_aspect_5,theo_aspect_6}.

Other issues come from astrophysics. For example,  the observed
Pioneer anomaly problem \cite{anderson} can be framed into the
same approach \cite{bertolami} and then, apart the cosmology and
quantum field theory, a systematic analysis of such theories urges
at small, medium and large scales.   In particular,  a delicate
point is to address the weak field limit of any extended theory of
gravity since two main issues are extremely relevant: $i)$
preserving the results of GR al local scales since they well fit
Solar System experiments and observations; $ii)$ enclosing in a
self-consistent and comprehensive picture phenomena as anomalous
acceleration  or dark matter at Galactic scales.

It is  straightforward to extend  GR to theories with additional
geometric degrees of freedom and several recent proposals focused
on the old idea  of modifying the gravitational Lagrangian in a
purely metric framework, leading to fourth-order and higher-order
field equations \cite{old_papers_fR,noethergae,stel,curvquin}.
Such an approach has become a sort of paradigm in the study of
gravitational interaction consisting, essentially, \emph{in adding
higher order curvature invariants and minimally or non-minimally
coupled scalar fields into dynamics which come out from the
effective action of some unification or quantum gravity theory}.

The idea to extend Einstein's theory of gravitation is fruitful
and economic also with respect to several attempts which try to
solve problems by adding new and, most of times, unjustified
ingredients in order to give  self-consistent pictures of
dynamics. The today observed accelerated expansion of the Hubble
flow and the missing matter at astrophysical  scales are primarily
enclosed in these considerations. Both the issues could be solved
by changing the gravitational sector, \emph{i.e.} the
\emph{l.h.s.} of field equations. The philosophy is alternative to
add new cosmic fluids (new components in the \emph{r.h.s.} of
field equations) which should give rise to clustered structures
(dark matter) or to accelerated dynamics (dark energy) thanks to
exotic equations of state. In particular, relaxing the hypothesis
that gravitational Lagrangian has to be a linear function of the
Ricci curvature scalar $R$, like in the Hilbert-Einstein
formulation, one can take into account an effective action where
the gravitational Lagrangian includes other scalar invariants.

Due to the increased complexity of the field equations, the main
body of theoretical works dealt with the effort to achieve some
formally equivalent theories which could be handled in a simpler
way. In this sense, a reduction of the differential order of the
field equations can be achieved by considering  metric and
connection as independent objects in the so called {\it Palatini
approach} \cite{francaviglia,torsion} and energy conditions have
to be carefully discussed \cite{condition}.

In addition, other authors exploited the formal relations to
Scalar-Tensor theories to make some statements about the weak
field regime \cite{olmo}, which was already worked out for
scalar-tensor theories \cite{Damour:Esposito-Farese:1992}. Also a
post-Newtonian parameterization with metric approach in the Jordan
Frame has been considered \cite{clifton}.

In this review paper, we want to address the general problem of
the weak field limit for theories of gravity where higher order
curvature invariants are present. In particular, we deal with
theories where Riemann tensor,  Ricci tensor, and Ricci scalar are
considered in the effective action. We deduce the field equations,
discuss the weak field limit and  derive the weak field potentials
with corrections to the Newtonian potential. The plan of the paper
is the following:

In \emph{section} \ref{GR}, we provide a short summary of GR and
discuss its extension to FOG. Besides we take into account the
conformal transformations showing how such theories can be
discussed under the standard of scalar-tensor gravity.

In \emph{section} \ref{perturb}, we give generalities on
spherically symmetric, Birkhoff theorem, Eddington parameters and
deviations from GR solutions. A general perturbation approach for
$f(R)$-gravity is discussed starting from  GR.

\emph{Section} \ref{PPN-paradigm} is devoted to  the technical
development of field equations with respect to Newtonian and
post-Newtonian approaches.

In \emph{section} \ref{TS} we take into account the debate on the
analogy  between $f(R)$- gravity and Scalar-Tensor gravity. We
show  that $f(R)$-models are dynamically equivalent to  O'Hanlon
models  which is a special case of Scalar-Tensor gravity
characterized by a self-interaction potential without a kinetic
term. We show that comparing the results of this theory with GR,
in the weak field limit, could lead to misleading conclusions.

In \emph{section} \ref{newtlimstandard}, we analyze the Newtonian
limit of $f(R)$-Gravity. We are going to focus exclusively on the
weak field limit within the metric approach. By using the
development for a generic analytic function $f(R)$, we find the
solution in the vacuum with standard coordinates. Besides, we show
that the Birkhoff theorem is not a general result for
$f(R)$-gravity since time-dependent evolution for spherically
symmetric solutions can be achieved according to the order of
perturbation. In other words, solutions could not be stable as
requested by the Birkhoff theorem.

In \emph{section} \ref{postnewtlimisitropic}, we report the
Newtonian and post-Newtonian limit of field equations by using the
Green function method for $f(R)$-gravity and the formal solutions
in the harmonic gauge. A general discussion about the mathematical
properties of equations, the solutions and their deviations with
respect to Gauss and Birkhoff theorem, and the Minkowskian
behavior of metric tensor are reported. We derive the general
solutions (at Newtonian and post-Newtonian levels) when an uniform
massive spherical source is considered. The point-like source
limit of the Newtonian solution is discussed and the compatibility
of $f(R)$-gravity with respect to GR is shown.

In \emph{section} \ref{newtlimitquadr}, we discuss the Newtonian
limit but for a  \emph{quadratic gravity Lagrangian} where
quadratic curvature invariants are presents. Also in this case, we
adopt  the Green function method. We find the internal and
external potential generated by an extended spherically symmetric
matter source. A particularly detailed discussion about the field
equations and their solutions is provided for a set of values of
the arbitrary constants present in the model.

In section \ref{fXYZgravity} the Newtonian limit of the most
general FOG is discussed with no gauge condition. The most general
theory with fourth order differential equations is obtained by
generalizing the $f(R)$-models in the action with a generic
function containing the other two curvature invariants:
\emph{Ricci square} ($R_{\alpha\beta}R^{\alpha\beta}$) and
\emph{Riemann square}
($R_{\alpha\beta\gamma\delta}R^{\alpha\beta\gamma\delta}$).
Considering the Gauss - Bonnet invariant, it is possible to show
that only two of these invariants are really necessary.

In section \ref{conclus},  we draw  the conclusions and discuss
the possible applications of the results.

\section{General Relativity and its extensions}\label{GR}

Any relativistic theory of gravity has to match some minimal
requirements to address gravitational dynamics. First of all, it
has to explain issues coming from Celestial Mechanics as the
planetary orbits,  the potential of self-gravitating systems, the
Solar System stability.

This means that it has to reproduce the Newtonian dynamics in the
weak field limit and then it has to pass the  Solar System
experiments which are all  well founded and constitute the test
bed of GR \cite{will}.

Besides, any theory of gravity has to be consistent with stellar
structures and galactic dynamics considering the observed baryonic
constituents (e.g. luminous components as stars, sub-luminous
components as planets, dust and gas), radiation and Newtonian
potential which is, by assumption, extrapolated to galactic
scales.

The third step is cosmology and  large scale structure  which
means to reproduce, in a self-consistent way, the cosmological
parameters as the expansion rate, the Hubble constant, the density
parameter and the clustering of galaxies. Observations  probe the
standard baryonic matter, the radiation and an attractive overall
interaction, acting at all scales and depending on distance. From
a phenomenological point of view this is  {\it gravity}.

GR is the simplest theory which partially  satisfies the above
requirements  \cite{einstein1}. It is  based on the assumption
that space and time are entangled into a single spacetime
structure, which, in the limit of no gravitational forces, has to
reproduce the Minkowski spacetime.  Besides,  the Universe is
assumed to be a curved  manifold and the curvature depends on
mass-energy distribution \cite{eisenhart}. In other words, the
distribution of matter  influences point by point the local
curvature of the spacetime structure.

Furthermore, GR is based on three first principles that are
\emph{Relativity}, \emph{Equivalence}, and \emph{General
Covariance}(see \cite{theo_aspect_3,theo_aspect_6,schroedinger}
for detailed discussions). Another requirement is the \emph{Principle of Causality} that means
 that each point of spacetime  admits a
notion of past, present and future.

Let us also recall that  the Newtonian theory, the weak field
limit of GR, requires absolute concepts of  space and time, that
particles move in a preferred inertial frame following curved
trajectories function of the sources (\emph{i.e.}, the "forces").

On these bases, GR postulates that  gravitational forces have to
be expressed by the curvature of a metric tensor field
$ds^2\,=\,g_{\alpha\beta}dx^{\alpha}dx^{\beta}$ on a
four-dimensional spacetime manifold, having the same signature of
Minkowski metric, here assumed to be $(+---)$.  Curvature is
locally determined by the distribution of the sources, that is,
being the spacetime a continuum, it is possible to define a
stress-energy tensor $T_{\mu\nu}$ which is the source of the
curvature.

Once a metric $g_{\mu\nu}$ is given, the inverse $g^{\mu\nu}$
satisfies the condition\footnote{The Greek index runs between $0$
and $3$; the Latin index between $1$ and $3$.}

\begin{eqnarray}\label{controvariant-metric-condition}
g^{\mu\alpha}g_{\alpha\beta}\,=\,\delta_\nu^\mu\
\end{eqnarray}
Its curvature is expressed by the \emph{Riemann tensor}
(curvature)

\begin{eqnarray}\label{rienmanntensor}
R^{\alpha}_{\,\,\,\,\,\mu\beta\nu}\,=\,\Gamma^{\alpha}_{\mu\nu,\beta}-\Gamma^{\alpha}_{\mu\beta,\nu}+
\Gamma^{\sigma}_{\mu\nu}\Gamma^{\alpha}_{\sigma\beta}-\Gamma^{\alpha}_{\sigma\nu}\Gamma^{\sigma}_{\mu\beta}
\end{eqnarray}
where the comas are partial derivatives. The
$\Gamma^{\alpha}_{\mu\nu}$ are the Christoffel symbols given by

\begin{eqnarray}\label{christoffel}
\Gamma^{\alpha}_{\mu\nu}\,=\,\frac{1}{2}g^{\alpha\sigma}(g_{\mu\sigma,\nu}+g_{\nu\sigma,
\mu}-g_{\mu\nu,\sigma})
\end{eqnarray}
if the Levi-Civita connection is assumed. The contraction of the Riemann tensor
(\ref{rienmanntensor})

\begin{eqnarray}\label{riccitensor}
R_{\mu\nu}\,=\,R^{\alpha}_{\,\,\,\,\,\mu\alpha\nu}=\,\Gamma^{\sigma}_{\mu\nu,\sigma}-\Gamma^{\sigma}_{\mu\sigma,\nu}+
\Gamma^{\sigma}_{\mu\nu}\Gamma^{\rho}_{\sigma\rho}-\Gamma^{\rho}_{\sigma\nu}\Gamma^{\sigma}_{\mu\rho}
\end{eqnarray}
is the \emph{Ricci tensor} and the scalar

\begin{eqnarray}\label{ricciscalar}
R\,=\,g^{\sigma\tau}R_{\sigma\tau}\,=\,R^{\sigma}_{\,\,\,\,\,\sigma}\,=\,g^{\tau\xi}\Gamma^{\sigma}_{\tau\xi,\sigma}-g^{\tau\xi}
\Gamma^{\sigma}_{\tau\sigma,\xi}+g^{\tau\xi}\Gamma^{\sigma}_{\tau\xi}\Gamma^{\rho}_{\sigma\rho}-g^{\tau\xi}\Gamma^{\rho}_
{\tau\sigma}\Gamma^{\sigma}_{\xi\rho}
\end{eqnarray}
is called the \emph{scalar curvature} of $g_{\mu\nu}$. The Riemann
tensor (\ref{rienmanntensor}) satisfies the so-called
\emph{Bianchi identities} and the \emph{contracted Bianchi
identities}, that is

\begin{eqnarray}\label{bianchi-identity}
\left\{\begin{array}{ll}
R_{\alpha\mu\beta\nu;\delta}+R_{\alpha\mu\delta\beta;\nu}+R_{\alpha\mu\nu\delta;\beta}\,=\,0
\\\\
R_{\alpha\mu\beta\nu}^{\,\,\,\,\,\,\,\,\,\,\,\,\,\,\,;\alpha}+R_{\mu\beta;\nu}-R_{\mu\nu;\beta}\,=\,0
\\\\
2R_{\alpha\beta}^{\,\,\,\,\,\,\,\,;\alpha}-R_{;\beta}\,=\,0
\\\\
2R_{\alpha\beta}^{\,\,\,\,\,\,\,\,;\alpha\beta}-\Box R\,=\,0
\end{array} \right.
\end{eqnarray}
where the covariant derivative is
$A^{\alpha\beta\dots\delta}_{\,\,\,\,\,\,\,\,\,\,\,\,\,\,\,\,;\mu}\,=\,\nabla_\mu
A^{\alpha\beta\dots\delta}\,=\,A^{\alpha\beta\dots\delta}_{\,\,\,\,\,\,\,\,\,\,\,\,\,\,\,\,,\mu}+\Gamma^\alpha_{\sigma\mu}
A^{\sigma\beta\dots\delta}+\Gamma^\beta_{\sigma\mu}
A^{\alpha\sigma\dots\delta}+\dots+\Gamma^\delta_{\sigma\mu}
A^{\alpha\beta\dots\sigma}$ and
$\nabla_\alpha\nabla^\alpha\,=\,\Box\,=\,\frac{\partial_\alpha(\sqrt{-g}g^{\alpha\beta}\partial_\beta)}{\sqrt{-g}}$
is  the d'Alembert operator with respect to the metric
$g_{\mu\nu}$ (see for the details \cite{landau}).

Assuming that matter is given as a perfect fluid, that is

\begin{eqnarray}\label{perfectfluid}
T_{\mu\nu}\,=\,(p+\rho)u_\mu u_\nu-p\,g_{\mu\nu}
\end{eqnarray}
where $u_\mu u_\nu$ defines a comoving observer (with the
conditions $g^{tt}{u_t}^2\,=\,1$, $u_i\,=\,0$ where $x^0=ct$ and
we are assuming natural units with $c=1$; $p$ is the pressure and
$\rho$ the mass-energy density of the fluid, then the continuity
equation requires $T_{\mu\nu}$ to be covariantly constant,
\emph{i.e.} it has to satisfy the conservation law

\begin{eqnarray}\label{conservationlaw}
T^{\mu\sigma}_{\,\,\,\,\,\,\,\,\,\,;\sigma}\,=\,0
\end{eqnarray}
that are nothing else but  contracted Bianchi identities. The GR
field equations are then

\begin{eqnarray}\label{fieldequationGR}
G_{\mu\nu}\,=\,\mathcal{X}\,T_{\mu\nu}
\end{eqnarray}
where

\begin{eqnarray}\label{einstein-tensor}
G_{\mu\nu}\,=\,R_{\mu\nu}-\frac{R}{2}g_{\mu\nu}
\end{eqnarray}
is the "\emph{Einstein tensor}" of $g_{\mu\nu}$. These equations
are both variational and satisfy the conservation laws
(\ref{conservationlaw}) since the following relation holds

\begin{eqnarray}
G^{\mu\sigma}_{\,\,\,\,\,\,;\sigma}\,=\,0
\end{eqnarray} as a byproduct of the above
Bianchi identities  \cite{ landau,weinberg}.

The Hilbert-Einstein Lagrangian that allows to obtain the field
equations (\ref{fieldequationGR}) is the sum of an ordinary
\emph{matter Lagrangian} $\mathcal{L}_m$ (minimally coupled) and
of the Ricci scalar:

\begin{eqnarray}\label{HElagrangian}
\mathcal{L}_{HE}\,=\,\sqrt{-g}(R+\mathcal{X}\mathcal{L}_m)
\end{eqnarray}
where $\sqrt{-g}$ denotes the square root of the value of the
determinant of the metric $g_{\mu\nu}$ and the coupling constant
is $\mathcal{X}=8\pi G$. The action of GR is

\begin{eqnarray}\label{HEaction}
\mathcal{A}^{GR}\,=\,\int d^4x\sqrt{-g}(R+\mathcal{X}\mathcal{L}_m)
\end{eqnarray}
From the action principle, we get the field equations
(\ref{fieldequationGR}) by the variation:

\begin{eqnarray}\label{variationprincipleGR}
\delta\mathcal{A}^{GR}\,=\,\delta\int d^4x\sqrt{-g}(R+\mathcal{X}\mathcal{L}_m)\,=\,
\int
d^4x\sqrt{-g}\biggr[R_{\mu\nu}-\frac{R}{2}g_{\mu\nu}-\mathcal{X}\,T_{\mu\nu}\biggr]\delta
g^{\mu\nu}+\int d^4x\sqrt{-g}g^{\mu\nu}\delta
R_{\mu\nu}=0
\end{eqnarray}
where $T_{\mu\nu}$ is energy momentum tensor of matter

\begin{eqnarray}\label{definitiontensormatter}
T_{\mu\nu}\,=\,-\frac{1}{\sqrt{-g}}\frac{\delta (\sqrt{-g}\mathcal{L}_m)}
{\delta g^{\mu\nu}}
\end{eqnarray}
The last term in (\ref{variationprincipleGR}) is a 4-divergence

\begin{eqnarray}
\int d^4x\sqrt{-g}g^{\mu\nu}\delta
R_{\mu\nu}\,=\,\int d^4x\sqrt{-g}[(-\delta
g^{\mu\nu})_{;\mu\nu}-\Box(g^{\mu\nu}\delta
g_{\mu\nu})]
\end{eqnarray}
then we can neglect it and we get the field equation
(\ref{fieldequationGR}). For the variational calculus
(\ref{variationprincipleGR}),  the following relations can be used

\begin{eqnarray}\label{variationalcalculus}
\left\{\begin{array}{ll}
\delta(g_{\mu\nu}g^{\mu\nu})\,=\,g_{\mu\nu}\delta g^{\mu\nu}+g^{\mu\nu}\delta g_{\mu\nu}\,=\,0
\\\\
\delta\sqrt{-g}\,=\,-\frac{1}{2}\,\sqrt{-g}\,g_{\alpha\beta}\,\delta
g^{\alpha\beta}
\\\\
\delta\,R\,=\,R_{\alpha\beta}\,\delta
g^{\alpha\beta}+g^{\alpha\beta}\,\delta R_{\alpha\beta}
\\\\
\delta\,R_{\alpha\beta}\,=\,\delta
g^\rho_{\,\,\,(\alpha;\beta)\rho}-\frac{1}{2}\Box\,\delta
g_{\alpha\beta}-\frac{1}{2}g^{\rho\sigma}\delta\,g_{\rho\sigma;\alpha\beta}
\end{array} \right.
\end{eqnarray}

The choice by Hilbert and Einstein was completely arbitrary  but
it was  the simplest one both from the mathematical and the
physical viewpoint. As it was later clarified by Levi-Civita
curvature is  not a "purely metric notion" but, rather, a notion
related to the "linear connection"  to which "parallel transport"
and "covariant derivation" refer \cite{levicivita}.

This is the precursor idea of the "gauge theoretical framework"
\cite{kle}, defined after the pioneering work by Cartan
\cite{cartan}.

It was later clarified that the  principles of relativity,
equivalence, covariance, and causality, just require that the
spacetime structure has to be determined by either one or both of
two fields, a Lorentzian metric  $g$  and a linear connection
$\Gamma$, assumed to be torsionless for the sake of simplicity.

The metric $g$ fixes the causal structure of spacetime (the light
cones) as well as its metric relations (clocks and rods); the
connection  $\Gamma$  fixes the free-fall, \emph{i.e.} the locally
inertial observers. They have, of course, to satisfy a number of
compatibility relations which amount to require that photons
follow the null geodesics of $\Gamma$, so that $\Gamma$ and $g$
can be independent, \emph{a priori}, but constrained, \emph{a
posteriori}, by some physical restrictions. These, however, do not
impose that $\Gamma$ has necessarily to be the Levi-Civita
connection of  $g$ \cite{palatini}.

This achievement justifies the fact, at least in principle,
"\emph{Extended Theories of Gravity}" can be considered starting
from the same points  by Einstein and Hilbert: they are theories
in which gravitation is described by either a metric (the
so-called "purely metric theories"), or by a linear connection
(the so-called "purely affine theories") or by both fields (the
so-called "metric-affine theories", also known as "first order
formalism theories"). In these theories,  the Lagrangian is a
scalar density of the curvature invariants constructed out of both
$g$ and $\Gamma$.

The choice (\ref{HElagrangian}) is by no means unique and it turns
out that the Hilbert-Einstein Lagrangian is in fact the only
choice that produces an invariant that is linear in second
derivatives of the metric (or first derivatives of the
connection). A Lagrangian that, unfortunately, is rather singular
from the Hamiltonian viewpoint, in much than same way as
Lagrangians, linear in canonical momenta, are rather singular in
Classical Mechanics (see \emph{e.g.} \cite{arnold}).

A number of attempts to generalize GR (and unify it to
Electromagnetism) along these lines were  followed by Einstein
himself and many others (Eddington, Weyl, Schrodinger, just to
quote the main contributors; see, \emph{e.g.}, \cite{app-fre}) but
they were  given up   because of a number of difficulties related
to the  complicated structure of a non-linear theory (where by
"non-linear" we mean here a theory that is based on non-linear
invariants of the curvature tensor), and also because of the new
understanding of physics that is currently based on four
fundamental forces and requires a general "gauge framework" to be
adopted (see \cite{kaku}).

Further curvature invariants or non-linear functions of them
should be also considered, especially in view of the fact that
they have to be included in both the semi-classical expansion of
any quantum Lagrangian or in the low-energy limit of a string
Lagrangian.

Moreover, it is clear from the recent astrophysical observations
that Einstein equations are no longer a good test for gravitation
at all scales,  that is at Solar System, galactic, extra-galactic
and cosmic scales, unless one does not admit that the {\it r.h.s.}
of Eqs.(\ref{fieldequationGR}) contains some kind of exotic
matter-energy density which are generically addressed as the "dark
matter" and "dark energy".

From our point of view, the philosophy of Extended Theories of
Gravity is much simpler \cite{theo_aspect_6}. Instead of changing
the matter side of Einstein Equations (\ref{fieldequationGR}) in
order to fit the "missing matter-energy" content of the currently
observed Universe by adding new forms of  matter and energy, we
assume that it is  more convenient to change the gravitational
side of the equations, admitting corrections coming from
non-linear Lagrangians. However such an approach should be
explored. Of course,  the Lagrangians should be  conveniently
tuned up  on the basis of their best fit with all possible
observational tests, at infrared scales (Solar, galactic,
extragalactic and cosmic) and on the basis of consistency with
fundamental theories at ultraviolet scales.

Let us take into account an important class of Extended Theories
of Gravity, the Fourth Order Gravity. It is the first
straightforward generalization of GR and can be directly related
to the ultraviolet  (quantum gravity) and infrared (cosmology)
issues of any final comprehensive theory of gravity.

\subsection{Fourth Order Gravity}

Let us start with the general class of \emph{Fourth Order Gravity}
(FOG) given by the action

\begin{eqnarray}\label{FOGaction}
\mathcal{A}^f=\int d^{4}x\sqrt{-g}\biggl[f(X,Y,Z)+\mathcal{X}\mathcal{L}_m\biggr]
\end{eqnarray}
where $f$ is an unspecified analytical function of curvature
invariants $X\,=\,R$ (Ricci scalar),
$Y\,=\,R_{\alpha\beta}R^{\alpha\beta}$ (Ricci tensor square)  and
$Z\,=\,R_{\alpha\beta\gamma\delta}R^{\alpha\beta\gamma\delta}$
(Riemann square). By varying the action (\ref{FOGaction}) in the
metric approach and by using the properties
(\ref{variationalcalculus}) with following additional properties

\begin{eqnarray}\label{variationalcalculus_2}
\left\{\begin{array}{ll}
\delta Y\,=\,\delta (R_{\alpha\beta}R^{\alpha\beta})\,=\,2R_\mu^{\,\,\,\,\alpha}R_{\alpha\nu}\delta
g^{\mu\nu}+2R^{\mu\nu}\delta R_{\mu\nu}
\\\\
\delta Z\,=\,\delta\,(R_{\alpha\beta\gamma\delta}R^{\alpha\beta\gamma\delta})\,=\,4R_{\mu\alpha\beta\gamma}R_{\nu}^{\,\,\,\,\,\alpha\beta\gamma}\delta
g^{\mu\nu}+2R^{\alpha\beta\gamma\delta}\delta
R_{\alpha\beta\gamma\delta}
\\\\
\delta\,R_{\alpha\beta\gamma\delta}\,=\,\delta(g_{\alpha\sigma}R^\sigma_{\,\,\,\,\,\beta\gamma
\delta})\,=\,R^\sigma_{\,\,\,\,\,\beta\gamma
\delta}\delta\,g_{\alpha\sigma}+g_{\alpha\sigma}\delta\,R^\sigma_{\,\,\,\,\,\beta\gamma
\delta}
\\\\
\delta R^\alpha_{\,\,\,\,\,\beta\gamma\delta}\,=\,\delta g^\alpha_{\,\,\,(\beta;\delta)\gamma}-\delta g^\alpha_{\,\,\,(\beta;\gamma)\delta}+\delta g_{\beta[\gamma\,\,\,\,\,\delta]}^{\,\,\,\,\,\,\,\,;\alpha}
\end{array} \right.
\end{eqnarray}
we have

\begin{eqnarray}
\delta\mathcal{A}^f\,=&&\delta\int d^4x\sqrt{-g}\,[f(X,Y,Z)+\mathcal{X}\mathcal{L}_m]
\nonumber\\\nonumber\\
=&&\int d^4x\sqrt{-g}
\biggr[\biggl(f_XR_{\mu\nu}+2f_YR_\mu^{\,\,\,\,\alpha}R_{\alpha\nu}+4f_ZR_{\mu\alpha\beta\gamma}R_{\nu}^{\,\,\,\,\,\alpha\beta\gamma}
-\frac{f}{2}g_{\mu\nu}-\mathcal{X}T_{\mu\nu}\biggr)\delta
g^{\mu\nu}
\nonumber\\
&&+\biggl(g^{\mu\nu}f_X+2f_YR^{\mu\nu}\biggr)\delta R_{\mu\nu}+2f_ZR^{\alpha\beta\gamma\delta}\delta
R_{\alpha\beta\gamma\delta}
\biggr]
\nonumber\\\nonumber\\
=&&\int
d^4x\sqrt{-g}\biggr\{\biggl(f_XR_{\mu\nu}+2f_YR_\mu^{\,\,\,\,\alpha}R_{\alpha\nu}+4f_ZR_{\mu\alpha\beta\gamma}R_{\nu}^{\,\,\,\,\,\alpha\beta\gamma}
-\frac{f}{2}g_{\mu\nu}-\mathcal{X}\,T_{\mu\nu}\biggr)\delta
g^{\mu\nu}
\nonumber\\
&&+f_X[-(\delta
g^{\mu\nu})_{;\mu\nu}-\Box(g^{\mu\nu}\delta
g_{\mu\nu})]+f_YR^{\mu\nu}[2\delta
g^\rho_{\,\,\,(\mu;\nu)\rho}-\Box\,\delta
g_{\mu\nu}-g^{\rho\sigma}\delta\,g_{\rho\sigma;\mu\nu}]
\\
&&+2f_ZR^{\alpha\beta\gamma\delta}[\delta g_{\alpha(\beta;\delta)\gamma}-\delta g_{\alpha(\beta;\gamma)\delta}+\delta g_{\beta[\gamma;\alpha\delta]}]
\biggr\}
\nonumber\\\nonumber\\
\sim&&\int
d^4x\sqrt{-g}\biggr\{f_XR_{\mu\nu}+2f_YR_\mu^{\,\,\,\,\alpha}R_{\alpha\nu}+2f_ZR_{\mu\alpha\beta\gamma}R_{\nu}^{\,\,\,\,\,\alpha\beta\gamma}
-\frac{f}{2}g_{\mu\nu}-\mathcal{X}\,T_{\mu\nu}-f_{X;\mu\nu}+g_{\mu\nu}\Box
f_X
\nonumber\\
&&-2[f_Y{R^\alpha}_{(\mu}]_{;\nu)\alpha}+\Box[f_YR_{\mu\nu}]+[f_YR_{\alpha\beta}]^{;\alpha\beta}g_{\mu\nu}
-4[f_Z{{R_\mu}^{\alpha\beta}}_\nu]_{;\alpha\beta}\biggr\}\delta g^{\mu\nu}
\nonumber\\\nonumber\\
=&&\int
d^4x\sqrt{-g}(H_{\mu\nu}-\mathcal{X}\,T_{\mu\nu})\delta
g^{\mu\nu}\,=\,0\nonumber
\end{eqnarray}
where $f_X\,=\,\frac{df}{dX}$, $f_Y\,=\,\frac{df}{dY}$,
$f_Z\,=\,\frac{df}{dZ}$  and the symbol $\sim$ means that we
neglected  pure divergences. Then the field equations of FOG are

\begin{eqnarray}\label{fieldequationFOG}
H_{\mu\nu}\,=\,&&f_XR_{\mu\nu}-\frac{f}{2}g_{\mu\nu}-f_{X;\mu\nu}+g_{\mu\nu}\Box
f_X+2f_Y{R_\mu}^\alpha
R_{\alpha\nu}-2[f_Y{R^\alpha}_{(\mu}]_{;\nu)\alpha}+\Box[f_YR_{\mu\nu}]+[f_YR_{\alpha\beta}]^{;\alpha\beta}g_{\mu\nu}
\nonumber\\\nonumber\\&&+2f_ZR_{\mu\alpha
\beta\gamma}{R_{\nu}}^{\alpha\beta\gamma}-4[f_Z{{R_\mu}^{\alpha\beta}}_\nu]_{;\alpha\beta}\,=\,
\mathcal{X}\,T_{\mu\nu}
\end{eqnarray}
The trace of field equations
(\ref{fieldequationFOG}) is the following

\begin{eqnarray}\label{tracefieldequationFOG}
H\,=\,g^{\alpha\beta}H_{\alpha\beta}\,=\,f_XR+2f_YR_{\alpha\beta}R^{\alpha\beta}+2f_ZR_{\alpha\beta\gamma\delta}
R^{\alpha\beta\gamma\delta}-2f+\Box[3
f_X+f_YR]+2[(f_Y+2f_Z)R^{\alpha\beta}]_{;\alpha\beta}\,=\,\mathcal{X}\,T
\end{eqnarray}
where $T\,=\,T^{\sigma}_{\,\,\,\,\,\sigma}$ is the trace of
energy-momentum tensor. Moreover the (\ref{fieldequationFOG})
satisfies the condition
$H^{\alpha\mu}_{\,\,\,\,\,\,\,\,;\alpha}\,=\,\mathcal{X}\,T^{\alpha\mu}_{\,\,\,\,\,\,\,\,;\alpha}\,=\,0$.
In fact it is easy to check (in the case of $f(R)$-Gravity) that

\begin{eqnarray}
H^{\alpha\mu}_{\,\,\,\,\,\,\,\,;\alpha}\,=&&f'_{;\alpha}R^{\alpha\mu}+f'R^{\alpha\mu}_{\,\,\,\,\,\,\,\,;
\alpha}-\frac{1}{2}f^{;\mu}-f'^{;\alpha\mu}_{\,\,\,\,\,\,\,\,\,\,\,\,\alpha}+f'^{;\alpha\,\,\,\,\,\,\mu}_{\,\,\,\,\,\,\,\,
\alpha}\,=\,f''R^{\alpha\mu}R_{;\alpha}-f'^{;\alpha\mu}_{\,\,\,\,\,\,\,\,\,\,\,\,\alpha}+f'^{;\alpha\,\,\,\,\,\,
\mu}_{\,\,\,\,\,\,\,\,\alpha}\,=\nonumber\\&&f''R^{\alpha\mu}R_{;\alpha}-f'^{;\alpha}R^{\,\,\,\,\,\mu}_{\alpha}\,=\,f''R^{\alpha\mu}R_{;\alpha}-
f''R^{;\alpha}R^{\,\,\,\,\,\mu}_{\alpha}\,=\,0
\end{eqnarray}
where we used the properties
$G^{\alpha\mu}_{\,\,\,\,\,\,\,\,\,;\alpha}\,=\,0$ and
$[\nabla^\mu,\nabla_\alpha]f'^{;\alpha}\,=\,-f'^{;\alpha}R_{\alpha}^{\,\,\,\,\mu}$.

\subsection{Conformal transformations}
The above theories can be easily viewed as scalar-tensor theories
of gravity (for a comprehensive discussion see
\cite{theo_aspect_6}). Let us now introduce conformal
transformations to show that any scalar-tensor theory, in absence
of ordinary matter, e.g. a perfect fluid, is conformally
equivalent to an Einstein theory plus minimally coupled scalar
fields. If standard matter is present, conformal transformations
allow to transfer non-minimal coupling to the matter component
\cite{mag-sok,cqg}. A general non-minimally coupled scalar-tensor
theory is

\begin{eqnarray}\label{TSaction}
\mathcal{A}^{ST}\,=\,\int d^4x \sqrt{-g}[F(\phi)R+\omega(\phi)\phi_{;\alpha}\phi^{;\alpha}+
V(\phi)+\mathcal{X}\mathcal{L}_m]
\end{eqnarray}
where $V(\phi)$ and $F(\phi)$ are generic functions describing
 the self-interaction potential and the coupling of a
scalar field $\phi$, respectively. The Brans-Dicke theory of
gravity is a particular case of the action (\ref{TSaction}) in
which  $V(\phi)\,=\,0$ and
$\omega(\phi)\,=\,-\frac{\omega_{BD}}{\phi}$. In fact we have

\begin{eqnarray}\label{BD-action}
\mathcal{A}^{BD}=\int d^4x\sqrt{-g}\left[\phi
R-\omega_{BD}\frac{\phi_{;\alpha}\phi^{;\alpha}}{\phi}+\mathcal{X}
\mathcal{L}_m\right]
\end{eqnarray}
The variation of (\ref{TSaction}) with respect to $g_{\mu\nu}$ and
$\phi$ gives the second-order field equations

\begin{eqnarray}\label{TSfieldequation}
\left\{\begin{array}{ll}
F(\phi)G_{\mu\nu}-\frac{1}{2}V(\phi)g_{\mu\nu}+\omega(\phi)\biggl[\phi_{;\mu}\phi_{;\nu}-\frac{1}{2}\phi_{;\alpha}\phi^
{;\alpha}g_{\mu\nu}\biggr]-F(\phi)_{;\mu\nu}+g_{\mu\nu}\Box
F(\phi)\,=\,\mathcal{X}\,T_{\mu\nu}
\\\\
2\omega(\phi)\Box\phi-\omega_{,\phi}(\phi)\phi_{;\alpha}\phi^{;\alpha}-[F(\phi)R+V(\phi)]_{,\phi}\,=\,0
\\\\
3\Box F(\phi)-F(\phi)R-2V(\phi)-\omega(\phi)\phi_{;\alpha}\phi^{;\alpha}\,=\,\mathcal{X}\,T
\\\\
2\omega(\phi)\Box\phi+3\Box F(\phi)-[\omega_{,\phi}(\phi)+\omega(\phi)]\phi_{;\alpha}\phi^{;\alpha}-[F(\phi)R+V(\phi)]_{,\phi}-F(\phi)R-2V(\phi)=\mathcal{X}\,T
\end{array} \right.
\end{eqnarray}
The third equation in (\ref{TSfieldequation}) is the trace of the
field equation for $g_{\mu\nu}$ and the last  is a combination of
the trace and of the field equation for $\phi$.

The conformal transformation on the metric $g_{\mu\nu}$ is

\begin{eqnarray}\label{transconf}\tilde{g}_{\mu\nu}\,=\,A(x^\lambda)g_{\mu\nu}\end{eqnarray}
with $A(x^\lambda)>0$. $A$ is the conformal factor. Obviously the
transformation rule for the contravariant metric tensor is
$\tilde{g}^{\mu\nu}=A^{-1}g^{\mu\nu}$. The  mathematical
quantities in the  Einstein frame (EF) (quantities referred to
$\tilde{g}_{\mu\nu}$) are linked to the ones in the  Jordan Frame
(JF) (quantities referred to $g_{\mu\nu}$ and action
(\ref{TSaction})) as follows

\begin{eqnarray}
\left\{\begin{array}{ll}
\tilde{\Gamma}^\alpha_{\mu\nu}=\Gamma^\alpha_{\mu\nu}+\phi_{,\mu}\delta^\alpha_\nu+\phi_{,\nu}\delta^\alpha_\mu
-\phi^{,\alpha}g_{\mu\nu}
\\\\
\tilde{R}^{\alpha}_{\,\,\,\,\,\mu\beta\nu}\,=\,R^{\alpha}_{\,\,\,\,\,\mu\beta\nu}-\delta^\alpha_\beta(\phi_{;\mu\nu}-\phi_{;\mu}
\phi_{;\nu}+g_{\mu\nu}\phi^{;\sigma}\phi_{;\sigma})+\delta^\alpha_\nu(\phi_{;\mu\beta}-\phi_{;\mu}\phi_{;\beta}+g_{\mu\beta}\phi^{;\sigma}\phi_
{;\sigma})\\\\\,\,\,\,\,\,
\,\,\,\,\,\,\,\,\,\,\,\,\,\,\,\,\,\,\,\,-g_{\mu\nu}(\phi^{;\alpha}_{\,\,\,\,\,\beta}
-\phi^{;\alpha}\phi_{;\beta})+g_{\mu\beta}(\phi^{;\alpha}_{\,\,\,\,\,\nu}
-\phi^{;\alpha}\phi_{;\nu})
\\\\
\tilde{R}_{\mu\nu}=R_{\mu\nu}-2\phi_{;\mu\nu}+2\phi_{;\mu}\phi_
{;\nu}-g_{\mu\nu}\Box\phi-2g_{\mu\nu}\phi_{;\sigma}\phi^{;\sigma}\\\\
\tilde{R}=e^{-2\phi}(R-6\Box\phi-6\phi_{;\sigma}\phi^{\sigma})
\\\\
\tilde{\phi_{;\mu\nu}}\,=\,\phi_{,\mu\nu}-\tilde{\Gamma}^\sigma_{\alpha\beta}\phi_{,\sigma}\,=\,\phi_{;\mu\nu}-2\phi_{;\mu}
\phi_{;\nu}+g_{\mu\nu}\phi^{;\sigma}\phi_{;\sigma}
\\\\
\tilde{G}_{\mu\nu}\,=\,G_{\mu\nu}-2\phi_{;\mu\nu}+2\phi_{;\mu}\phi_{;\nu}+2g_{\mu\nu}\,\Box\phi+
g_{\mu\nu}\phi^{;\sigma}\phi_{;\sigma}
\end{array} \right.
\end{eqnarray}
where $\phi\,\doteq\,\ln A^{1/2}$.  The inverse relations are

\begin{eqnarray}
\left\{\begin{array}{ll}
\Gamma^\alpha_{\mu\nu}=\tilde{\Gamma}^\alpha_{\mu\nu}-\phi_{,\mu}\delta^\alpha_\nu-\phi_{,\nu}\delta^\alpha_\mu
+\tilde{\phi^{,\alpha}}\tilde{g}_{\mu\nu}
\\\\
R^{\alpha}_{\,\,\,\,\,\mu\beta\nu}\,=\,\tilde{R}^{\alpha}_{\,\,\,\,\,\mu\beta\nu}+\delta^\alpha_\beta(\tilde{\phi_{;\mu\nu}}+
\phi_{;\mu}\phi_{;\nu})-\delta^\alpha_\nu(\tilde{\phi_{;\mu\beta}}+\phi_{;\mu}\phi_{;\beta})+\tilde{g}_{\mu\nu}(\tilde{\phi^{;\alpha}_
{\,\,\,\,\,\,\,\beta}}-\tilde{\phi^{;\alpha}}\phi_{;\beta})-
\tilde{g}_{\mu\beta}(\tilde{\phi^{;\alpha}_{\,\,\,\,\,\,\,\nu}}-\tilde{\phi^{;\alpha}}\phi_{;\nu})
\\\\
R_{\mu\nu}=\tilde{R}_{\mu\nu}+2\tilde{\phi_{;\mu\nu}}+2\phi_{;\mu}\phi_{;\nu}+\tilde{g}_{\mu\nu}\,\tilde{\Box}\phi-2\tilde{g}_
{\mu\nu}\tilde{\phi^{;\sigma}\phi_{;\sigma}}\\\\
R=e^{2\phi}(\tilde{R}+6\tilde{\Box}\phi-6\tilde
{\phi^{;\sigma}\phi_{;\sigma}})
\\\\
\phi_{;\mu\nu}\,=\,\phi_{,\mu\nu}-\Gamma^\sigma_{\alpha\beta}\phi_{,\sigma}\,=\,\tilde{\phi_{;\mu\nu}}+2\phi_{;\mu}\phi_{;\nu}
-\tilde{g}_{\mu\nu}\tilde{\phi^{;\sigma}\phi_{;\sigma}}
\\\\
G_{\mu\nu}\,=\,\tilde{G}_{\mu\nu}+2\tilde{\phi_{;\mu\nu}}+2\phi_{;\mu}\phi_{;\nu}-2\tilde{g}_{\mu\nu}\,\tilde{\Box}\phi\,
+\tilde{g}_{\mu\nu}\tilde
{\phi^{;\sigma}\phi_{;\sigma}}
\end{array} \right.
\end{eqnarray}
where $\Box$ and $\tilde{\Box}$ are the d'Alembert operators  with
respect to the metric $g_{\mu\nu}$ and $\tilde{g}_{\mu\nu}$. The
transformation between the d'Alembert operators is
$\Box\,=\,e^{2\phi}\tilde{\Box}-2\phi^{;\nu}\partial_{\nu}$.

Under these transformations, the action (\ref{TSaction}) can be
reformulated as follows

\begin{eqnarray}\label{TS-EF-action}
\mathcal{A}_{EF}^{ST}\,=\,\int d^4x \sqrt{-\tilde{g}}
\biggl[\Lambda\,\tilde{R}
+\Omega(\varphi)\varphi_{;\alpha}\varphi^{;\alpha}+W(\varphi)+\mathcal{X}
\tilde{\mathcal{L}}_{m}\biggr]
\end{eqnarray}
in which $\tilde{R}$ is the Ricci scalar relative to the metric
$\tilde{g}$ and $\Lambda$ is a generic constant. The relations
between the quantities in two frames are

\begin{eqnarray}\label{transconfTS}
\left\{\begin{array}{ll}\Omega(\varphi){d\varphi}^2\,=\,\Lambda\biggl[\frac{\omega(\phi)}{F(\phi)}-\frac{3}{2}\biggl(\frac{
d\ln F(\phi)}{d\phi}\biggr)^2\biggr]{d\phi}^2\\\\W(\varphi)=\frac{\Lambda^2}{F(\phi(\varphi))^2}V(\phi(\varphi))\\\\
\tilde{\mathcal{L}}_m=\frac{\Lambda^2}{F(\phi(\varphi))^2}\mathcal{L}_m\biggl(\frac{\Lambda\,\tilde{g}_{\rho\sigma}}{F(\phi(
\varphi))}\biggr)\\\\F(\phi){A(x^\lambda)}^{-1}=\Lambda\end{array}\right.
\end{eqnarray}
The field equations for the new fields $\tilde{g}_{\mu\nu}$ and
$\varphi$ are

\begin{eqnarray}\label{fe-EF-conformal-transformed}
\left\{\begin{array}{ll}
\Lambda\tilde{G}_{\mu\nu}-\frac{1}{2}W(\varphi)\tilde{g}_{\mu\nu}+\Omega(\varphi)\biggl[\varphi_{;\mu}\varphi_{;\nu}-
\frac{1}{2}\varphi_{;\alpha}\varphi^{;\alpha}\tilde{g}_{\mu\nu}\biggr]=\mathcal{X}\,\tilde{T}^\varphi_{\mu\nu}
\\\\
2\Omega(\varphi)\tilde{\Box}\varphi-\Omega_{,\varphi}(\varphi)\varphi_{;\alpha}\varphi^{;\alpha}-W_{,\varphi}(\varphi)=
\mathcal{X}\tilde{\mathcal{L}}_{m,\varphi}
\\\\
\tilde{R}=-\frac{1}{\Lambda}\biggl(\mathcal{X}\tilde{T}^\varphi+2W(\varphi)+\Omega(\varphi)\tilde{g}^{\sigma\tau}\varphi_
{;\sigma}\varphi_{;\tau}\biggr)
\end{array} \right.
\end{eqnarray}

Therefore, any non-minimally coupled scalar-tensor theory, in
absence of ordinary matter,  is conformally equivalent to an
Einstein theory, being the conformal transformation and the
potential suitably defined by (\ref{transconfTS}). The converse is
also true: for a given $F(\phi)$, such that the relations
(\ref{transconfTS})hold, we can transform a standard Einstein
theory plus scalar field into a non-minimally coupled
scalar-tensor theory. This means that, in principle, if we are
able to solve the field equations in the framework of the Einstein
theory in presence of a scalar field with a given potential, we
should be able to get  solutions for the scalar-tensor theories,
assigned by the coupling $F(\phi)$, via the conformal
transformation (\ref{transconf}) with the constraints given by
(\ref{transconfTS}). Following the standard terminology, the
``Einstein frame'' is the framework of the Einstein theory with
the minimal coupling and the ``Jordan frame'' is the framework of
the non-minimally coupled theory \cite{mag-sok,cqg}. These
considerations will be extremely useful for the discussion below.

\section{Spherical symmetry}\label{perturb}
Spherical symmetry is the first step necessary  to develop the
Newtonian and the post-Newtonian limits of any alternative theory
of gravity. Here, we will discuss the basic features related to
such a symmetry starting from GR and then we develop a
perturbation approach for spherically symmetric solutions of
$f(R)$-gravity. This machinery will be extremely useful to develop
the FOG weak-field limit.

\subsection{Generalities on spherical symmetry}

Since we are interested to understand the modifications of GR
predictions  when one takes into account  concentrations of matter
in the space, it is fundamental to discuss  properties of the
metric $g_{\mu\nu}$. As a first step,  we will study the
gravitational potential generated by spherically symmetric
distributions of matter. The most general spherically symmetric
metric\footnote{The metric is spherically symmetric if it depends
only on $\textbf{x}$ and $d\textbf{x}$ only through the rotational
invariants $d\textbf{x}^2$, $\textbf{x}\cdot d\textbf{x}$ and
$\textbf{x}^2$.} can be written as

\begin{eqnarray}\label{me0}
{ds}^2\,=\,g_1(t,|\textbf{x}|)\,dt^2+g_2(t,|\textbf{x}|)\,dt\,\textbf{x}\cdot
d\textbf{x}+g_3(t,|\textbf{x}|) (\textbf{x}\cdot
d\textbf{x})^2+g_4(t,|\textbf{x}|){d|\textbf{x}|}^2
\end{eqnarray}
where $g_i$ are functions of the spatial distance $|\textbf{x}|$
and of the time $t$. The set of coordinates is
$x^\mu\,=\,(t,x^1,x^2,x^3)$. The scalar product is defined as
$\textbf{x}\cdot d\textbf{x}=x^1dx^1+x^2dx^2+x^3dx^3$. By the
spherically symmetric form of (\ref{me0}), it is convenient to
replace $\textbf{x}$ with spherical  coordinates $r, \theta, \phi$
defined as

\begin{eqnarray}\label{polar}
x^1\,=\,r \sin\theta\cos\phi\,,\,\,\,\,x^2\,=\,r \sin\theta\sin\phi\,,\,\,\,\,x^3\,=\,r\cos\phi
\end{eqnarray}
The line element (\ref{me0}) then becomes

\begin{eqnarray}\label{me1}
{ds}^2\,=\,g_1(t,r)\,dt^2+rg_2(t,r)\,dtdr+r^2g_3(t,r)
dr^2+g_4(t,r)(dr^2+r^2d\Omega)
\end{eqnarray}
where $d\Omega=d\theta^2+\sin^2\theta d\phi^2$ is the solid angle.
We are free to reset our clocks by defining the time coordinate

\begin{eqnarray}
t\,=\,t'+\epsilon(t',r)
\end{eqnarray}
with $\epsilon(t',r)$ an arbitrary function of $t'$ and $r$. This
allows us to eliminate the off-diagonal element $g_{tr}$ in the
metric (\ref{me1}) by setting

\begin{eqnarray}
\frac{d\epsilon(t',r)}{dr}\,=\,-\frac{rg_2(t',r)}{2g_1(t',r)}
\end{eqnarray}
The metric (\ref{me1}) becomes

\begin{eqnarray}\label{me2}
{ds}^2\,=\,g_1(t',r)\biggl[1+\frac{d\epsilon(t',r)}{dt'}\biggr]^2\,dt'^2+\biggl[r^2g_3(t',r)-\frac{r^2g_2(t',r)^2}{4g_1
(t',r)}+g_4(t',r)\biggr]{dr}^2+g_4(t',r)r^2d\Omega
\end{eqnarray}
where  introducing a new metric coefficients $g_{tt}(t',r)$,
$g_{rr}(t',r)$ and $g_{\Omega\Omega}(t',r)$, we can recast Eq.
(\ref{me2}) as follows

\begin{eqnarray}\label{me3}
{ds}^2\,=\,g_{tt}(t',r)\,dt'^2-g_{rr}(t',r){dr}^2-g_{\Omega\Omega}(t',r)d\Omega
\end{eqnarray}
Introducing a new radial coordinate $(r')$ by considering the
further transformation

\begin{eqnarray}
r'\,=\,(const)e^{\int dr\sqrt{\frac{g_{rr}(t',r)}{g_{\Omega\Omega}(t',r)}}}
\end{eqnarray}
it is possible to recast Eq.(\ref{me3}) into the isotropic form
(isotropic coordinates)

\begin{eqnarray}\label{me4}
{ds}^2\,=\,g_{tt}(t',r')\,dt^2-g_{ij}(t',r')dx^idx^j
\end{eqnarray}
and then it is possible  to choose
$g_{\Omega\Omega}(t',r)\,=\,r''^2$ (this condition allows  to
obtain the standard definition of the circumference with radius
$r''$) and to have the metric (\ref{me3}) in the standard form
(standard coordinates)

\begin{eqnarray}\label{me5}
{ds}^2\,=\,g_{tt}(t',r'')\,dt^2-g_{rr}(t',r''){dr''}^2-r''^2d\Omega
\end{eqnarray}
Obviously the functions $g_{tt}(t',r'')$ and $g_{rr}(t',r'')$ are
not the same of (\ref{me3}). If we suppose
$g_{ij}(t',r')=Y(t',r')\delta_{ij}$, it is worth noticing that it
is possible to pass from (\ref{me4}) to (\ref{me5}) by the
coordinate transformation

\begin{eqnarray}\label{transf-stan-isotr-coord}
r'\,=\,r'(r'')\,=\,(const)e^{\int dr'' \frac{\sqrt{\tilde{Y}(r'')}}{r''}}
\end{eqnarray}
We can, then, state that the expressions (\ref{me3}), (\ref{me4})
and (\ref{me5}) are equivalent to the metric (\ref{me0}) and we
can consider them without loss of generality as the most general
definitions of a spherically symmetric metric compatible with a
pseudo\,-\,Riemannian manifold without torsion. The choice of the
metric form is only a practical issue useful to develop
calculations.

\subsection{The Birkhoff theorem in General Relativity}

The Birkhoff theorem is an important result of GR which
essentially states that stationary solutions are also static and
stable. Specifically, the theorem holds for spherically symmetric
solutions.

Let us start our considerations by  rewriting the metric
(\ref{me5}) as follows

\begin{eqnarray}
ds^2\,=\,e^{\nu(t,r)}\,dt^2-e^{\mu(t,r)}{dr}^2-r^2d\Omega
\end{eqnarray}
where we redifined the radial coordinate. The only non-vanishing
components of metric tensor $g_{\mu\nu}$ are

\begin{eqnarray}
g_{tt}\,=\,e^{\nu(t,r)}\,,\,\,\,\,\,\,\,g_{rr}\,=\,-e^{\mu(t,r)}\,,\,\,\,\,\,\,\,g_{\theta\theta}\,=\,-r^2
\,,\,\,\,\,\,\,\,g_{\phi\phi}\,=\,-r^2\sin^2\theta
\end{eqnarray}
with functions $\mu(t,r)$ and $\nu(t,r)$ that are to be determined
by solving the field equations in GR (\ref{fieldequationGR}).
Since $g_{\mu\nu}$ is diagonal, it is easy to write  all the
non-vanishing inverse components:

\begin{eqnarray}
g^{tt}\,=\,e^{-\nu(t,r)}\,,\,\,\,\,\,\,\,g^{rr}\,=\,-e^{-\mu(t,r)}\,,\,\,\,\,\,\,\,g^{\theta\theta}\,=\,
-r^{-2}\,,\,\,\,\,\,\,\,g^{\phi\phi}\,=\,-r^{-2}\sin^{-2}\theta
\end{eqnarray}
Furthermore, the determinant of the metric tensor is

\begin{eqnarray}
g\,=\,-e^{\mu(t,r)+\nu(t,r)}r^4\sin^2\theta
\end{eqnarray}
so the invariant volume element is

\begin{eqnarray}
\sqrt{-g}\,dr\,d\theta\,d\phi\,=\,r^2e^{-\frac{\mu(t,r)+\nu(t,r)}{2}}\,\sin\theta\,dr\,d\theta\,d\phi
\end{eqnarray}
The only non-vanishing components of  Christoffel symbols
(\ref{christoffel}) are

\begin{eqnarray}
\left\{\begin{array}{ll} \begin{array}{ccccc}
  \Gamma^{t}_{tt}\,=\,\frac{\dot{\nu}(t,r)}{2} & \,\,\, & \Gamma^r_{rr}\,=\,\frac{\mu'(t,r)}{2} & \,\,\, & \Gamma^r_{tt}
  \,=\,\frac{\nu'(t,r)}{2}e^{\nu(t,r)-\mu(t,r)} \\
  & & & \\
  \Gamma^t_{rr}\,=\,\frac{\mu'(t,r)}{2}e^{\mu(t,r)-\nu(t,r)}& \,\,\, & \Gamma^t_{tr}\,=\,\frac{\nu'(t,r)}{2} & \,\,\, &
  \Gamma^r_{tr}\,=\,\frac{\nu(t,r)}{2} \\
  & & & \\
  \Gamma^r_{\theta\theta}\,=\,-r\,e^{-\mu(t,r)}& \,\,\, & \Gamma^\theta_{r\theta}\,=\,\Gamma^\phi_{r\phi}\,=\,\frac{1}{r}
   & \,\,\, &
  \Gamma^r_{\phi\phi}\,=\,-r\,e^{-\mu(t,r)}\,\sin^2\theta \\
  & & & \\
  \Gamma^\theta_{\phi\phi}\,=\,-\sin\theta\,\cos\theta& \,\,\, & \Gamma^\phi_{\theta\phi}\,=\cot\theta & \,\,\, & \\
  \end{array}
\end{array}\right.
\end{eqnarray}
and the field equations (\ref{fieldequationGR}) become

\begin{eqnarray}\label{fe-schwarz-sol}
\left\{\begin{array}{ll}
  \frac{1}{r^2}-e^{-\mu(t,r)}\biggl[\frac{1}{r^2}-\frac{\mu'(t,r)}{r}\biggr]\,=\,\mathcal{X}\,T^t_{\,\,\,t} \\
  \\
  \frac{\dot{\mu}(t,r)}{r}e^{-\mu(t,r)}\,=\,\mathcal{X}\,T^r_{\,\,\,\,t} \\
  \\
  \frac{1}{r^2}-e^{-\mu(t,r)}\biggl[\frac{\nu'(t,r)}{r}+\frac{1}{r^2}\biggr]\,=\,\mathcal{X}\,T^r_{\,\,\,\,r} \\
  \\
  \frac{e^{-\nu(t,r)}}{2}\biggl[\ddot{\mu}(t,r)+\frac{\dot{\mu}^2(t,r)}{2}-\frac{\dot{\mu}(t,r)\,\dot{\nu}(t,r)}{2}\biggr]
  \\
  \,\,\,\,\,\,\,\,\,\,\,\,\,\,\,\,-\frac{e^{-\mu(t,r)}}{2}\biggl[\nu''(t,r)+\frac{\nu'^2(t,r)}{2}+\frac{\nu'(t,r)-\mu'(t,r)}
  {r}-\frac{\nu'(t,r)\,\mu'(t,r)}{2}\biggr]\,=\,\mathcal{X}\,T^\theta_{\,\,\,\,\theta}\,=\,\mathcal{X}\,T^\phi_{\,\,\,\,
  \phi}
\end{array}\right.
\end{eqnarray}
If we suppose a stress-energy tensor  (\ref{perfectfluid}) induced
by a point-like source with mass $M$

\begin{eqnarray}\label{point_like}
\left\{\begin{array}{ll}
T_{\mu\nu}\,=\,\rho(\mathbf{x})u_\mu u_\nu
\\\\
T\,=\,\rho(\mathbf{x})\end{array}\right.
\end{eqnarray}
where $\rho\,=\,M\,\delta(\textbf{x})$, we obtain the
\emph{Schwarzschild solution} in standard
coordinates\footnote{$\delta(\mathbf{x})$ is the 3-dimensional
Dirac $\delta$-function.}

\begin{eqnarray}\label{schwarz-solution-stand-coord}
ds^2\,=\,\biggl[1-\frac{r_g}{r''}\biggr]dt^2-\frac{dr''^2}{1-\frac{r_g}
{r''}}-r''^2d\Omega
\end{eqnarray}
where $r_g\,=\,2GM$ is the  \emph{Schwarzschild radius}.

Metric (\ref{schwarz-solution-stand-coord}) determines completely
the gravitational field in the vacuum generated by a spherical
matter density distribution. Furthermore the Schwarzschild
solution is valid also when we consider a moving source with a
spherical distribution. The spatial metric is determined by
expression of spatial distance element

\begin{eqnarray}
dl^2\,=\,\frac{dr''^2}{1-\frac{r_g}
{r''}}+r''^2d\Omega
\end{eqnarray}
We have to note that, while the length of circumference with "radius" $r''$ is the usual one
$2\pi r''$, the distance between two points on the same radius is
given by the integral

\begin{eqnarray}\label{dist-curved-space}
\int_{r''_1}^{r''_2}\frac{dr''}{\sqrt{1-\frac{r_g}{r''}}}\,>\,r''_2-r''_1
\end{eqnarray}
this means that the space is curved. Besides we note that
$g_{tt}\,\leq\,1$, then, by the relation between the time
coordinate $t$ and the proper time $\tau$
($d\tau\,=\,\sqrt{g_{tt}}\,dt$), we get the condition

\begin{eqnarray}\label{proper-time}
d\tau\leq dt
\end{eqnarray}
At infinity, the  coordinate time coincides with the physical
time. We can state that when we are at a finite distance from the
the mass, there is a slowdown of the time with respect to the time
measured at infinity.

In presence of matter the situation is the following. From the
first equation in the (\ref{fe-schwarz-sol}), when $r\rightarrow
0$, $\mu(t,r)$ has to vanish as $r^2$; otherwise $T^t_{\,\,\,\,t}$
could have a singular point in the origin. By formally integrating
the equation with the condition $\mu(t,r)|_{r\,=\,0}\,=\,0$, one
gets

\begin{eqnarray}
\mu(t,r)\,=\,-\ln\biggl[1-\frac{\mathcal{X}}{r}\int_0^rT^t_{\,\,\,\,t}\,\hat{r}^2\,d\hat{r}\biggr]
\end{eqnarray}
Beside the point-like case, it is easy to demonstrate that the
proprieties (\ref{dist-curved-space}), (\ref{proper-time}) and
$\mu(t,r)+\nu(t,r)\,\leq\,0$ hold  also in more general matter
distributions with spherical symmetry \cite{landau}. If the
gravitational field is generated by a spherical body with "radius"
$\xi$, we have $T^t_{\,\,\,\,t}\,=\,0$ outside the body
($r\,>\,\xi$) and we can write

\begin{eqnarray}
\mu_\xi(t,r)\,=\,-\ln\biggl[1-\frac{\mathcal{X}}{r}\int_0^\xi T^t_{\,\,\,\,t}\,\hat{r}^2\,d\hat{r}\biggr]
\end{eqnarray}
 obtaining the analogous expression of
(\ref{schwarz-solution-stand-coord}) in the matter

\begin{eqnarray}\label{schwarz-solution-stand-coord-matter}
ds^2\,=,\biggl[1-\frac{r_g(r'')}{r''}\biggr]dt^2-\frac{dr''^2}{1-
\frac{r_g(r'')}{r''}}-r''^2d\Omega
\end{eqnarray}
where we have introduced the Schwarzschild radius related to the
quantity of matter included in the sphere with radius $r''$:

\begin{eqnarray}
r_g(r'')\,=\,\mathcal{X}\int_0^{r''}T^t_{\,\,\,\,t}\,\hat{r}^2\,d\hat{r}
\end{eqnarray}
Obviously when the distance is bigger than the radius of the body,
the metric (\ref{schwarz-solution-stand-coord-matter}) is equal to
(\ref{schwarz-solution-stand-coord}).

If we consider the transformation (\ref{transf-stan-isotr-coord}),
which in the case of Schwarzschild solution is

\begin{eqnarray}
r'\,=\,\frac{2r''-r_g+2\sqrt{r''^2-r_g\,r''}}{4}
\end{eqnarray}
it is possible to obtain the Schwarzschild solution
(\ref{schwarz-solution-stand-coord}) in isotropic coordinates

\begin{eqnarray}\label{schwarz-solution-isot-coord}
ds^2\,=\,\biggl[\frac{1-\frac{r_g}{4r'}}{1+\frac{r_g}{4r'}}
\biggr]^2dt^2-\biggl[1+\frac{r_g}{4r'}\biggr]^{4}(dr'^2+r'^2d\Omega)
\end{eqnarray}

In both cases,  solutions (\ref{schwarz-solution-stand-coord}) and
(\ref{schwarz-solution-stand-coord-matter}) obviously satisfy the
trace  of field equations: $R\,=\,-\mathcal{X}\,T$. Since in the
vacuum the trace of matter tensor is vanishing (except in the
origin, where the trace is proportional to $\delta(\textbf{x})$)
we can state that the Schwarzschild solution is "Ricci flat":
$R\,=\,0$.

If we add in the Hilbert - Einstein Lagrangian
(\ref{HElagrangian}) a term like ($-2\sqrt{-g}\,\Lambda$), with
$\Lambda$ a generic constant, the field equations
(\ref{fieldequationGR}) are modified as follows

\begin{eqnarray}\label{fe-equationGR-lambda}
G_{\mu\nu}+\Lambda g_{\mu\nu}\,=\,\mathcal{X}\,T_{\mu\nu}
\end{eqnarray}
In the case of  a point-like source, we find the
\emph{Schwarzschild - de Sitter solution}

\begin{eqnarray}\label{schwarz-desitt-solution}
ds^2\,=\,\biggl[1-\frac{r_g}{r''}+\frac{\Lambda}{3}r''^2\biggr]dt^2-\frac{
dr''^2}{1-\frac{r_g}{r''}+\frac{\Lambda}{3}r''^2}-r''^2d\Omega
\end{eqnarray}
The trace of (\ref{fe-equationGR-lambda}) is

\begin{eqnarray}
R\,=\,4\Lambda-\mathcal{X}\,T
\end{eqnarray}
from which we note that this solution does not admit solution in
vacuum, since also in absence of ordinary matter
($T_{\mu\nu}\,=\,0$) we have a non-vanishing scalar curvature. The
contribution is given by the cosmological constant $\Lambda$. Also
in this case,   analogous considerations as in
(\ref{schwarz-solution-stand-coord-matter}) hold.

Finally let us consider as source a radial and static electric
field $\textbf{E}\,=\,Q\,\textbf{x}\,/|\textbf{x}|^3$. We know
that the Lagrangian of electromagnetic field is
$-\frac{1}{4\pi}F_{\alpha\beta}F^{\alpha\beta}$ where
$F_{\alpha\beta}$ is the electromagnetic tensor. Then, the Hilbert
- Einstein Lagrangian is

\begin{eqnarray}
\mathcal{L}_{HE}\,=\,\sqrt{-g}(R-\frac{1}{4\pi}\,F_{\alpha\beta}F^{\alpha\beta})
\end{eqnarray}
and the Einstein equation (\ref{fieldequationGR}) becomes

\begin{eqnarray}
G_{\mu\nu}\,=\,-\frac{1}{8\pi}\,(g_{\mu\nu}F_{\alpha\beta}F^{\alpha\beta}-4F_{\mu\alpha}F^\alpha_{\,\,\,\,\nu})
\end{eqnarray}
The solution for a spherically symmetric
system is the \emph{Reissner - Nordstrom solution}

\begin{eqnarray}\label{reis-nord-solution}
ds^2\,=\,\biggl[1-\frac{r_g}{r''}+\frac{Q^2}{r''^2}\biggr]dt^2-\frac{
dr''^2}{1-\frac{r_g}{r''}+\frac{Q^2}{r''^2}}-r''^2d\Omega
\end{eqnarray}

In all the above cases, the Birkhoff theorem holds: \emph{The
metric tensor generated in vacuum by a matter density distribution
with a spherical symmetry is time-independent}. Also a
time-dependent source with a spherical symmetry produces a static
metric. The curvature of spacetime in the matter, at a distance
$r$ from the origin, is proportional only to the matter inside the
sphere of radius $r$. This conclusion is compatible with the Gauss
theorem of Classical Mechanics.

\subsection{The Schwarzschild solution and the Eddington parameterization}

The Schwarzschild radius $r_g$ in the solution
(\ref{schwarz-solution-isot-coord}) is a scale-length induced by
theory and represents a natural parameter to study the
gravitational interaction with respect to the generating source.
At radial distance $r'$ where $r_g/r'\ll 1$, solution
(\ref{schwarz-solution-isot-coord}) becomes

\begin{eqnarray}\label{schwarz-isotropic-pertur}
ds^2\simeq\biggl[1-\frac{r_g}{r'}+\frac{1}{2}\biggl(\frac{r_g}{r'}\biggr)^2+\dots\biggr]dt^2-\biggl[1+
\frac{r_g}{r'}+\dots\biggr]\biggl[dr'^2+r'^2d\Omega\biggr]
\end{eqnarray}
that is the standard situation presented by any self-gravitating
system far from its critical radius.

Since we are interested to investigate the deviations, induced by
FOG, from the behavior (\ref{schwarz-isotropic-pertur}). it is
useful to introduce the method taking into account such deviations
with respect to GR. The standard approach is the
Parameterized-Post-Newtonian (PPN) expansion of the Schwarzschild
metric (\ref{schwarz-solution-isot-coord}). Eddington
parameterized deviations with respect to GR considering a Taylor
expansion in term of $r_g/r'$ and assumed that, in Solar System,
the limit $r_g/r'\ll 1$ holds \cite{will}. The resulting metric is

\begin{eqnarray}\label{schwarz-isotropic-PPN}
ds^2\simeq\biggl[1-\alpha\frac{r_g}{r'}+\frac{\beta}{2}\biggl(\frac{r_g}{r'}\biggr)^2+\dots\biggr]dt^2-\biggl[1+\gamma
\frac{r_g}{r'}+\dots\biggr]\biggl[dr'^2+r'^2d\Omega\biggr]
\end{eqnarray}
where $\alpha$, $\beta$ and $\gamma$ are  dimensionless parameters
(Eddington parameters) which parameterize deviations with respect
to GR. The reason to carry out this expansion up to the order
$(r_g /r')^2$ in $g_{tt}$ and only to the order $(r_g/r')$ in
$g_{ij}$ is that, in applications to Celestial Mechanics, $g_{ij}$
always appears multiplied by an extra factor $v^2\backsim M/r')$.
It is evident that the standard GR solution for a spherically
symmetric gravitational system in vacuum, is obtained for
$\alpha\,=\,\beta\,=\,\gamma\,=\,1$ giving again the
"approximated" Schwarzschild solution
(\ref{schwarz-isotropic-pertur}). Actually, the parameter $\alpha$
can be settled to the unity due to the mass definition of the
system itself \cite{will}. As a consequence, the expanded metric
(\ref{schwarz-isotropic-PPN}) can be recast in the form\,:

\begin{eqnarray}
ds^2\simeq\biggl[1-\frac{r_g}{r''}+\frac{\beta-\gamma}{2}\biggl(\frac{r_g}{r''}\biggr)^2+\dots\biggr]dt^2-\biggl[1+\gamma
\frac{r_g}{r''}+\dots\biggr]dr''^2-r''^2d\Omega
\end{eqnarray}
where we have restored the standard spherical coordinates by means
of the transformation
$r''\,=\,r'\biggl[1+\frac{r_g}{4r'}\biggr]^2$. The  parameters
$\beta,\,\gamma$ have a physical interpretation. The parameter
$\gamma$ measures the amount of curvature of space generated by a
body of mass $M$ at radius $r'$. In fact, the spatial components
of the Riemann curvature tensor are, at post-Newtonian order,

\begin{eqnarray}
R_{ijkl}=\frac{3}{2}\gamma\frac{r_g}{r'^3}N_{ijkl}
\end{eqnarray}
independently of the gauge choice, where $N_{ijkl}$ represents the
geometric tensor properties (e.g. symmetries of the Riemann tensor
and so on). On the other hand, the parameter $\beta$ measures the
amount of non-linearity ($\sim (r_g/r')^2 $) in the $g_{tt}$
component of the metric. However, this statement is valid only in
the standard post-Newtonian gauge.

These considerations can be developed for any relativistic theory
of gravity but, as we shall see below, the above results, valid in
GR, cannot be simply extrapolated to any modified theory of
gravity since misleading conclusions could be achieved. For
example, in literature, there are some papers stating that
$f(R)\neq R$ are not viable models in Solar System since, by
recasting the $f(R)$-gravity as  the O'Hanlon model, $\gamma 1/2$
\cite{chietall} in evident contrast with GR measurement giving
$\gamma \simeq 1$. On the other hand, assuming
$f(R)=R^{1+\epsilon}$ with $\epsilon\rightarrow 0$, the standard
GR would be hard to be recovered in the weak field limit and the
Cauchy boundary conditions \cite{wald,vignolo} would be highly
violated if the theory results switched from $\gamma=1$, for
$\epsilon=0$, to $\gamma=1/2$, for $\epsilon\neq 0$ . As shown in
\cite{TS-fR-analogy}, the shortcoming can be solved by considering
separately the weak field limits of $f(R)$ and O'Hanlon theory
which have to be  confronted "after" the PPN-approximation. The
conceptual reason is clear: the invariance properties of the
theories are lost in the approximation process so the confront has
to be performed assuming the same gauge and formalism. In other
words, different theories have to be carefully confronted at
general level and at approximated level without incautious
extrapolations. This argument will be considered in detail below.

\subsection{Perturbing the spherically symmetric solutions of
$f(R)$-gravity}

Let us focus now on the differences of perturbing a  spherically
symmetric solution coming from a generic  $f(R)$-gravity model
with respect to the standard spherically symmetric solutions
coming from the GR case $f(R)=R$.  We neglect, for the moment, the
contributions of Ricci and Riemann square \cite{spher_symm_fR}
which will be considered below. The search for solutions in FOG
can be faced by means of a perturbation theory.

The general approach is to start from analytical $f(R$-function
assuming that the background model slightly deviates from  GR.
Such a method can provide interesting results on  astrophysical
scales where spherically symmetric solutions, characterized by
small values of the scalar curvature, can be taken into account.
Below, we will consider the perturbing approach assuming that the
background metric matches, at zero order, the GR solutions.

The field equations (\ref{fieldequationFOG}) and
(\ref{tracefieldequationFOG}) in $f(R)$-gravity, if we develop
also the covariant derivatives, become

\begin{eqnarray}\label{fieldequationfR}
\left\{\begin{array}{ll}
H_{\mu\nu}\,=\,f'R_{\mu\nu}-\frac{f}{2}g_{\mu\nu}+\mathcal{H}_{\mu\nu}\,=\,\mathcal{X}\,T_{\mu\nu}\\\\
H\,=\,f'R-2f+\mathcal{H}\,=\,\mathcal{X}\,T
\end{array} \right.
\end{eqnarray}
where the two quantities $\mathcal{H}_{\mu\nu}$ and $\mathcal{H}$
read

\begin{eqnarray}\label{highterms1}
\left\{\begin{array}{ll}
\mathcal{H}_{\mu\nu}\,=\,-f''\biggl\{R_{,\mu\nu}-\Gamma^\sigma_{\mu\nu}R_{,\sigma}-g_{\mu\nu}\biggl[\biggl({g^{\sigma\tau}}
_{,\sigma}+g^{\sigma\tau}\ln\sqrt{-g}_{,\sigma}\biggr)R_{,\tau}+g^{\sigma\tau}R_{,\sigma\tau}\biggr]\biggr\}-f'''\biggl(R_{,\mu}R_{,\nu}-g_{\mu\nu}
g^{\sigma\tau}R_{,\sigma}R_{,\tau}\biggr)
\\\\
\mathcal{H}\,=\,3f''\biggl[\biggl({g^{\sigma\tau}}_{,\sigma}+g^{\sigma\tau}
\ln\sqrt{-g}_{,\sigma}\biggr)R_{,\tau}+g^{\sigma\tau}R_{,\sigma\tau}
\biggr]+3f'''g^{\sigma\tau}R_{,\sigma}R_{,\tau}
\end{array} \right.
\end{eqnarray}
where $f^i$ is the $i$-th derivative of $f$ with respect to Ricci scalar $R$.

In general, searching for solutions by a perturbation technique
means to perturb the metric

\begin{eqnarray}\label{approx-metric}
g_{\mu\nu}\,=\,g^{(0)}_{\mu\nu}+g^{(1)}_{\mu\nu}
\end{eqnarray}
This implies that the field equations (\ref{fieldequationfR})
split, up to first order, in two parts. Besides, a perturbation on
the metric acts also on the Ricci scalar $R$ defined in
Eq.(\ref{ricciscalar})

\begin{eqnarray}
R\,\sim\,R^{(0)}+R^{(1)}
\end{eqnarray}
and then we can Taylor expand the analytic function $f(R)$ about
the background value of $R$, \emph{i.e.}:

\begin{eqnarray}\label{approx}
\left\{\begin{array}{ll}
f\,=\,\sum_{n}\frac{f^n(R^{(0)})}{n!}\biggl[R-R^{(0)}\biggr]^n\,=\,\sum_{n}\frac{{f^n}^{(0)}}{n!}{R^{(1)}}^n\,\sim\,f^{(0)}+f'^{(0)}
R^{(1)}+\frac{f''^{(0)}}{2}{R^{(1)}}^2+\frac{f'''^{(0)}}{6}{R^{(1)}}^3+
\frac{{f^{IV}}^{(0)}}{24}{R^{(1)}}^4
\\\\
f'\,=\,\sum_{n}\frac{f^{n+1}(R^{(0)})}{n!}\biggl[R-R^{(0)}\biggr]^n\,\sim\,f'^{(0)}
+f''^{(0)}R^{(1)}+\frac{f'''^{(0)}}{2}{R^{(1)}}^2+
\frac{{f^{IV}}^{(0)}}{6}{R^{(1)}}^3\,=\,\frac{df}{dR^{(1)}}
\\\\
f''\,=\,\sum_{n}\frac{f^{n+2}(R^{(0)})}{n!}\biggl[R-R^{(0)}\biggr]^n\,\sim\,
f''^{(0)}+f'''^{(0)}R^{(1)}+
\frac{{f^{IV}}^{(0)}}{2}{R^{(1)}}^2\,=\,\frac{df'}{dR^{(1)}}
\\\\
f'''\,=\,\sum_{n}\frac{f^{n+3}(R^{(0)})}{n!}\biggl[R-R^{(0)}\biggr]^n\,\sim\,f'''^{(0)}+
{f^{IV}}^{(0)}R^{(1)}\,=\,\frac{df''}{dR^{(1)}}
\end{array} \right.
\end{eqnarray}

The zero order of field equations (\ref{fieldequationfR}) reads

\begin{equation}\label{eqp0}
{f'}^{(0)}R^{(0)}_{\mu\nu}-\frac{1}{2}g^{(0)}_{\mu\nu}f^{(0)}+\mathcal{H}^{(0)}_{\mu\nu}\,=\,\mathcal{X}\,T^{(0)}_{\mu\nu}
\end{equation}
where

\begin{eqnarray}
\mathcal{H}^{(0)}_{\mu\nu}\,&=&\,-{f''}^{(0)}\biggl\{R^{(0)}_{,\mu\nu}-{\Gamma^{(0)}}^{\rho}_{\mu\nu}R^{(0)}_{,\rho}-g^{(0)}_
{\mu\nu}\biggl({g^{(0)\rho\sigma}}_{,\rho}R^{(0)}_{,\sigma}+g^{(0)\rho\sigma}R^{(0)}_{,\rho\sigma}+g^{(0)\rho\sigma}\ln\sqrt{-g}_{,\rho}R^{(0)}_{,\sigma}\biggl)\biggl\}
\nonumber\\&&-{f'''}^{(0)}\biggl\{R^{(0)}_{,\mu}R^{(0)}_{,\nu}-g^{(0)}
_{\mu\nu}g^{(0)\rho\sigma}R^{(0)}_{,\rho}R^{(0)}_{,\sigma}\biggl\}
\end{eqnarray}
At first order, one has\,:

\begin{eqnarray}\label{eqp1}
{f'}^{(0)}\biggl\{R^{(1)}_{\mu\nu}-\frac{1}{2}g^{(0)}_{\mu\nu}R^{(1)}\biggr\}+{f''}^{(0)}R^{(1)}R^{(0)}_{\mu\nu}
-\frac{1}{2}f^{(0)}g^{(1)}_{\mu\nu}+\mathcal{H}^{(1)}_{\mu\nu}\,=\,\mathcal{X}\,T^{(1)}_{\mu\nu}
\end{eqnarray}
with

\begin{eqnarray}
\mathcal{H}^{(1)}_{\mu\nu}\,&=&\,-{f''}^{(0)}\biggl\{R^{(1)}_{,\mu\nu}-{\Gamma^{(0)}}^{\rho}_{\mu\nu}R^{(1)}_{,\rho}-
{\Gamma^{(1)}}^{\rho}_{\mu\nu}R^{(0)}_{,\rho}-g^{(0)}_{\mu\nu}\biggl[{g^{(0)\rho\sigma}}_{,\rho}R^{(1)}_{,\sigma}+{g^{(1)
\rho\sigma}}_{,\rho}R^{(0)}_{,\sigma}
\nonumber\\\nonumber\\&&
+g^{(0)\rho\sigma}R^{(1)}_{,\rho\sigma}+g^{(1)\rho\sigma}
R^{(0)}_{,\rho\sigma}+g^{(0)\rho\sigma}\biggl(\ln\sqrt{-g}^{(0)}_{,\rho}R^{(1)}_{,\sigma}+\ln\sqrt{-g}^{(1)}_{,\rho}R^{(0)}
_{,\sigma}\biggl)
\nonumber\\\nonumber\\&&
+g^{(1)\rho\sigma}\ln\sqrt{-g}^{(0)}_{,\rho}R^{(0)}_{,\sigma}\biggr]-g^{(1)}_
{\mu\nu}\biggl({g^{(0)\rho\sigma}}_{,\rho}R^{(0)}_{,\sigma}+g^{(0)\rho\sigma}R^{(0)}_{,\rho\sigma}\\&&
+g^{(0)\rho\sigma}\ln\sqrt{-g}^{(0)}_{,\rho}R^{(0)}_{,\sigma}\biggr)\biggr\}-{f'''}^{(0)}\biggl\{R^{(0)}_{,\mu}R^{(1)}
_{,\nu}+R^{(1)}_{,\mu}R^{(0)}_{,\nu}-g^{(0)}_{\mu\nu}g^{(0)\rho\sigma}\biggl(R^{(0)}_{,\rho}R^{(1)}_{,\sigma}
\nonumber\\\nonumber\\&&
+R^{(1)}_{,\rho}R^{(0)}_{,\sigma}\biggr)-g^{(0)}_{\mu\nu}g^{(1)\rho\sigma}R^{(0)}_{,\rho}R^{(0)}_{,\sigma}-g^{(1)}
_{\mu\nu}g^{(0)\rho\sigma}R^{(0)}_{,\rho}R^{(0)}_{,\sigma}\biggr\}-{f'''}^{(0)}R^{(1)}\biggl\{R^{(0)}_{,\mu\nu}
\nonumber\\\nonumber\\&&
-{\Gamma^{(0)}}^{\rho}_{\mu\nu}R^{(0)}_{,\rho}-g^{(0)}_{\mu\nu}\biggl({g^{(0)\rho\sigma}}_{,\rho}R^{(0)}_
{,\sigma}+g^{(0)\rho\sigma}R^{(0)}_{,\rho\sigma}+g^{(0)\rho\sigma}\ln\sqrt{-g}^{(0)}_{,\rho}R^{(0)}_{,\sigma}\biggl)\biggl\}
\nonumber\\\nonumber\\&&
-{f^{IV}}^{(0)}R^{(1)}\biggl\{R^{(0)}_{,\mu}R^{(0)}_{,\nu}-g^{(0)}_{\mu\nu}g^{(0)\rho\sigma}R^{(0)}_
{,\rho}R^{(0)}_{,\sigma}\biggl\}
\end{eqnarray}
A part the analyticity, no hypothesis has been invoked here on the
form of the function $f(R)$. As a matter of fact, $f(R)$ can be
completely general. At this level, to solve the problem, it is
required the zero order solution of (\ref{eqp0}) which, in
general, \emph{could not be a GR solution}. This problem can be
overcome assuming the same order of perturbation on the $f(R)$,
that is:

\begin{eqnarray}\label{theapprox}
f(R)\,=\,R+\mathcal{F}(R)
\end{eqnarray}
where $\mathcal{F}(R)$ is a generic function of the Ricci scalar
(this means to consider $\mathcal{F}(R)\ll R$). Then we have

\begin{eqnarray}\label{approx2}
f\,=\,R^{(0)}+R^{(1)}+\mathcal{F}^{(0)}\,,\ \ \
f'\,=\,1+\mathcal{F}'^{(0)}\,,\ \ \
f''\,=\,\mathcal{F}''^{(0)}\,,\ \ \ \ \ \
f'''\,=\,\mathcal{F}'''^{(0)}
\end{eqnarray}

However the condition $\mathcal{F}(R)\ll R$ has to imply the
validity of the linear approximation
$\frac{f''(R^{(0)})R^{(1)}}{f'(R^{(0)})}\ll 1$. This is
demonstrated by assuming $f'\,=\,1+\mathcal{F}'$ and
$f''\,=\,\mathcal{F}''$. Immediately, we obtain that the condition
is fulfilled for

\begin{eqnarray}\label{strong-condition}
\frac{\mathcal{F}''(R^{(0)})R^{(1)}}{1+\mathcal{F}'(R^{(0)})}\ll1
\end{eqnarray}
For example, given a Lagrangian of the form
$f(R)\,=\,R+\frac{R_0}{R}$, Eq.(\ref{strong-condition}) becomes

\begin{eqnarray}
\frac{2R_0R^{(1)}}{R^{(0)}({R^{(0)}}^2-R_0)}\ll1
\end{eqnarray}
while, for $f(R)\,=\,R+\alpha R^2$,  Eq.(\ref{strong-condition})
becomes

\begin{eqnarray}
\frac{2\alpha R^{(1)}}{1+2\alpha R^{(0)}}\ll1
\end{eqnarray}
This means that the validity of the approximation strictly depends
on the form of the models and the value of the parameters, in the
previous cases, $R_0$ and $\alpha$. For the considerations below,
we will assume that condition (\ref{strong-condition}) always
holds.

Eqs.(\ref{eqp0}) reduce to the GR form

\begin{eqnarray}
R^{(0)}_{\mu\nu}-\frac{1}{2}R^{(0)}g^{(0)}_{\mu\nu}\,=\,G^{(0)}_{\mu\nu}\,=\,\mathcal{X}\,T^{(0)}_{\mu\nu}
\end{eqnarray}
On the other hand, Eqs. (\ref{eqp1}) reduce to

\begin{eqnarray}
R^{(1)}_{\mu\nu}-\frac{1}{2}g^{(0)}_{\mu\nu}R^{(1)}-\frac{1}{2}g^{(1)}_{\mu\nu}R^{(0)}-\frac{1}{2}g^{(0)}_{\mu\nu}
\mathcal{F}^{(0)}+\mathcal{F}'^{(0)}R^{(0)}_{\mu\nu}+\mathcal{H}^{(1)}_{\mu\nu}\,=\,\mathcal{X}\,T^{(1)}_{\mu\nu}
\end{eqnarray}
where

\begin{eqnarray}
\mathcal{H}^{(1)}_{\mu\nu}\,&=&\,-\mathcal{F}'''^{(0)}\biggl\{R^{(0)}_{,\mu}R^{(0)}_{,\nu}-g^{(0)}_{\mu\nu}g^{(0)\sigma\tau}
R^{(0)}_{,\sigma}R^{(0)}_{,\tau}\biggr\}\nonumber\\\nonumber\\&&-\mathcal{F}''^{(0)}\biggl\{R^{(0)}_{,\mu\nu}-{\Gamma^{(0)}}^\sigma_{\mu\nu}R^{(0)}
_{,\sigma}-g^{(0)}_{\mu\nu}\biggl({g^{(0)}}^{\sigma\tau}_{\,\,\,\,\,\,,\sigma}R^{(0)}_{,\tau}+
{g^{(0)}}^{\sigma\tau}R^{(0)}_{,\sigma\tau}+{g^{(0)}}^{\sigma\tau}\ln\sqrt{-g}^{(0)}_{,\sigma}R^{(0)}_{,\tau}\biggl)\biggr\}
\end{eqnarray}
The new system of field equations is evidently
simpler than the starting one and once  the zero order solution is
obtained, the solutions at the first order correction can be
easily achieved.

By assuming the time-independent metric (\ref{me5})

\begin{eqnarray}\label{metimeind}
ds^2\,=\,a(r)dt^2-b(r)dr^2-r^2d\Omega^2
\end{eqnarray}
it is straightforward to obtain new spherically symmetric
solutions by substituting into the above perturbed field
equations. In Tables \ref{table-solutions-1} and
\ref{table-solutions-2}, a list of solutions, obtained with this
perturbation method, is given considering different classes of
$f(R)$-models. Some remarks on these solutions are in order at
this point. In the case of $f(R)$-models which  are evidently
corrections to the Hilbert-Einstein Lagrangian as
$\Lambda+R+\epsilon R \ln R$ and $R+\epsilon R^n$, with
$\epsilon\ll 1$, one obtains  exact solutions for the
gravitational potentials $a(r)$ and $b(r)$ related by
$a(r)\,=\,b(r)^{-1}$. The first order expansion is straightforward
as in GR. If the functions $a(r)$ and $b(r)$ are not related, for
$f(R)\,=\,\Lambda+R+\epsilon R \ln R$, the first order system is
directly solved  without any prescription on the perturbation
functions $x(r)$ and $y(r)$ indicated in the Tables. This is not
the case for the models $f(R)\,=\,R+\epsilon R^n$ since, in this
case, one can obtain an explicit constraint on the perturbation
function. In such a case, no corrections are found with respect to
the standard solution. The models $f(R)\,=\,R^n$ and
$f(R)\,=\,\frac{R}{(R_0+R)}$ show similar behaviors. The case
$f(R)\,=\,R^2$ is peculiar and it has to be dealt independently.
For details, see \cite{newtonian_limit_fR}.

\begin{table}
\begin{center}
\begin{tabular}{cccc}
  \hline\hline\hline
  & & & \\
  & $f$ - theory: & $\Lambda+R+\epsilon R \ln R$ & \\
  & & & \\
  & spherical potentials: & $a(r)=b(r)^{-1}=1+\frac{k_1}{r}-\frac{\Lambda r^2}{6}+\delta x(r)$ & \\
  & & & \\
  & solutions: & $x(r)=\frac{k_2}{r}+\frac{\epsilon\Lambda[\ln(-2\Lambda)-1]r^2}{6\delta}$ & \\
  & & & \\
  & first order metric: & $a(r)=1-\frac{\Lambda r^2}{6}+\delta x(r)$,\,\,\,\,\,$b(r)=\frac{1}{1-\frac{\Lambda r^2}{6}}
  +\delta y(r)$ & \\
  & & & \\
  & solutions: & $\begin{cases} x(r)=(\Lambda r^2-6)\biggl\{k_1+\int\frac{dr}{36r\delta(\Lambda r^2-6)}\biggl[4\delta(2
  \Lambda^2r^4-15\Lambda r^2+18)y(r)+r\{36r\epsilon\Lambda[\log(-2\Lambda)-1]
  \\\,\,\,\,\,\,\,\,\,\,\,\,\,\,\,\,\,\,\,+\delta(\Lambda r^2-6)^2y'(r)\}\biggr]\biggr\}\\\\
  y(r)=\frac{k_2\delta-6r^3\epsilon\Lambda[\ln(-2\Lambda)-1]}{r\delta(r^2\Lambda-6)^2}  \end{cases} $& \\
  & & & \\
  & & & \\
  & $f$ - theory: & $R+\epsilon R^n$ & \\
  & & & \\
  & spherical potentials: & $a(r)=b(r)^{-1}=1+\frac{k_1}{r}+\delta x(r)$ & \\
  & & & \\
  & solutions: & $x(r)=\frac{k_2}{r}$ & \\
  & & & \\
  & first order metric: & $a(r)=1+\delta\frac{x(r)}{r}$,\,\,\,\,\,$b(r)=1+\delta\frac{y(r)}{r}$ & \\
  & & & \\
  & solutions: & $x(r)=k_1+k_2r$,\,\,\,\,\,$y(r)=k_3$ & \\
  & & & \\
  & & & \\
  & $f$ - theory: & $R/(R_0+R)$ & \\
  & & & \\
  & first order metric: & $a(r)=1+\delta\frac{x(r)}{r}$,\,\,\,\,\,$b(r)=1+\delta\frac{y(r)}{r}$ & \\
  & & & \\
  & solutions: & $\begin{cases} x(r)=-\frac{4e^{-\frac{R_0^{1/2}r}{\sqrt{6}}}}{R_0}k_1-\frac{2\sqrt{6}e^{\frac{R_0^{1/2}r}
  {\sqrt{6}}}}{R_0^{3/2}}k_2+k_3r \\ y(r)=-\frac{2e^{-\frac{R_0^{1/2}r}{\sqrt{6}}}(6R_0^{1/2}+\sqrt{6}R_0\,r)}{3b^{3/2}}
  k_1-\frac{2e^{\frac{R_0^{1/2}r}{\sqrt{6}}}(\sqrt{6}-R_0^{1/2}r)}{R_0^{3/2}}k_2
  \end{cases}$ & \\
  & & & \\
  \hline\hline\hline
\end{tabular}
\end{center}
\caption{A list of exact solutions obtained \emph{via} the
perturbation approach for some classes of $f(R)$-models; $k_i$ are
integration constants; the potentials $a(r)$ and $b(r)$ are
defined by the metric (\ref{metimeind}).\label{table-solutions-1}}
\end{table}

\begin{table}
\begin{center}
\begin{tabular}{cccc}
  \hline\hline\hline
  & & & \\
  & $f$ - theory: & $R^n$ & \\
  & & & \\
  & spherical potentials: & $a(r)=b(r)^{-1}=1+\frac{k_1}{r}+\frac{R_0r^2}{12}+\delta x(r)$ & \\
  & & & \\
  & solutions: & $\begin{cases} n=2,\,\,\,\,\,\,R_0\neq 0 \,\,\text{and}\,\,x(r)=\frac{3k_2-k_3}{3r}+\frac{k_3r^2}{12} +\frac{k_4}{r}\int dr\,r^2\biggl\{\int dr\frac{\exp\biggl[\frac{R_0r_0^2\ln(r-r_0)}{8+3R_0r_0^2}
  \biggr]}{r^5}\biggr\} \\
  \,\,\,\,\,\,\,\,\,\,\,\,\,\,\,\,\,\,\,\,\,\,\,\, \text{with}\,\,r_0\,\,$satisfying the condition$ \,\, 6k_1+8r_0+R_0r_0^3=0
  \\\\ n\geq 2,\,\,\,\,\,\,
  \text{System solved only whit $R_0=0$} \,\,\text{and no prescriptions on
  $x(r)$} \end{cases}$ & \\
  & & & \\
  & first order metric: & $a(r)=1+\delta\frac{x(r)}{r}$,\,\,\,\,\,$b(r)=1+\delta\frac{y(r)}{r}$ & \\
  & & & \\
  & solutions: & $\begin{cases} n=2 & y(r)=-\frac{R_0r^3}{6}-\frac{x(r)}{2}+\frac{1}{2}rx'(r)+k_1, \\
  & \text{with}\,R(r)=\delta R_0 \\\\ n\neq 2 &
  y(r)=-\frac{1}{2}\int dr\,r^2R(r)-\frac{x(r)}{2}+\frac{1}{2}rx'(r)+k_1 \\ & \text{with $R(r)$ whatever}
  \end{cases}$ & \\
  & & & \\
  & first order metric: & $a(r)=1-\frac{r_g}{r}+\delta x(r)$,\,\,\,\,\,$b(r)=\frac{1}{1-\frac{r_g}{r}}+\delta y(r)$ & \\
  & & & \\
  & solutions: & $\begin{cases} n=2 & y(r)=\frac{rk_1}{3r_g^2-7r_gr+4r^2}+\frac{r^2k_2}{3(3r_g^2-7r_gr+4r^2)} +\frac{r_gr^2x(r)+2(r_gr^3-r^4)x'(r)}{(3r_g-4r)(r_g-r)^2} \\\\ n\neq 2 &
  \text{whatever functions $x(r)$\,, $y(r)$ and $R(r)$} \end{cases}$ & \\
  & & & \\
  \hline\hline\hline
\end{tabular}
\end{center}
\caption{A list of exact solutions obtained \emph{via} the
perturbation approach for some classes of $f(R)$-models; $k_i$ are
integration constants; the potentials $a(r)$ and $b(r)$ are
defined by the metric (\ref{metimeind}).\label{table-solutions-2}}
\end{table}

\section{General remarks on  Newtonian and  Post-Newtonian approximations}\label{PPN-paradigm}

At this point, it is worth discussing  some general issues on the
Newtonian and post-Newtonian limits. Basically there are some
general features one has to take into account when approaching
these limits, whatever the underlying theory of gravitation is. In
fact here we are not interested in entering the theoretical
discussion on how to formulate a mathematically well founded
Newtonian and post-Newtonian limits  of general relativistic field
theories, nevertheless we point the interested reader to
\cite{trautman, friedrichs, kilmister, dautcourt, kuenzle, ehlers,
ehlers1}. In this section, we provide the explicit form of the
various quantities needed to compute the approximations in the
field equations in GR theory and any metric theory of gravity. We
only mention that there is also  a discussion on alternative ways
to define the Newtonian and post-Newtonian limits of FOG in the
recent literature, see for example \cite{dick}. In this work, the
Newtonian and post-Newtonian limits are identified by the
maximally symmetric solutions, which are not necessarily the
Minkowski spacetime in $f(R)$-gravity which could be singular.

Let us start with some very general considerations on the
Newtonian gravitating systems. If one takes into account a system
of gravitationally interacting particles of mass $\bar{M}$, the
kinetic energy $\frac{1}{2}\bar{M}\bar{v}^2$ will be, roughly, of
the same order of magnitude as the typical potential energy
$U\,=\,G\bar{M}^2/\bar{r}$, with $\bar{M}$, $\bar{r}$, and
$\bar{v}$ the typical average values of  masses, separations, and
velocities of these particles. As a consequence:

\begin{eqnarray}
\bar{v^2}\sim \frac{G\bar{M}}{\bar{r}}
\end{eqnarray}
(for instance, a test particle in a circular orbit of radius $r$
about a central mass $M$ has velocity $v$, given in Newtonian
Mechanics, by the exact formula $v^2\,=\,GM/r$.)

The post-Newtonian approximation can be described as a method for
obtaining the motion of the system to higher approximations than
the first order\footnote{This approximation  coincides with the
Newtonian Mechanics.} with respect to the quantities
$G\bar{M}/\bar{r}$ and $\bar{v}^2$ assumed small with respect to
the squared light speed. This approximation is sometimes referred
to as an expansion in inverse powers of the light speed.

The typical values of the Newtonian gravitational potential $\Phi$
are nowhere larger (in modulus) than $10^{-5}$ in the Solar System
(in geometrized units, $\Phi$ is dimensionless). On the other
hand, planetary velocities satisfy the condition
$\bar{v}^2\lesssim-\Phi$, while the matter pressure $p$,
experienced inside the Sun and the planets, is generally smaller
than the matter gravitational energy density $-\rho\Phi$, in other
words \footnote{Typical values of $p/\rho$ are $\sim 10^{-5}$ in
the Sun and  $\sim 10^{-10}$ in the Earth \cite{will}.}
$p/\rho\lesssim -\Phi$. Furthermore one has to consider that even
other forms of energy in the Solar System (compressional energy,
radiation, thermal energy, etc.) have small intensities and the
specific energy density $\Pi$ (the ratio of the energy density to
the rest-mass density) is related to $U$ by $\Pi\lesssim U$ ($\Pi$
is $\sim 10^{-5}$ in the Sun and $\sim 10^{-9}$ in the Earth
\cite{will}). As matter of fact, one can consider that these
quantities, as function of the velocity, give second order
contributions\,:

\begin{eqnarray}
-\Phi\sim v^2\sim p/\rho\sim \Pi\sim\mathcal{O}(2)
\end{eqnarray}
Therefore, the velocity $v$ gives $\mathcal{O}$(1) terms in the velocity
expansions, $U^2$ is of order $\mathcal{O}$(4), $Uv$ of $\mathcal{O}$(3), $U\Pi$ is of
$\mathcal{O}$(4), and so on. Considering these approximations, one has

\begin{eqnarray}
\frac{\partial}{\partial t}\sim\textbf{v}\cdot\nabla
\end{eqnarray}
and

\begin{eqnarray}
\frac{|\partial/\partial t|}{|\nabla|}\sim\mathcal{O}(1)
\end{eqnarray}

The paradigm of post-Newtonian limit is to start from a
series-developed  metric tensor (and  all additional fields in the
theory) with respect to the dimensionless quantity $v$ (in natural
units!). A system of moving bodies radiates gravitational waves
and thus loses energy. This loss appears only at the fifth-order
in the approximation of $v$. In the first four approximations, the
energy of the system remains constant. From this, it follows that
a system of gravitating bodies can be correctly described by a
Lagrangian up to terms of order $v^4$, in the absence of
electromagnetic fields. We thus find the equations of motion of
the system in the next approximation after the Newtonian.

To solve the problem, we have to start from the determination of
the metric tensor $g_{\mu\nu}$ in the same approximation of the
gravitational field
\cite{newtonian_limit_fR,postnewtonian_limit_fR}.

Considering that particles move along the geodesics

\begin{eqnarray}
\frac{d^2x^\mu}{ds^2}+\Gamma^\mu_{\sigma\tau}\frac{dx^\sigma}{ds}\frac{dx^\tau}{ds}\,=\,0
\end{eqnarray}
these can be written in details as

\begin{eqnarray}
\frac{d^2x^i}{dt^2}\,=\,-\Gamma^i_{tt}-2\Gamma^i_{tm}\frac{dx^m}{dt}-
\Gamma^i_{mn}\frac{dx^m}{dt}\frac{dx^n}{dt}+\biggl[\Gamma^t_{tt}+
2\Gamma^t_{tm}\frac{dx^m}{dt}+2\Gamma^t_{mn}\frac{dx^m}{dt}\frac{dx^n}{dt}\biggr]\frac{dx^i}{dt}
\end{eqnarray}
In the Newtonian approximation, that is vanishingly small
velocities and only  first-order terms in the difference between
$g_{\mu\nu}$ and the Minkowski metric $\eta_{\mu\nu}$, one obtains
that the particle motion equations reduce to the standard
result

\begin{eqnarray}
\frac{d^2x^i}{dt^2}\,\simeq\,-\Gamma^i_{tt}\simeq-\frac{1}{2}\frac{\partial
g_{tt}}{\partial x^i}
\end{eqnarray}

The quantity $1-g_{tt}$ is of order $G\bar{M}/\bar{r}$, so that
the Newtonian approximation gives
$\displaystyle\frac{d^2x^i}{dt^2}$ to the order
$G\bar{M}/\bar{r}^2$, that is, to the order $\bar{v}^2/r$. As a
consequence, if we would like to search for the post-Newtonian
approximation, we need to compute
$\displaystyle\frac{d^2x^i}{dt^2}$ to the order
$\bar{v}^4/\bar{r}$. Due to the Equivalence Principle and the
differentiability of spacetime manifold, we expect that it should
be possible to find out a coordinate system in which the metric
tensor is nearly equal to the Minkowski one $\eta_{\mu\nu}$, the
correction being expandable in powers of
$G\bar{M}/\bar{r}\sim\bar{v}^2$. In other words, one has to
consider the metric developed as follows\,:

\begin{eqnarray}\label{PPN-metric}
\left\{\begin{array}{ll}
g_{tt}(t,\textbf{x})\simeq 1+g^{(2)}_{tt}(t,\textbf{x})+g^{(4)}_{tt}(t,\textbf{x})+\mathcal{O}(6)
\\\\
g_{ti}(t,\textbf{x})\simeq g^{(3)}_{ti}(t,\textbf{x})+\mathcal{O}(5)
\\\\
g_{ij}(t, \textbf{x})\simeq - \delta_{ij}+g^{(2)}_{ij}(t,\textbf{x})+\mathcal{O}(4)
\end{array}\right.
\end{eqnarray}
where $\delta_{ij}$ is the Kronecker delta, and for the
controvariant form of $g_{\mu\nu}$, one has

\begin{eqnarray}\label{PPN-metric-contro}
\left\{\begin{array}{ll}g^{tt}(t,\textbf{x})\simeq 1+g^{(2)tt}(t,
\textbf{x})+g^{(4)tt}(t, \textbf{x})+\mathcal{O}(6)
\\\\
g^{ti}(t,\textbf{x})\simeq g^{(3)ti}(t,\textbf{x})+\mathcal{O}(5)\\\\
g^{ij}(t,\textbf{x})\simeq-\delta_{ij}+g^{(2)ij}(t,{\textbf{x}})+\mathcal{O}(4)
\end{array}\right.
\end{eqnarray}
The inverse of the metric tensor (\ref{PPN-metric}) is defined by
(\ref{controvariant-metric-condition}). The relations among the
higher than first order terms turn out to be

\begin{eqnarray}\label{PPN-metric-contro-cov}
\left\{\begin{array}{ll}g^{(2)tt}(t,\textbf{x})=-g^{(2)}_{tt}(t,\textbf{x})\\\\g^{(4)tt}(t,\textbf{x})={g^{(2)}_{tt}
(t,\textbf{x})}^2-g^{(4)}_{tt}(t,\textbf{x})\\\\g^{(3)ti}=g^{(3)}_{ti}\\\\g^{(2)ij}(t,\textbf{x})=-g^{(2)}_{ij}(t,\textbf{x})
\end{array}\right.
\end{eqnarray}
In evaluating $\Gamma^\mu_{\alpha\beta}$ we must take into account
that the scale of distance and time, in our systems, are
respectively set by $\bar{r}$ and $\bar{r}/\bar{v}$, thus the
space and time derivatives should be regarded as being of order

\begin{eqnarray}
\frac{\partial}{\partial x^i}\sim\frac{1}{\bar{r}}\,, \ \ \ \ \ \
\ \frac{\partial}{\partial t}\sim\frac{\bar{v}}{\bar{r}}
\end{eqnarray}
Using the above approximations (\ref{PPN-metric}),
(\ref{PPN-metric-contro}) and (\ref{PPN-metric-contro-cov}) we
have, from the definition (\ref{christoffel}),

\begin{eqnarray}\label{PPN-christoffel}
\left\{\begin{array}{ll} \begin{array}{ccc}
  {\Gamma^{(3)}}^t_{tt}=\frac{1}{2}g^{(2)}_{tt,t}\, & \,\, & {\Gamma^{(2)}}^{i}_{tt}=\frac{1}{2}g^{(2)}_{tt,i} \\
  & & \\
  {\Gamma^{(2)}}^{i}_{jk}=\frac{1}{2}\biggl(g^{(2)}_{jk,i}-g^{(2)}_{ij,k}-g^{(2)}_{ik,j}\biggr)\,
  & \,\, & {\Gamma^{(3)}}^{t}_{ij}=\frac{1}{2}\biggl
  (g^{(3)}_{ti,j}+g^{(3)}_{jt,i}-g^{(2)}_{ij,t}\biggr) \\
  & & \\
  {\Gamma^{(3)}}^{i}_{tj}=\frac{1}{2}\biggl(g^{(3)}_{tj,i}-g^{(3)}_{it,j}-g^{(2)}_{ij,t}\biggr)\,
  & \,\, & {\Gamma^{(4)}}^{t}_{ti}=\frac{1}{2}\biggl
  (g^{(4)}_{tt,i}-g^{(2)}_{tt}g^{(2)}_{tt,i}\biggr) \\
  & & \\
  {\Gamma^{(4)}}^{i}_{tt}=\frac{1}{2}\biggl(g^{(4)}_{tt,i}+g^{(2)}_{im}g^{(2)}_{tt,m}-2g^{(3)}_{it,t}\biggr)\, & \,\, &
  {\Gamma^{(2)}}^{t}_{ti}= \frac{1}{2}g^{(2)}_{tt,i}
  \end{array}
\end{array}\right.
\end{eqnarray}
The Ricci tensor components (\ref{riccitensor}) are

\begin{eqnarray}\label{PPN-ricci-tensor}
\left\{\begin{array}{ll}R^{(2)}_{tt}=\frac{1}{2}g^{(2)}_{tt,mm}\\\\R^{(4)}_{tt}=\frac{1}{2}g^{(4)}_{tt,mm}+\frac{1}{2}g^{(2)}
_{mn,m}g^{(2)}_{tt,n}+\frac{1}{2}g^{(2)}_{mn}g^{(2)}_{tt,mn}+\frac{1}{2}g^{(2)}_{mm,tt}-
\frac{1}{4}g^{(2)}_{tt,m}g^{(2)}_{tt,m}-\frac{1}{4}g^{(2)}_{mm,n}g^{(2)}_{tt,n}-g^{(3)}_{tm,tm}\\\\
R^{(3)}_{ti}=\frac{1}{2}g^{(3)}_{ti,mm}-\frac{1}{2}g^{(2)}_{im,mt}-\frac{1}{2}g^{(3)}_{mt,mi}+\frac{1}{2}g^{(2)}_{mm,ti}\\\\
R^{(2)}_{ij}=\frac{1}{2}g^{(2)}_{ij,mm}-\frac{1}{2}g^{(2)}_{im,mj}-\frac{1}{2}g^{(2)}_{jm,mi}-\frac{1}{2}g^{(2)}_{tt,ij}+
\frac{1}{2}g^{(2)}_{mm,ij}
\end{array}\right.
\end{eqnarray}
and the Ricci scalar (\ref{ricciscalar}) is

\begin{eqnarray}\label{PPN-ricci-scalar}
\left\{\begin{array}{ll}R^{(2)}=R^{(2)}_{tt}-R^{(2)}_{mm}=g^{(2)}_{tt,mm}-g^{(2)}_{nn,mm}+g^{(2)}_{mn,mn}\\\\
R^{(4)}=R^{(4)}_{tt}-g^{(2)}_{tt}R^{(2)}_{tt}-g^{(2)}_{mn}R^{(2)}_{mn}=\\\\\,\,\,\,\,\,\,\,\,\,
\,\,\,=\,\frac{1}{2}g^{(4)}_{tt,mm}+\frac{1}{2}g^{(2)}_{mn,m}g^{(2)}_{tt,n}+\frac{1}{2}g^{(2)}_{mn}g^{(2)}_{tt,mn}+\frac{1}{2}
g^{(2)}_{mm,tt}-\frac{1}{4}g^{(2)}_{tt,m}g^{(2)}_{tt,m}+\\\\\,\,\,\,\,\,\,\,\,\,\,\,\,\,\,\,\,\,\,-\frac{1}{4}g^{(2)}_{mm,n}
g^{(2)}_{tt,n}-g^{(3)}_{tm,tm}-\frac{1}{2}g^{(2)}_{tt}g^{(2)}_{tt,mm}-\frac{1}{2}g^{(2)}_{mn}\biggl(g^{(2)}_{mn,ll}-g^{(2)}_
{ml,ln}-g^{(2)}_{nl,lm} -g^{(2)}_{tt,mn}+g^{(2)}_{ll,mn}\biggr)
\end{array}\right.
\end{eqnarray}
The Einstein tensor components (\ref{einstein-tensor}) are

\begin{eqnarray}\label{PPN-einstein-tensor}
\left\{\begin{array}{ll}G^{(2)}_{tt}=R^{(2)}_{tt}-\frac{1}{2}R^{(2)}=\frac{1}{2}g^{(2)}_{mm,nn}+\frac{1}{2}g^{(2)}_{mn,mn}\\\\
G^{(4)}_{tt}=R^{(4)}_{tt}-\frac{1}{2}R^{(4)}-\frac{1}{2}g^{(2)}_{tt}R^{(2)}=...\\\\
G^{(3)}_{ti}=R^{(3)}_{ti}=\frac{1}{2}g^{(3)}_{ti,mm}-\frac{1}{2}g^{(2)}_{im,mt}-\frac{1}{2}g^{(3)}_{mt,mi}+\frac{1}{2}g^{(2)}
_{mm,ti}\\\\G^{(2)}_{ij}=R^{(2)}_{ij}+\frac{\delta_{ij}}{2}R^{(2)}=\frac{1}{2}g^{(2)}_{ij,mm}-\frac{1}{2}g^{(2)}_{im,mj}-
\frac{1}{2}g^{(2)}_{jm,mi}-\frac{1}{2}g^{(2)}_{tt,ij}+\frac{1}{2}g^{(2)}_{mm,ij}+\frac{\delta_{ij}}{2}\biggl[g^{(2)}_{tt,mm}-g^{(2)}_{nn,mm}+g^{(2)}_{mn,mn}\biggr]
\end{array}\right.
\end{eqnarray}

By assuming the harmonic gauge \footnote{The gauge transformation
is $\tilde{h}_{\mu\nu}=h_{\mu\nu}-\zeta_{\mu,\nu}-\zeta_{\nu,\mu}$
when we perform a coordinate transformation as
$x'^\mu=x^\mu+\zeta^\mu$ with O($\zeta^2$)$\ll 1$. To obtain our
gauge and the validity of the field equations for both
perturbation $h_{\mu\nu}$ and $\tilde{h}_{\mu\nu}$, the vector
$\zeta_\mu$ has to satisfy the harmonic condition
$\Box\zeta^\mu=0$.}

\begin{eqnarray}\label{gauge-harmonic}
g^{\rho\sigma}\Gamma^\mu_{\rho\sigma}\,=\,0
\end{eqnarray}
it is possible to simplify the components of Ricci tensor
(\ref{PPN-ricci-tensor}). In fact for $\mu\,=\,0$ one has

\begin{eqnarray}\label{gau1}
2g^{\sigma\tau}\Gamma^{t}_{\sigma\tau}\approx g^{(2)}_{tt,t}
-2g^{(3)}_{tm,m}+g^{(2)}_{mm,t}\,=\,0
\end{eqnarray} and for
$\mu=i$

\begin{eqnarray}\label{gau2}
2g^{\sigma\tau}\Gamma^{i}_{\sigma\tau}\approx
g^{(2)}_{tt,i}+2g^{(2)}_{mi,m}-g^{(2)}_{mm,i}\,=\,0
\end{eqnarray}
Differentiating (\ref{gau1}) with respect to $t$, $x^j$ and
(\ref{gau2}) and with respect to $t$, one obtains

\begin{eqnarray}\label{gau3}
g^{(2)}_{tt,tt}-2g^{(3)}_{tm,mt}+g^{(2)}_{mm,tt}\,=\,0
\end{eqnarray}

\begin{eqnarray}\label{gau4}
g^{(2)}_{tt,tj}-2g^{(3)}_{mt,jm}+g^{(2)}_{mm,tj}\,=\,0
\end{eqnarray}

\begin{eqnarray}\label{gau5}
g^{(2)}_{tt,ti}+2g^{(2)}_{mi,tm}-g^{(2)}_{mm,ti}\,=\,0
\end{eqnarray}
On the other side, combining (\ref{gau4}) and (\ref{gau5}),
we get

\begin{eqnarray}\label{gau6}
g^{(2)}_{mm,ti}-g^{(2)}_{mi,tm}-g^{(3)}_{mt,mi}\,=\,0
\end{eqnarray}
Finally, differentiating (\ref{gau2}) with respect to $x^j$,
one has\,:

\begin{eqnarray}\label{gau7}
g^{(2)}_{tt,ij}+2g^{(2)}_{mi,jm}-g^{(2)}_{mm,ij}\,=\,0
\end{eqnarray}
and redefining indexes as $j\rightarrow i$, $i\rightarrow j$ since
these are mute indexes, we get

\begin{eqnarray}\label{gau8}
g^{(2)}_{tt,ij}+2g^{(2)}_{mj,im}-g^{(2)}_{mm,ij}\,=\,0
\end{eqnarray}
Combining  (\ref{gau7}) and  (\ref{gau8}), we obtain

\begin{eqnarray}\label{gau9}
g^{(2)}_{tt,ij}+g^{(2)}_{mi,jm}+g^{(2)}_{mj,im}-g^{(2)}_{mm,ij}\,=\,0
\end{eqnarray}
Relations (\ref{gau3}), (\ref{gau6}), (\ref{gau9}) guarantee  to
rewrite the Ricci tensor components (\ref{PPN-ricci-tensor}) at
higher orders as

\begin{eqnarray}\label{PPN-ricci-tensor-HG}
\left\{\begin{array}{ll}R^{(2)}_{tt}|_{HG}=\frac{1}{2}\triangle
g^{(2)}_{tt}\\\\R^{(4)}_{tt}|_{HG}=\frac{1}{2}\triangle
g^{(4)}_{tt}+\frac{1}{2}g^{(2)}_{mn}g^{(2)}_{tt,mn}-\frac{1}{2}g^{(2)}_{tt,tt}-\frac{1}{2}|\bigtriangledown
g^{(2)}_{tt}|^2\\\\R^{(3)}_{ti}|_{HG}=\frac{1}{2} \triangle
g^{(3)}_{ti}\\\\R^{(2)}_{ij}|_{HG}=\frac{1}{2}\triangle
g^{(2)}_{ij}\end{array}\right.
\end{eqnarray}
and the Ricci scalar (\ref{PPN-ricci-scalar}) becomes

\begin{eqnarray}\label{PPN-ricci-scalar-HG}
\left\{\begin{array}{ll}R^{(2)}|_{HG}=\frac{1}{2}\triangle
g^{(2)}_{tt}-\frac{1}{2}\triangle
g^{(2)}_{mm}\\\\R^{(4)}|_{HG}=\frac{1}{2}\triangle
g^{(4)}_{tt}+\frac{1}{2}g^{(2)}_{mn}g^{(2)}_{tt,mn}-\frac{1}{2}g^{(2)}_{tt,tt}-\frac{1}{2}|\bigtriangledown
g^{(2)}_{tt}|^2-\frac{1}{2}g^{(2)}_{tt}\triangle
g^{(2)}_{tt}-\frac{1}{2}g^{(2)}_{mn}\triangle
g^{(2)}_{mn}\end{array}\right.
\end{eqnarray}
where $\nabla$ and $\triangle$ are, respectively, the gradient and
the Laplacian in flat space. The Einstein tensor components
(\ref{einstein-tensor}) in the harmonic gauge are

\begin{eqnarray}\label{PPN-einstein-tensor-HG}
\left\{\begin{array}{ll}G^{(2)}_{tt}|_{HG}=\frac{1}{4}\triangle
g^{(2)}_{tt}+\frac{1}{4}\triangle g^{(2)}_{mm}\\\\G^{(4)}_{tt}|_{HG}=...\\\\
G^{(3)}_{ti}|_{HG}=\frac{1}{2}\triangle
g^{(3)}_{ti}\\\\G^{(2)}_{ij}|_{HG}= \frac{1}{2}\triangle
g^{(2)}_{ij}+\frac{\delta_{ij}}{4}\biggl[\triangle
g^{(2)}_{tt}-\triangle g^{(2)}_{mm}\biggr]
\end{array}\right.
\end{eqnarray}

On the matter side, \emph{i.e.}\ right-hand side of the field equations
(\ref{fieldequationGR}), we start with the general definition of
the energy-momentum tensor of a perfect fluid (\ref{perfectfluid}) with additional energy $\Pi$

\begin{eqnarray}
T_{\mu\nu}\,=\,(\rho+\Pi\rho+p)u_\mu u_\nu-p\,g_{\mu\nu}
\end{eqnarray}
The explicit form of the energy-momentum tensor can be derived as
follows

\begin{eqnarray}\label{PPN-tensor-matter}
\left\{\begin{array}{ll}
T_{tt}\,=\,\rho+\rho(v^2-2U+\Pi)+\rho\biggl[v^2\biggl(\frac{p}{\rho}+v^2+2V+\Pi\biggr)+\sigma-2\Pi
U\biggr]
\\\\
T_{ti}\,=\,-\rho v^i+\rho\biggl[-v^i\biggl(\frac{p}{\rho}+2V+v^2+\Pi\biggl)+h_{ti}\biggr]
\\\\
T_{ij}\,=\,\rho v^iv^j+p\delta_{ij}+\rho\biggl[v^iv^j\biggl(\Pi+\frac{p}{\rho}+4V+v^2+2U\biggr)
-2v^c\delta_{c(i}h_{0|j)}+2\frac{p}{\rho}V\delta_{ij}\biggr]
\end{array}\right.
\end{eqnarray}
As a first application of these results, let us now take into
account the simplest case, that is GR.

\subsection{The Newtonian and Post-Newtonian Limit of General Relativity}

Einstein Eqs. (\ref{fieldequationGR})  can be rewritten as

\begin{eqnarray}\label{fieldequationGR-2}
R_{\mu\nu}=\mathcal{X}\biggl[T_{\mu\nu}-\frac{T}{2}g_{\mu\nu}\biggr]
\end{eqnarray}
From the interpretation of stress-energy tensor components as
energy density, momentum density and momentum flux,  we have
$T_{tt}$, $T_{ti}$ and $T_{ij}$ at the various orders

\begin{eqnarray}\label{PPN-matter-density}
\left\{\begin{array}{ll}T_{tt}\,=\,T^{(0)}_{tt}+T^{(2)}_{tt}+\mathcal{O}(4)\\\\
T_{ti}\,=\,T^{(1)}_{ti}+\mathcal{O}(3)\\\\T_{ij}=T^{(2)}_{ij}+\mathcal{O}(4)
\end{array}\right.
\end{eqnarray}
where $T^{(N)}_{\mu\nu}$ denotes the term in $T_{\mu\nu}$ of order
$\bar{M}/\bar{r}^3\,\,\bar{v}^{N}$. In particular $T^{(0)}_{tt}$
is the density of rest-mass, while $T^{(2)}_{tt}$ is the
non-relativistic part of the energy density. What we need is the
tensor
\begin{eqnarray}
S_{\mu\nu}\,=\,T_{\mu\nu}-\frac{T}{2}g_{\mu\nu}
\end{eqnarray}
$G\bar{M}/\bar{r}$ is of order $\bar{v}^2$, so (\ref{PPN-metric})
and (\ref{PPN-matter-density}) give

\begin{eqnarray}\label{PPN-matter-density-2}
\left\{\begin{array}{ll}S_{tt}\,=\,S^{(0)}_{tt}+S^{(2)}_{tt}+\mathcal{O}(6)\\\\
S_{ti}\,=\,S^{(1)}_{ti}+\mathcal{O}(3)\\\\S_{ij}\,=\,S^{(0)}_{ij}+\mathcal{O}(2)
\end{array}\right.
\end{eqnarray}
where $S^{(N)}_{\mu\nu}$ denotes the term in $S_{\mu\nu}$ of order
$\bar{M}/\bar{r}^3\,\,\bar{v}^{N}$. In particular

\begin{eqnarray}\label{PPN-matter-density-3}
\left\{\begin{array}{ll}
S^{(0)}_{tt}\,=\,\frac{1}{2}T^{(0)}_{tt}\\\\
S^{(2)}_{tt}\,=\,\frac{1}{2}T^{(2)}_{tt}+\frac{1}{2}T^{(2)}_{mm}\\\\
S^{(1)}_{ti}=T^{(1)}_{ti}\\\\
S^{(0)}_{ij}=\frac{1}{2}\delta_{ij}T^{(0)}_{tt}
\end{array}\right.
\end{eqnarray}
Substituting Eqs.(\ref{PPN-ricci-tensor-HG}) and
(\ref{PPN-matter-density-2}) in  Eqs.(\ref{fieldequationGR-2}), we
find that the field equations in harmonic coordinates are indeed
consistent with the expansions we are using, and give

\begin{eqnarray}
\left\{\begin{array}{ll}R^{(2)}_{tt}\,=\,\mathcal{X}S^{(0)}_{tt}\\\\
R^{(4)}_{tt}\,=\,\mathcal{X}S^{(2)}_{tt}\\\\
R^{(3)}_{ti}\,=\,\mathcal{X}S^{(0)}_{ti}\\\\
R^{(2)}_{ij}\,=\,\mathcal{X}S^{(0)}_{ij}
\end{array}\right.
\end{eqnarray}
and, in particular,

\begin{eqnarray}\label{PPN-fieldequationGR}
\left\{\begin{array}{ll}\triangle
g^{(2)}_{tt}\,=\,\mathcal{X}\,T^{(0)}_{tt}\\\\
\triangle
g^{(4)}_{tt}\,=\,\mathcal{X}\,\biggl[T^{(2)}_{tt}+T^{(2)}_{mm}\biggr]-g^{(2)}_{mn}g^{(2)}_{tt,mn}+g^{(2)}_{tt,tt}+
|\bigtriangledown g^{(2)}_{tt}|^2\\\\
\triangle g^{(3)}_{ti}\,=\,2\,\mathcal{X}\,T^{(1)}_{ti}\\\\
\triangle
g^{(2)}_{ij}\,=\,\mathcal{X}\,\delta_{ij}\,T^{(0)}_{tt}\end{array}\right.
\end{eqnarray}
From the first one of (\ref{PPN-fieldequationGR}), we find, as
expected, the Newtonian result:

\begin{eqnarray}\label{gravitational-potential-metric}
g^{(2)}_{tt}\,=\,-\frac{\mathcal{X}}{4\pi}\int
d^3\mathbf{x}'\frac{T^{(0)}_{tt}(\mathbf{x}')}{|\mathbf{x}
-\mathbf{x}'|}\,=\,-2G\int
d^3\mathbf{x}'\frac{T^{(0)}_{tt}(\mathbf{x}')}{|\mathbf{x}
-\mathbf{x}'|}\,\doteq\,2\Phi(\mathbf{x})
\end{eqnarray}
where $\Phi(\mathbf{x})$ is the gravitational potential. In fact if we consider a
point-like source (\ref{point_like}) we find

\begin{eqnarray}
\Phi(\mathbf{x})\,=\,-\frac{GM}{|\mathbf{x}|}
\end{eqnarray}
From the third and fourth equations of system
(\ref{PPN-fieldequationGR}), we find that

\begin{eqnarray}
\left\{\begin{array}{ll}
g^{(3)}_{ti}\,=\,-\frac{\mathcal{X}}{2\pi}\,\int
d^3\mathbf{x}'\frac{T^{(1)}_{ti}(\mathbf{x}')}{|\mathbf{x}
-\mathbf{x}'|}\,\doteq\,Z_i(\mathbf{x})\\\\
g^{(2)}_{ij}\,=\,-\frac{\mathcal{X}}{4\pi}\,\delta_{ij}\,\int
d^3\mathbf{x}'\frac{T^{(0)}_{tt}(\mathbf{x}')}{|\mathbf{x}
-\mathbf{x}'|}\,=\,2\delta_{ij}\Phi(\mathbf{x})\end{array}\right.
\end{eqnarray}
The second equation of (\ref{PPN-fieldequationGR}) can be
rewritten as follows

\begin{eqnarray}
\triangle\biggl[g^{(4)}_{tt}-2\Phi^2\biggr]\,=\,\mathcal{X}\,\biggl[T^{(2)}_{tt}+T^{(2)}_{mm}\biggr]
-8\Phi\triangle\Phi+2\Phi_{,tt}
\end{eqnarray}
and the solution for $g^{(4)}_{tt}$ is

\begin{eqnarray}
g^{(4)}_{tt}\,=2\Phi^2-\frac{\mathcal{X}}{4\pi}\int
d^3\textbf{x}'\frac{T^{(2)}_{tt}(\mathbf{x}')+T^{(2)}_{mm}(\mathbf{x}')}{|\textbf{x}-\textbf{x}'|}+\frac{2}{\pi}\int
d^3\textbf{x}'\frac{\Phi(\mathbf{x}')\triangle_{\mathbf{x}'}\Phi(\mathbf{x}')}{|\textbf{x}-\textbf{x}'|}-
\frac{1}{2\pi}\int
d^3\textbf{x}'\frac{\Phi_{,tt}(\mathbf{x}')}{|\textbf{x}-\textbf{x}'|}\doteq
2\,\Upsilon(\mathbf{x})
\end{eqnarray}
By using the equations at second order, we obtain the final
expression for the correction at fourth order in the time-time
component of the metric:

\begin{eqnarray}
\Upsilon(\textbf{x})\,=\,\Phi(\textbf{x})^2-\frac{\mathcal{X}}{8\pi}\int
d^3\textbf{x}'\frac{T^{(2)}_{tt}(\mathbf{x}')+T^{(2)}_{mm}(\mathbf{x}')}{|\textbf{x}-\textbf{x}'|}+\frac{\mathcal{X}}{\pi}
\int
d^3\textbf{x}'\frac{\Phi(\mathbf{x}')\,\,T^{(0)}_{tt}(\textbf{x}')}{|\textbf{x}-\textbf{x}'|}-
\frac{1}{4\pi}\partial^2_{tt}\int d^3\textbf{x}'\frac{\Phi
(\mathbf{x}')}{|\textbf{x}-\textbf{x}'|}
\end{eqnarray}
We can rewrite the metric expression (\ref{PPN-metric}) as follows

\begin{eqnarray}\label{PPN-metric-potential-GR}
  g_{\mu\nu}\sim \begin{pmatrix}
  1+2\Phi+2\,\Upsilon & \vec{Z}^T \\
  \vec{Z} & -\delta_{ij}(1-2\Phi)
\end{pmatrix}
\end{eqnarray}
where $\vec{Z}$ are higher order terms  that can be assumed null
at this approximation level.

Finally the Lagrangian of a particle in presence of a
gravitational field can be expressed as proportional to the
invariant distance $ds^{1/2}$, thus we have\,:

\begin{eqnarray}
L\,=\,\biggl(g_{\rho\sigma}\frac{dx^\rho}{dt}\frac{dx^\sigma}{dt}\biggr)^{1/2}\,=\,\biggl(g_{tt}+2g_{tm}v^m+g_{mn}v^m
v^n\biggr)^{1/2}\,=\,\biggl(1+g^{(2)}_{tt}+g^{(4)}_{tt}+2g^{(3)}_{tm}v^m-\textbf{v}^2+g^{(2)}_{mn}v^mv^n\biggr)^{1/2}
\end{eqnarray}
which, to the $\mathcal{O}$(2) order, reduces to the classic Newtonian
Lagrangian of a test particle
$L_{\text{New}}\,=\,\biggl(1+2\Phi-\textbf{v}^2\biggr)^{1/2}$, where
$v^m\,=\,\frac{dx^m}{dt}$ and $|\mathbf{v}|^2\,=\,v^mv^m$. As matter of
fact, post-Newtonian physics has to involve higher than $\mathcal{O}$(2) order
terms in the Lagrangian. In fact we obtain

\begin{eqnarray}
L\,&\sim&\,1+\biggl[\Phi-\frac{1}{2}\mathbf{v}^2\biggr]+\frac{3}{4}\biggl[\Upsilon+Z_m
v^m+\Phi\,\textbf{v}^2\biggr]
\end{eqnarray}

An important remark concerns the odd-order perturbation terms $\mathcal{O}$(1)
or $\mathcal{O}$(3). Since, these terms contain  odd powers of velocity
$\textbf{v}$ or of time derivatives, they are related to the
energy dissipation or absorption by the system. Nevertheless, the
mass-energy conservation prevents the energy and mass losses and,
as a consequence, prevents, in the Newtonian limit, terms of $\mathcal{O}$(1)
and $\mathcal{O}$(3) orders in the  Lagrangian. If one takes into account
contributions  higher than $\mathcal{O}$(4) order, different theories give
different predictions. GR, for example, due to the conservation of
post-Newtonian energy, forbids terms of $\mathcal{O}$(5) order; on the other
hand, terms of $\mathcal{O}$(7) order can appear and are related to the energy
lost by means of the gravitational radiation.

\section{The Newtonian limit of $f(R)$-gravity by the O'Hanlon theroy analogy}\label{TS}

Let us start our analysis of Newtonian and post-Newtonian limits
of Extended Theories of Gravity discussing  the possible
shortcomings related to the use of analogies in the weak field
limit approximation.
 As briefly pointed out above, some authors have  claimed
that FOG models are characterized by an ill defined behavior in
the Newtonian regime. In particular, in a series of papers
\cite{chietall} it is addressed that post-Newtonian corrections of
the gravitational potential violate experimental constraints since
these quantities can be recovered by a direct analogy with
Brans-Dicke Gravity \cite{bransdicke} simply supposing the
Brans-Dicke  parameter $\omega_{BD}$ in Eq.  (\ref{BD-action}) vanishing for $f(R)$-gravity.
Actually ,despite the calculations of the
Newtonian and the post-Newtonian limit of $f(R)$-gravity,
performed in a rigorous manner, have showed that this is not the
case \cite{dick,cap-tro, faraoni4,  nav-van1}, it remains to
clarify why the analogy with Brans-Dicke Gravity seems to fail its
predictions. The issue is easily overcame once the correct analogy
between $f(R)$-gravity and the Brans-Dicke theory is taken into
account.

It can be easily shown that, $f(R)$-gravity models can be rewritten
in term of a scalar-field Lagrangian non minimally coupled with
gravity but without any kinetic term implying
$\omega_{BD}\,=\,0$. Actually, the
simplest case of Scalar-Tensor Gravity models has been
introduced some decades ago by Brans and Dicke in order to give a
general mechanism capable of explaining the inertial forces by
means of a background gravitational interaction. The explicit
expression of such gravitational action is (\ref{BD-action}),
while the general action of $f(R)$-gravity is (\ref{FOGaction}) when $f(X,Y,Z,)\,=\,f(R)$. As
said above, $f(R)$-gravity can be recast as a Scalar-Tensor theory by
introducing a suitable scalar field $\phi$ which non-minimally
couples with the gravity sector. It is important to remark that
such an analogy holds in a formalism in which the scalar field
displays no kinetic term but it is characterized by means of a
self-interaction potential which determines the whole dynamics
(\emph{O'Hanlon Lagrangian}) \cite{ohanlon}. We can resume the
actions as follow

\begin{eqnarray}\label{assdf}
\left\{\begin{array}{ll} \mathcal{A}_{JF}^{f(R)}=\int
d^4x\sqrt{-g}\biggl[f(R)+\mathcal{X}\mathcal{L}_m\biggr]\\\\
\mathcal{A}_{JF}^{BD}=\int d^4x\sqrt{-g}\biggl[\phi
R-\omega_{BD}\frac{\phi_{;\alpha}\phi^{;\alpha}}{\phi}+\mathcal{X}\mathcal{L}_m\biggr]\\\\
\mathcal{A}_{JF}^{OH}=\int d^4x\sqrt{-g}\biggl[\phi
R+V(\phi)+\mathcal{X}\mathcal{L}_m\biggr]
\end{array}\right.
\end{eqnarray}
where $JF$ means that we are considering all theories in the Jordan frame.
This consideration, therefore, implies that the scalar field
Lagrangian equivalent to the purely geometrical $f(R)$-gravity turns
out to be quite different with respect to the ordinary Brans-Dicke
action in  (\ref{BD-action}). This point represents a
crucial aspect of our analysis. In fact, as we will
show, such a difference implies completely different results in
the Newtonian limit of the two models and, consequently, shows that it is misleading to extend predictions from the PPN approximation of
Brans-Dicke models to $f(R)$-gravity. Considering natural units, the
O'Hanlon Lagrangian \cite{ohanlon} is the third of (\ref{assdf}).
The field equations are obtained by varying the action with respect to both
$g_{\mu\nu}$ and $\phi$ which now represent the dynamical variables (the same field
equations are given setting $\omega(\phi)\,=\,0$ and
$F(\phi)\,=\,\phi$ in Eqs. (\ref{TSfieldequation})). Thus, one
obtains

\begin{eqnarray}\label{ohanlon-field-equation}
\left\{\begin{array}{ll} \phi
G_{\mu\nu}-\frac{1}{2}V(\phi)g_{\mu\nu}-\phi_{;\mu\nu}+g_{\mu\nu}\Box\phi\,=\,\mathcal{X}\,T_{\mu\nu}\\\\
R+\frac{dV(\phi)}{d\phi}=0\\\\
\Box\phi+\frac{1}{3}\phi\frac{dV(\phi)}{d\phi}-\frac{2}{3}V(\phi)=\frac{\mathcal{X}}{3}T
\end{array}\right.
\end{eqnarray}
where the second line of (\ref{ohanlon-field-equation}) is the field
equation for $\phi$. While the third equation is a combination of
the trace of the first  and  the second ones. Field equations for
$f(R)$-gravity are obtained from (\ref{FOGaction})

\begin{eqnarray}\label{fieldequationfR2}
\left\{\begin{array}{ll}
f'R_{\mu\nu}-\frac{f}{2}g_{\mu\nu}-f'_{;\mu\nu}+g_{\mu\nu}\Box f'\,=\,\mathcal{X}\,T_{\mu\nu}\\\\
3\Box f'+f'R-2f\,=\,\mathcal{X}\,T
\end{array} \right.
\end{eqnarray}
The two approaches
can be mapped one into the other considering the following
equivalences

\begin{eqnarray}\label{TS-fR-relation}
\left\{\begin{array}{ll}
\phi\,=\,f'\\\\
V(\phi)\,=\,f-f'R\\\\
\phi\frac{d V(\phi)}{d\phi}-2V(\phi)\,=\,f'R-2f
\end{array}\right.
\end{eqnarray}
where the Jacobian matrix of the transformation
$\phi\,\Longleftrightarrow\,f'$ has to be non-vanishing. Henceforth we
can consider instead of (\ref{fieldequationfR2}) a new set of
field equations determined by the equivalence of $f(R)$-gravity with
the O'Hanlon approach  \cite{TS-fR-analogy}

\begin{eqnarray}\label{fets}
\left\{\begin{array}{ll}
\phi R_{\mu\nu}+\frac{1}{6}\biggl(V(\phi)+\phi\frac{d V(\phi)}{d
\phi}\biggr)g_{\mu\nu}-\phi_{;\mu\nu}\,=\,\mathcal
{X}\Sigma_{\mu\nu}\\\\
\Box\phi+\frac{1}{3}\biggl(\phi\frac{d V(\phi)}{d
\phi}-2V(\phi)\biggr)=\frac{\mathcal{X}}{3}T
\end{array}\right.
\end{eqnarray}
where $\Sigma_{\mu\nu}=T_{\mu\nu}-\frac{1}{3}Tg_{\mu\nu}$.
Let us, now, calculate the Newtonian limit of field equations (\ref{fets}).
We take into account the perturbations of  metric tensor $g_{\mu\nu}$ in Eqs.(\ref{PPN-metric})
up to $\mathcal{O}$(2)-order and also for scalar field $\phi$ an analogous
perturbation with respect to the background value

\begin{eqnarray}
\phi\sim\phi^{(0)}+\phi^{(2)}
\end{eqnarray}
The differential operators turn out to be approximated as

\begin{eqnarray}
\Box\approx\partial^2_t-\Delta\,\,\,\,\,\,\,\,\text{and}\,\,\,\,\,\,\,\,\,\nabla_\mu\nabla_\nu\approx\partial^2_{\mu\nu}
\end{eqnarray}
Actually in order to simplify calculations we can exploit the
intrinsic gauge freedom  in the metric definition. In
particular, we choose the harmonic gauge (\ref{gauge-harmonic})
and the expressions of Ricci tensor components are given by (\ref{PPN-ricci-tensor-HG}).
In relation to the adopted approximation we coherently develop the self-interaction potential
at second order. In particular, the quantities in (\ref{fets})
read

\begin{eqnarray}
\left\{\begin{array}{ll} \phi V(\phi)+\phi\frac{d V(\phi)}{d
\phi}\simeq V(\phi^{(0)})+\phi^{(0)}\frac{d V(\phi^{(0)})}{d
\phi}+\biggr[\phi^{(0)} \frac{d^2V(\phi^{(0)})}{d\phi^2}+2\frac{d
V(\phi^{(0)})}{d\phi}\biggl]\phi^{(2)}\\\\
\phi\frac{d V(\phi)}{d \phi}-2V(\phi)\simeq\phi^{(0)}\frac{d
V(\phi^{(0)})}{d \phi}-2V(\phi^{(0)})+\biggr[\phi^{(0)}
\frac{d^2V(\phi^{(0)})}{d\phi^2}-\frac{d
V(\phi^{(0)})}{d\phi}\biggl]\phi^{(2)}
\end{array}\right.
\end{eqnarray}
The field equations (\ref{fets}), solved at $\mathcal{O}$(0)-order of
approximation, provide the two solutions

\begin{eqnarray}
V(\phi^{(0)})=0\,\,\,\,\,\,\,\,\text{and}\,\,\,\,\,\,\,\,\,\frac{dV(\phi^{(0)})}{d\phi}\,=\,0
\end{eqnarray}
which fix the $\mathcal{O}$(0)-order terms in the development of the
self-interaction potential; therefore we have

\begin{eqnarray}
\left\{\begin{array}{ll} V(\phi)+\phi\frac{d V(\phi)}{d
\phi}\simeq\phi^{(0)}
\frac{d^2V(\phi^{(0)})}{d\phi^2}\phi^{(2)}\doteq3m^2\phi^{(2)}\\\\
\phi\frac{d V(\phi)}{\delta \phi}-2V(\phi)\simeq\phi^{(0)}
\frac{\delta^2V(\phi^{(0)})}{d\phi^2}\phi^{(2)}\doteq
3m^2\phi^{(2)}
\end{array}\right.
\end{eqnarray}
where constant factors $\phi^{(0)}\frac{d^2 V(\phi^{(0)})}{d
\phi^2}$ have been condensed within the quantity
$3m^2$\footnote{The factor 3 is introduced to simplify an
analogous factor present in the field equations (\ref{fets}).}.
Such a constant can be easily interpreted as a mass term as will
become clearer in the following. Now, taking into account the
above simplifications, we can rewrite field Eqs. (\ref{fets})
at the at $\mathcal{O}$(2)-order in the form

\begin{eqnarray}\label{fets1.2t}
\triangle
g^{(2)}_{tt}\,=\,\frac{2\mathcal{X}}{\phi^{(0)}}\Sigma_{tt}^{(0)}-m^2\frac{\phi^{(2)}}{
\phi^{(0)}}
\end{eqnarray}
\begin{eqnarray}\label{fets1.2r}
\triangle
g^{(2)}_{ij}\,=\,\frac{2\mathcal{X}}{\phi^{(0)}}\Sigma_{ij}^{(0)}+m^2\frac{\phi^{(2)}}{
\phi^{(0)}}\delta_{ij}+2\frac{\phi^{(2)}_{,ij}}{\phi^{(0)}}
\end{eqnarray}
\begin{eqnarray}\label{fetstr1.2}
\triangle\phi^{(2)}-m^2\phi^{(2)}\,=\,-\frac{\mathcal{X}}{3}T^{(0)}
\end{eqnarray}
The scalar field solution can be easily obtained from Eq. (\ref{fetstr1.2}) as

\begin{eqnarray}
\phi(\mathbf{x})=\phi^{(0)}+\frac{\mathcal{X}}{3}\int\frac{d^3\mathbf{k}}{(2\pi)^{3/2}}\frac{\tilde{T}^{(0)}(\mathbf{k})e^
{i\mathbf{k}\cdot \mathbf{x}}}{\mathbf{k}^2+m^2}
\end{eqnarray}
where $\tilde{T}^{(0)}(\mathbf{k})$ is the Fourier transform of the trace $T^{(0)}$.
While for $g^{(2)}_{tt}$ and $g^{(2)}_{ij}$ we have

\begin{eqnarray}
g^{(2)}_{tt}(\mathbf{x})=-\frac{\mathcal{X}}{2\pi\phi^{(0)}}\int
d^3\mathbf{x}'\frac{\Sigma^{(0)}_{tt}(\mathbf{x}')}{|\mathbf{x}-\mathbf{x}'
|}+\frac{m^2}{4\pi\phi^{(0)}}\int
d^3\mathbf{x}'\frac{\phi^{(2)}(\mathbf{x}')}{|\mathbf{x}-\mathbf{x}'|}
\end{eqnarray}
\begin{eqnarray}
g^{(2)}_{ij}(\mathbf{x})=&-&\frac{\mathcal{X}}{2\pi\phi^{(0)}}\int
d^3\mathbf{x}'\frac{\Sigma^{(0)}_{ij}(\mathbf{x}')}{|\mathbf{x}-\mathbf
{x}'|}-\frac{m^2\delta_{ij}}{4\pi\phi^{(0)}}\int
d^3\mathbf{x}'\frac{\phi^{(2)}(\mathbf{x}')}{|\mathbf{x}-\mathbf{x}'|}\nonumber\\\nonumber\\&+&\frac{2}{\phi^{(0)}}\biggl
[\frac{x_ix_j}{\mathbf{x}^2}\phi^{(2)}(\mathbf{x})+\biggl(\delta_{ij}-\frac{3x_ix_j}{\mathbf{x}^2}\biggr)\frac{1}{|\mathbf{x}|
^3}\int_0^{|\mathbf{x}|}d|\mathbf{x}'||\mathbf{x}|'^2\phi^{(2)}(\mathbf{x}')\biggr]
\end{eqnarray}
The above three solutions represent a completely general result.
In particular adopting  transformations
(\ref{TS-fR-relation}), one can straightforwardly obtain the
solutions in the  $f(R)$-scheme.

Let us analyze the above results with an example. We can
consider a FOG Lagrangian of the form

\begin{eqnarray}\label{gravity_quadratic}
f(R)\,=\,a_1R+a_2R^2
\end{eqnarray}
so that the scalar field reads $\phi\,=\,a_1+2a_2R$ ($a_1$ and $a_2$ are arbitrary constants). The
self-interaction potential turns out the be
$V(\phi)\,=\,-\frac{(\phi-a_1)^2}{4a_2}$ satisfying the conditions
$V(a_1)\,=\,0$ and $V'(a_1)\,=\,0$. In relation with the
definition of the scalar field, we can opportunely identify $a_1$
with a constant value $\phi^{(0)}\,=\,a_1$. Furthermore, the
scalar field "mass" can be expressed in term of the Lagrangian
parameters as follows

\begin{eqnarray}\label{mass_definition_TS}
m^2\,=\,\frac{1}{3}\phi^{(0)}
\frac{\delta^2V(\phi^{(0)})}{\delta\phi^2}\,=\,-\frac{a_1}{6a_2}
\end{eqnarray}
Since the Ricci scalar at the second order reads

\begin{eqnarray}
R\simeq
R^{(2)}\,=\,\frac{\phi^{(2)}}{2a_2}\,=\,\frac{\mathcal{X}}{6a_2}\int\frac{d^3\mathbf{k}}{(2\pi)^{3/2}}\frac{\tilde{T}^{(0)}
(\mathbf{k})e^{i\mathbf{k}\cdot
\mathbf{x}}}{\mathbf{k}^2+m^2}
\end{eqnarray}
if we consider a point-like source (\ref{point_like}) therefore we obtain

\begin{eqnarray}\label{ricci-scalar-solution-ohanlon}
R^{(2)}\,=\,\frac{GM}{3\pi^2a_2}\int
d^3\mathbf{k}\frac{e^{i\mathbf{k}\cdot
\mathbf{x}}}{\mathbf{k}^2+m^2}\,=\,-\sqrt{\frac{\pi}{2}}\frac{r_gm^2}{a_1}\frac{e^{-m|\mathbf{x}|}}{|\mathbf{x}|}
\end{eqnarray}
The immediate consequence is that the solution for the scalar
field $\phi$ at second order is

\begin{eqnarray}
\phi^{(2)}\,=\,2a_2R^{(2)}\,=\,\sqrt{\frac{\pi}{2}}\frac{r_g}{3}\frac{e^{-m|\mathbf{x}|}}{|\mathbf{x}|}
\end{eqnarray}
while the complete scalar field solution up to the second order of
perturbation is given by

\begin{eqnarray}
\phi\,=\,a_1+\sqrt{\frac{\pi}{2}}\frac{r_g}{3}\frac{e^{-m|\mathbf{x}|}}{|\mathbf{x}|}
\end{eqnarray}
Once the behavior of the scalar field has been obtained up to the
second order of perturbation, in the same way, one can deduce the
expressions for $g^{(2)}_{tt}$ and $g^{(2)}_{ij}$, where
$\Sigma^{(0)}_{tt}\,=\,\frac{2}{3}\rho$ and
$\Sigma^{(0)}_{ij}\,=\,\frac{1}{3}\rho\,\delta_{ij}\,=\,\frac{1}{2}\Sigma^{(0)}_{tt}\delta_{ij}$.
As matter of fact the metric solutions at the second order of
perturbation are

\begin{eqnarray}\label{gsol1}
\left\{\begin{array}{ll}
g_{tt}\,=\,1-\frac{2}{3a_1}\frac{r_g}{|\mathbf{x}|}-\sqrt{\frac{\pi}{2}}\frac{1}{3a_1}\frac{r_ge^{-m|\mathbf{x}|}}{|\mathbf{x}|}
\\\\
g_{ij}\,=\,-\biggl\{1+\frac{1}{3a_1}\frac{r_g}{|\mathbf{x}|}-\sqrt{\frac{\pi}{2}}\frac{r_g}{3a_1}\biggl[\biggl(\frac{1}{|\mathbf{x}
|}-\frac{2}{m|\textbf{x}|^2}-\frac{2}{m^2|\textbf{x}|^3}\biggr)e^{-m|\mathbf{x}|}-\frac{2}{m^2
|\textbf{x}|^3}\biggr]\biggr\}\delta_{ij}\\\\\,\,\,\,\,\,\,\,\,\,\,\,\,\,\,\,+\frac{(2\pi)^{1/2}r_g}{3a_1}\biggl[\biggl(\frac{1}
{|\textbf{x}|}+\frac{3}{m|\textbf{x}|^2}+\frac{3}{m^2|\textbf{x}|^3}\biggr)e^{-m|\textbf{x}|}-\frac{3}
{m^2|\textbf{x}|^3}\biggr]\frac{x_ix_j}{|\textbf{x}|^2}
\end{array}\right.
\end{eqnarray}
This quantity, which is directly related to the gravitational
potential, shows that the gravitational potential of the O'Hanlon
Lagrangian is non-Newtonian. Such a behavior prevents from
obtaining a natural definition of the PPN parameters as
corrections to the Newtonian potential. As matter of fact, since it
is indeed not true that a generic $f(R)$-gravity model corresponds
to a Brans-Dicke model with $\omega_{BD}\,=\,0$ coherently to its
post-Newtonian approximation. In particular it turns out to be
wrong considering the PPN parameter
$\gamma\,=\,\frac{1+\omega_{BD}}{2+\omega_{BD}}$ (see, for
example, \cite{will}) of Brans - Dicke gravity and evaluating this
at $\omega_{BD}\,=\,0$ so that one gets $\gamma\,=\,1/2$ as
derived in \cite{chietall}.

Differently, because of the presence of the self-interaction
potential $V(\phi)$, in the O'Hanlon Lagrangian, a Yukawa like
correction appears in the Newtonian limit. As matter of fact, one obtains a
different gravitational potential with respect to the standard
Newtonian one and, as matter of fact, the fourth order corrections
in term of the $v/c$ ratio (Newtonian level), have to be evaluated
in a complete new  way. In other words, considering a
Brans-Dicke Lagrangian and an O'Hanlon one, despite their similar
structure,  implies different predictions in the weak
field and small velocity limits. Such a result represents a
significant argument against the claim that FOG models can be
ruled out only on the bases of the analogy with Brans-Dicke PPN
parameters.

An important consideration is in order at this point on  the meaning of  PPN-parameters
$\gamma$ and $\beta$, defined as a correction to the Newtonian-like behavior of the
gravitational potentials (\ref{schwarz-isotropic-PPN}). Actually,
if we consider the limit $f(R)\rightarrow R$, from (\ref{gsol1}) and set $a_1\,=\,2/3$
($a_1$ is an arbitrary constant), we have

\begin{eqnarray}
\left\{\begin{array}{ll}
g_{tt}\,=\,1-\frac{r_g}{|\mathbf{x}|}\\\\
g_{ij}\,=\,-\biggl(1+\frac{1}{2}\frac{r_g}{|\mathbf{x}|}\biggr)\delta_{ij}
\end{array}\right.
\end{eqnarray}
which suggest that the PPN parameter $\gamma$, in this limit,
results $1/2$ which is in striking contrast with GR predictions
($\gamma\sim1$). Such a result is  not surprising. In
fact, the GR limit of the O'Hanlon Lagrangian requires $\phi\sim$
const and $V(\phi)\rightarrow 0$ but such approximations induce
mathematical inconsistencies in the field equations of $f(R)$-gravity,
once these have been obtained by a given O'Hanlon
Lagrangian. Actually this is a general issue of O'Hanlon
Lagrangian. In fact it can be demonstrated that  field
Eqs. (\ref{fets}) do not reduce to the standard GR ones (for
$V(\phi)\rightarrow0$ and $\phi\sim\text{const}$) since we
have

\begin{eqnarray}
\left\{\begin{array}{ll}
\phi R_{\mu\nu}+\frac{1}{6}\biggl(V(\phi)+\phi\frac{d V(\phi)}{d
\phi}\biggr)g_{\mu\nu}-\phi_{;\mu\nu}\,=\,\mathcal
{X}\Sigma_{\mu\nu}\\\\
\Box\phi+\frac{1}{3}\biggl(\phi\frac{d V(\phi)}{d
\phi}-2V(\phi)\biggr)=\frac{\mathcal{X}}{3}T
\end{array}\right.
\,\,\,\,\,\,\,\,\,\,\,\rightarrow\,\,\,\,\,\,\,\,\,\,\,\
\left\{\begin{array}{ll}
R_{\mu\nu}=\frac{\mathcal{X}}{a_1}\Sigma_{\mu\nu}\\\\
0=\frac{\mathcal{X}}{3}T
\end{array}\right.
\end{eqnarray}
In fact  $\Sigma_{\mu\nu}$ components read
$\Sigma_{tt}\,=\,\frac{2}{3}\rho$ and
$\Sigma_{ij}\,=\,\frac{1}{3}\rho\,\delta_{ij}\,=\,\frac{1}{2}\Sigma_{tt}\,\delta_{ij}$
in place of $S_{tt}\,=\,\frac{1}{2}\rho$ and
$S_{ij}\,=\,\frac{1}{2}\rho\,\delta_{ij}\,=\,S_{tt}\,\delta_{ij}$ as it should be for
 the GR field  Eqs. (\ref{fieldequationGR-2}).
Such a pathology emerges also when the GR limit is performed
from a pure Brans-Dicke Lagrangian. In such a case, in order to
match the Hilbert-Einstein Lagrangian, one needs $\phi\sim
const$ and $\omega_{BD}\,=\,0$, the immediate consequence is that the
PPN parameter $\gamma$ turns out to be $1/2$, while it is well
known that Brans-Dicke model fulfils low energy limit
prescriptions in the limit $\omega\rightarrow \infty$. Even in
this case, the problem, with respect to the GR predictions, is that
the GR limit of the model introduces inconsistencies in the field
equations. In other words, it is not possible to impose the same
transformation which leads the Brans-Dicke theory into GR at the
Lagrangian level on the solutions obtained by solving the field
equations descending from the general Lagrangian. The relevant
aspect of this discussion is that considering a $f(R)$-model, in
analogy with the O'Hanlon Lagrangian and supposing that the
self-interaction potential is negligible, introduces a
pathological behaviour in  dynamics which results in
obtaining a PPN parameter $\gamma\,=\,1/2$. This is what happens when
an effective approximation scheme is introduced in the field
equations in order to calculate the weak field limit of FOG by
means of Brans-Dicke model. Such a result seems, from another
point of view, to enforce the claim that FOG models have to be
carefully investigated in this limit and their analogy with
scalar-tensor gravity should be opportunely considered.

\subsection{Scalar-Tensor Gravity in  Jordan and Einstein frames}

Up to now, we have discussed the weak field
and small velocity limit of FOG in term of Brans-Dicke like
Lagrangian remaining in the Jordan frame. We have considered
 the weak field and small velocity limit when a
conformal transformation (\ref{transconf}) is applied to the
O'Hanlon Lagrangian.  Now we want to analyze the differences of considering a Scalar-Tensor theory
in the Jordan frame and in the Einstein frame.The
Scalar-Tensor action $\mathcal{A}_{JF}^{ST}$ in the Jordan frame
(\ref{TSaction}) is linked to the action
$\mathcal{A}_{EF}^{ST}$ in the Einstein frame (\ref{TS-EF-action})
via the transformations (\ref{transconfTS}) between the quantities
in the two frames. The O'Hanlon theory in the
Jordan frame is recovered  from Eq.(\ref{TSaction}) imposing  $F(\phi)\,=\,\phi$ and
$\omega{(\phi)}\,=\,0$.  Action (\ref{TS-EF-action}), in the
Einstein frame results  simplified and the transformation between the
two scalar fields reads

\begin{equation}\label{phivarphi}
\Omega(\varphi){d\varphi}^2\,=\,-\frac{3\Lambda}{2}\frac{{d\phi}^2}{\phi^2}
\end{equation}
If we suppose $\Omega(\varphi)\,=\,-\Omega_{0}<0$ we have

\begin{equation}\label{phivarphisol}
\phi\,=\,k\,e^{\pm \lambda\varphi}
\end{equation}
where $\lambda=\sqrt{\frac{2\Omega_0}{3\Lambda}}$ and $k$ is an
integration constant. The O'Hanlon theory, transformed
in the Einstein frame, is

\begin{eqnarray}\label{ohanlon-EF}
\mathcal{A}_{EF}^{OH}=\int
d^4x\sqrt{-\tilde{g}}\biggl[\Lambda\tilde{R}-\Omega_0\varphi_
{;\alpha}\varphi^{;\alpha}+\frac{\Lambda^2}{k^2}e^{\mp
2\lambda\varphi}V(k\,e^{\pm
\lambda\varphi})+\frac{\mathcal{X}\Lambda^2}{k^2}e^{\mp
2\lambda\varphi}\mathcal{L}_m\biggl(\frac{\Lambda}{k}e^{\mp
\lambda\varphi}\tilde{g}_{\rho\sigma}\biggr)\biggr]
\end{eqnarray}
The field equations are

\begin{eqnarray}
\left\{\begin{array}{ll}
\Lambda\tilde{G}_{\mu\nu}-\frac{1}{2}\frac{\Lambda^2}{k^2}e^{\mp
2\lambda\varphi}V(k\,e^{\pm
\lambda\varphi})\tilde{g}_{\mu\nu}-\Omega_0\biggl(\varphi_{;\mu}\varphi_{;\nu}-\frac{1}{2}
\varphi_{;\alpha}\varphi^{;\alpha}\tilde{g}_{\mu\nu}\biggr)\,=\,\mathcal{X}\,\tilde{T}^\varphi_{\mu\nu}
\\\\
2\Omega_0\tilde{\Box}\varphi+\frac{\Lambda^2}{k^2}e^{\mp
2\lambda\varphi}[\frac{\delta V}{\delta\phi}(k\,e^{\pm \lambda\varphi})\mp 2 \lambda
V(k\,e^{\pm
\lambda\varphi})]+\mathcal{X}\tilde{\mathcal{L}}_{m,\varphi}\,=\,0
\\\\
\tilde{R}\,=\,-\frac{\mathcal{X}}{2\Lambda}\tilde{T}^\varphi+\frac{\Omega_0}{\Lambda}\varphi_{;\alpha}\varphi^{;\alpha}-\frac{2
\Lambda}{k^2}e^{\mp 2\lambda\varphi}V(ke^{\pm \lambda\varphi})
\end{array} \right.
\end{eqnarray}
where the matter tensor, which now coupled with the scalar field
$\varphi$, in the Einstein frame reads

\begin{eqnarray}
\tilde{T}^\varphi_{\mu\nu}\,=\,\frac{-1}{\sqrt{-\tilde{g}}}\frac{\delta(\sqrt{-\tilde{g}}\tilde{\mathcal{L}}
_m)}{\delta\tilde{g}^{\mu\nu}}=\frac{\Lambda^2}{2k^2}e^{\mp
2\lambda\varphi}\biggl[\mathcal{L}_m\biggl(\frac{\Lambda}{k}e^{\mp \lambda
\varphi}\tilde{g}_{\rho\sigma}\biggr)\tilde{g}_{\mu\nu}-2\frac{\delta}{\delta\tilde{g}^{\mu\nu}}\mathcal{L}_m
\biggl(\frac{\Lambda}{k}e^{\mp
\lambda\varphi}\tilde{g}_{\rho\sigma}\biggr)\biggr]
\end{eqnarray}
and
\begin{eqnarray}
\tilde{\mathcal{L}}_{m,\varphi}\,=\,\mp\frac{\Lambda^2\lambda}{k^2}e^{\mp
2\lambda\varphi}\biggl[2\mathcal{L}_m\biggl(\frac{\Lambda}{k}e^{\mp
\lambda\varphi}\tilde{g}_{\rho\sigma}\biggr)+\frac{\Lambda}{k}e^{\mp
\lambda\varphi}\tilde{g}_{\rho\sigma}\frac{\delta\mathcal{L}_m}{\delta
g_{\rho\sigma}}\biggl(\frac{\Lambda}{k}e^{\mp \lambda
\varphi}\tilde{g}_{\rho\sigma}\biggr)\biggr]
\end{eqnarray}
Actually, in order to calculate the weak field and small velocity
limit of the model in the Einstein frame, we can develop the two
scalar fields at the second order $\phi\sim \phi^{(0)}+\phi^{(2)}$
and $\varphi\sim \varphi^{(0)}+\varphi^{(2)}$ with respect to a
background value. This choice gives the relations

\begin{eqnarray}
\left\{\begin{array}{ll}\varphi^{(0)}\,=\,\pm \lambda^{-1}\ln{\frac{\phi^{(0)}}{k}}\\\\
\varphi^{(2)}\,=\,\pm
\lambda^{-1}\frac{\phi^{(2)}}{\phi^{(0)}}\end{array}\right.
\end{eqnarray}
Let us consider the conformal transformation
$\tilde{g}_{\mu\nu}\,=\,\frac{\phi}{\Lambda}g_{\mu\nu}$. From this relation, considering Eq.(\ref{phivarphisol}),
 one obtains for  $\phi^{(0)}\,=\,\Lambda$

\begin{eqnarray}\label{confg}
\left\{\begin{array}{ll}\tilde{g}^{(2)}_{tt}\,=\,g^{(2)}_{tt}+\frac{\phi^{(2)}}{\phi^{(0)}}\\\\
\tilde{g}^{(2)}_{ij}\,=\,g^{(2)}_{ij}-\frac{\phi^{(2)}}{\phi^{(0)}}\delta_{ij}\end{array}\right.
\end{eqnarray}
As matter of fact, since $g_{tt}^{(2)}\,=\,2\,\Phi^{JF}$,
$g_{ij}^{(2)}\,=\,2\,\Psi^{JF}\,\delta_{ij}$ and
$\tilde{g}_{tt}^{(2)}\,=\,2\,\Phi^{EF}$,
$\tilde{g}_{ij}^{(2)}\,=\,2\,\Psi^{EF}\,\delta_{ij}$ from
Eqs.(\ref{confg}),  relevant relations emerge  linking the
gravitational potentials between Jordan and Einstein frames

\begin{eqnarray}\label{gravpotconf1}
\left\{\begin{array}{ll}
\Phi^{EF}\,=\,\Phi^{JF}+\frac{\phi^{(2)}}{2\phi^{(0)}}\,=\,\Phi^{JF}\pm\frac{\lambda}{2}\varphi^{(2)}\\\\
\Psi^{EF}\,=\,\Psi^{JF}-\frac{\phi^{(2)}}{2\phi^{(0)}}\,=\,\Psi^{JF}\mp\frac{\lambda}{2}\varphi^{(2)}\end{array}\right.
\end{eqnarray}
If we introduce the variations of two potentials,
$\Delta\Phi\,=\,\Phi^{JF}-\Phi^{EF}$ and
$\Delta\Psi\,=\,\Psi^{JF}-\Psi^{EF}$, we obtain

\begin{eqnarray}
\Delta\Phi\,=\,-\,\Delta\Psi\,=\,-\,\frac{\phi^{(2)}}{2\phi^{(0)}}\,=\,\mp\,\frac{\lambda}{2}\varphi^{(2)}\,
\propto\,a_2\,\propto\,f''(0)
\end{eqnarray}
here specified in the case of Lagrangian (\ref{gravity_quadratic}).

From the above expressions, one can notice that there is an
evident difference between the behavior of the two gravitational
potentials in the two frames \cite{TS-fR-analogy}. Such a result suggests that, at the
Newtonian level, it is possible to discriminate between the two
 frames thus one can deduce what is the true physical
one. In particular, once, the gravitational potential is
calculated in the Jordan frame and the dynamical evolution of
$\phi$ is taken into account at the suitable perturbation level,
this can be substituted in the first of (\ref{gravpotconf1})  to obtain its Einstein frame evolution. The final step is
that the two potentials have to be matched with experimental data
in order to select what is the true physical solution.

\section{The Newtonian limit of $f(R)$-gravity  in standard coordinates}\label{newtlimstandard}

The Newtonian limit of FOG can be worked out by comparing its viability
with respect to the standard results of GR. Here, we
investigate the limit in the metric approach, refraining from
exploiting the formal equivalence of FOG with
 specific Scalar-Tensor theories,
\emph{i.e.} we work in the Jordan frame in order to avoid possible
misleading interpretations of the results \cite{newtonian_limit_fR}.

Considering the Taylor expansion of a generic $f(R)$-gravity model, it is possible to obtain
general solutions in term of the metric coefficients up to the
third order of approximation. Furthermore, it is possible to show that the
Birkhoff theorem is not a general result for $f(R)$-gravity since
time-dependent evolution of spherically symmetric solutions can
be achieved depending on the order of perturbations.

Exploiting the formalism of Newtonian and post-Newtonian
approximations previously described, we can develop a systematic
analysis in the limits of weak field and small velocities for
$f(R)$-gravity. We are going to assume, as background, a spherically
symmetric spacetime and we are going to investigate the vacuum
case. Considering the metric (\ref{me5}), we have, for a given
$g_{\mu\nu}$

\begin{eqnarray}\label{definexpans}
\left\{\begin{array}{ll}
g_{tt}(t,r)\,\simeq\,1+g^{(2)}_{tt}(t,r)+g^{(4)}_{tt}(t,r)
\\\\
g_{rr}(t,r)\,\simeq\,-1+g^{(2)}_{rr}(t,r)\\\\
g_{\theta\theta}(t,r)\,=\,-r^2\\\\
g_{\phi\phi}(t,r)\,=\,-r^2\sin^2\theta
\end{array}\right.
\end{eqnarray}
while considering Eqs. (\ref{PPN-metric-contro}), it is

\begin{eqnarray}
\left\{\begin{array}{ll}
g^{tt}\simeq
1-g^{(2)}_{tt}+[{g^{(2)}_{tt}}^2-g^{(4)}_{tt}]
\\\\
g^{rr}\simeq-1-g^{(2)}_{rr}
\end{array} \right.
\end{eqnarray}
The determinant reads

\begin{eqnarray}
g\simeq r^4\sin^2\theta\{-1+[g^{(2)}_{rr}-g^{(2)}_{tt}]+[g^{(2)}_{tt}g^{(2)}_{rr}-g^{(4)}_{tt}]\}
\end{eqnarray}
 Christoffel symbols (\ref{PPN-christoffel}) are

\begin{eqnarray}
\left\{\begin{array}{ll} \begin{array}{ccc}
  {\Gamma^{(3)}}^{t}_{tt}=\frac{g^{(2)}_{tt,t}}{2}\, & \,\,\, & {\Gamma^{(2)}}^{r}_{tt}+{\Gamma^{(4)}}^{r}_{tt}=
  \frac{g^{(2)}_{tt,r}}{2}+\frac{g^{(2)}_{rr}g^{(2)}_{tt,r}+g^{(4)}_{tt,r}}{2} \\
  & & \\
  {\Gamma^{(3)}}^{r}_{tr}=-\frac{g^{(2)}_{rr,t}}{2}\, & \,\,\, & {\Gamma^{(2)}}^{t}_{tr}+{\Gamma^{(4)}}^{t}_{tr}
  =\frac{g^{(2)}_{tt,r}}{2}+\frac{g^{(4)}_{tt,r}-g^{(2)}_{tt}g^{(2)}_{tt,r}}{2} \\
  & & \\
  {\Gamma^{(3)}}^{t}_{rr}=-\frac{g^{(2)}_{rr,t}}{2}\, & \,\,\, & {\Gamma^{(2)}}^{r}_{rr}+{\Gamma^{(4)}}^{r}_{rr}
  =-\frac{g^{(2)}_{rr,r}}{2}-\frac{g^{(2)}_{rr}g^{(2)}_{rr,r}}{2} \\
  & & \\
  \Gamma^{r}_{\phi\phi}=\sin^2\theta \Gamma^{r}_{\theta\theta}\, & \,\,\, & {\Gamma^{(0)}}^{r}_{\theta\theta}+
  {\Gamma^{(2)}}^{r}_{\theta\theta}+{\Gamma^{(4)}}^{r}_{\theta\theta}=-r-rg^{(2)}_{rr}-r{g^{(2)}_{rr}}^2 \\
\end{array}
\end{array} \right.
\end{eqnarray}
while the Ricci tensor components (\ref{PPN-ricci-tensor})  are

\begin{eqnarray}
\left\{\begin{array}{ll}
R^{(2)}_{tt}=\frac{rg^{(2)}_{tt,rr}+2g^{(2)}_{tt,r}}{2r}\\\\
R^{(4)}_{tt}=\frac{-r{g^{(2)}_{tt,r}}^2+4g^{(4)}_{tt,r}+rg^{(2)}_{tt,r}g^{(2)}_{rr,r}+2g^{(2)}_{rr}[2g^{(2)}_{tt,r}+
rg^{(2)}_{tt,rr}]+2rg^{(4)}_{tt,rr}+2rg^{(2)}_{rr,tt}}{4r}\\\\
R^{(3)}_{tr}=-\frac{g^{(2)}_{rr,t}}{r}\\\\
R^{(2)}_{rr}=-\frac{rg^{(2)}_{tt,rr}+2g^{(2)}_{rr,r}}{2r}\\\\
R^{(2)}_{\theta\theta}=-\frac{2g^{(2)}_{rr}+r[g^{(2)}_{tt,r}+g^{(2)}_{rr,r}]}{2}\\\\
R^{(2)}_{\phi\phi}=\sin^2\theta R^{(2)}_{\theta\theta}
\end{array} \right.
\end{eqnarray}
and, finally, the Ricci scalar expression is

\begin{eqnarray}
\left\{\begin{array}{ll}
R^{(2)}\,=\,\frac{2g^{(2)}_{rr}+r[2g^{(2)}_{tt,r}+2g^{(2)}_{rr,r}+rg^{(2)}_{tt,rr}]}{r^2}\\\\
R^{(4)}\,=\,\frac{1}{2r^2}\biggr[4{g^{(2)}_{rr}}^2+2rg^{(2)}_{rr}[2g^{(2)}_{tt,r}+4g^{(2)}_{rr,r}+rg^{(2)}_{tt,rr}]+r\{-r{g^{(2)}
_{tt,r}}^2+4g^{(4)}_{tt,r}+\\\,\,\,\,\,\,\,\,\,\,\,\,\,\,\,\,\,\,\,\,\,\,\,\,\,\,\,\,\,\,\,\,\,\,\,\,\,\,\,\,\,+rg^{(2)}_
{tt,r}g^{(2)}_{rr,r}-2g^{(2)}_{tt}[2g^{(2)}_{tt,r}+rg^{(2)}_{tt,rr}]+2rg^{(4)}_{tt,rr}+2rg^{(2)}_{rr,tt}\}\biggr]\end{array}
\right.\end{eqnarray}
By metric tensor (\ref{definexpans}) and by inserting it into the field equations
(\ref{fieldequationfR2}), one obtains

\begin{eqnarray}\label{asd}
\left\{\begin{array}{ll}
H_{\mu\nu}\,=\,f'R_{\mu\nu}-\frac{1}{2}fg_{\mu\nu}-f''\biggl\{R_{,\mu\nu}-\Gamma^t_{\mu\nu}R_{,t}-\Gamma^r_{\mu\nu}R_{,r}-
g_{\mu\nu}\biggl[\biggl({g^{tt}}_{,t}+g^{tt}\ln\sqrt{-g}_{,t}\biggr)R_{,t}\\
\,\,\,\,\,\,\,\,\,\,\,\,\,\,\,\,\,\,
+\biggl({g^{rr}}_{,r}+g^{rr}\ln\sqrt{-g}_{,r}\biggr)R_{,r}+g^{tt}R_{,tt}
+g^{rr}R_{,rr}\biggr]\biggr\}-f'''\biggl[R_{,\mu}R_{,\nu}-g_{\mu\nu}\biggl(g^{tt}{R_{,t}}^2+g^{rr}{R_{,r}}^2\biggr)\biggr]
\\\\
H\,=\,f'R-2f+3f''\biggl[\biggl({g^{tt}}_{,t}+g^{tt}\ln\sqrt{-g}_{,t}\biggr)R_{,t}+\biggl({g^{rr}}_{,r}+g^{rr}\ln\sqrt{-g}_{,r}
\biggr)R_{,r}+g^{tt}R_{,tt}+g^{rr}R_{,rr}\biggr]\\\,\,\,\,\,\,\,\,\,\,\,\,\,\,+3f'''\biggl[g^{tt}{R_{,t}}^2
+g^{rr}{R_{,r}}^2\biggr]
\end{array} \right.
\end{eqnarray}

In order to derive the Newtonian and post-Newtonian approximations
for a generic $f(R)$-function , one should specify the
$f(R)$\,-\,Lagrangian into the field Eqs.(\ref{asd}). This is a
crucial point because once a certain Lagrangian is chosen, one
will obtain a particular approximation referred to such a choice.
This means to lose any general prescription and to obtain
corrections to the Newtonian potential, $\Phi(\mathbf{x})$, which
refer "univocally" to the considered $f(R)$-function.
Alternatively, one can restrict to analytic $f(R)$-functions
expandable with respect to a certain value $R\,=\,R_0\,=$ constant
or, at least, its non-analytic part, if exists at all, goes to
zero faster than $R^n$, with $n\geq 2$ at $R\rightarrow 0$. In
general, such theories are physically interesting and allow to
recover the GR results and the correct boundary and asymptotic
conditions. Then we assume

\begin{eqnarray}\label{sertay}
f(R)\,=\,\sum_{n}\frac{f^n(R_0)}{n!}(R-R_0)^n\simeq
f_0+f_1R+f_2R^2+f_3R^3+\dots
\end{eqnarray}
One has to note that the expansion (\ref{sertay}), also if
similar to (\ref{approx}),  presents some differences. In fact $R^{(0)}$ is a
general space-time function linked to the background metric
$g^{(0)}_{\mu\nu}$  given in (\ref{approx-metric}). Here $R_0$ is a
constant value of scalar curvature, which can be negligible in weak field limit approximation.
Besides, the coefficients
$f_0$, $f_1$, $f_2$, $f_3$ are not proportional, respectively, to
zero-th, first, second, third coefficient of the Taylor expansion of
$f(R)$. In fact, we have

\begin{eqnarray}\label{TS-function-gravity}
\left\{\begin{array}{ll}
f_0\,=\,f(R_0)-R_0f'(R_0)+\frac{1}{2}R_0^2f''(R_0)-\frac{1}{6}R_0^3f'''(R_0)
\\\\
f_1\,=\,f'(R_0)-R_0f''(R_0)+\frac{1}{2}R_0^2f'''(R_0)
\\\\
f_2\,=\,\frac{1}{2}f''(R_0)-\frac{1}{2}R_0f'''(R_0)
\\\\
f_3\,=\,\frac{1}{6}f'''(R_0)
\end{array} \right.
\end{eqnarray}
If we consider a flat background, then $R_0\,=\,0$ and the
coefficients $f_0$, $f_1$, $f_2$, $f_3$ are the terms of Taylor
series. But if we are searching for solutions at Newtonian and
(possibility) post-Newtonian level, we have to consider a vanishing
background scalar curvature. It is possible to obtain the
Newtonian and post-Newtonian approximation of $f(R)$-gravity
considering such an expansion (\ref{sertay}) into the field
Eqs. (\ref{asd}) and to expand the system up to the orders
$\mathcal{O}$(0), $\mathcal{O}$(2), $\mathcal{O}$(3) and $\mathcal{O}$(4).
This approach provides general results and specific (analytic) Lagrangians are selected by  the
coefficients $f_i$ in Eq.(\ref{sertay}). Developing the equations in
the case of vanishing matter, \emph{i.e.} $T_{\mu\nu}\,=\,0$, we have

\begin{eqnarray}\label{sys1}
\left\{\begin{array}{ll}
H^{(0)}_{\mu\nu}\,=\,0,\,\,&\,\,H^{(0)}\,=\,0
\\\\
H^{(2)}_{\mu\nu}\,=\,0,\,\,&\,\,H^{(2)}\,=\,0
\\\\
H^{(3)}_{\mu\nu}\,=\,0,\,\,&\,\,H^{(3)}\,=\,0
\\\\
H^{(4)}_{\mu\nu}\,=\,0,\,\,&\,\,H^{(4)}\,=\,0
\end{array} \right.
\end{eqnarray}
and, in particular, from the $\mathcal{O}$(0) order approximation, one obtains

\begin{eqnarray}\label{eq0}
f_0\,=\,0
\end{eqnarray}
which trivially follows from the above assumption that the
space-time is asymptotically Minkowski  (asymptotically flat
background). This result suggests a first  consideration. \emph{If
the Lagrangian is expanded around a vanishing value of the
Ricci scalar ($R_0\,=\,0$), the relation (\ref{eq0})  implies that
the cosmological constant contribution has to be zero whatever is
the $f(R)$-gravity model}.

If we now consider the $\mathcal{O}$(2)-order approximation,
system (\ref{sys1}), in the vacuum case, results to be

\begin{eqnarray}\label{eq23}
\left\{\begin{array}{ll}
f_1rR^{(2)}-2f_1g^{(2)}_{tt,r}+8f_2R^{(2)}_{,r}-f_1rg^{(2)}_{tt,rr}+4f_2rR^{(2)}\,=\,0
\\\\
f_1rR^{(2)}-2f_1g^{(2)}_{rr,r}+8f_2R^{(2)}_{,r}-f_1rg^{(2)}_{tt,rr}\,=\,0
\\\\
2f_1g^{(2)}_{rr}-r[f_1rR^{(2)}-f_1g^{(2)}_{tt,r}-f_1g^{(2)}_{rr,r}+4f_2R^{(2)}_{,r}+4f_2rR^{(2)}_{,rr}]\,=\,0
\\\\
f_1rR^{(2)}+6f_2[2R^{(2)}_{,r}+rR^{(2)}_{,rr}]\,=\,0
\\\\
2g^{(2)}_{rr}+r[2g^{(2)}_{tt,r}-rR^{(2)}+2g^{(2)}_{rr,r}+rg^{(2)}_{tt,rr}]\,=\,0
\end{array} \right.
\end{eqnarray}
The last equation of the system (\ref{eq23}) is the definition of
Ricci scalar (\ref{ricciscalar}) at $\mathcal{O}$(2)-order. The trace
equation (the fourth line in Eqs. (\ref{eq23})), in particular,
provides a differential equation with respect to the Ricci scalar
which allows to solve, if $\text{sign}[f_1]\,=\,-\text{sign}[f_2]$,
the system (\ref{eq23}) at $\mathcal{O}$(2)-order. The solutions are

\begin{eqnarray}\label{yukawa-solution-O(2)-order}
\left\{\begin{array}{ll}
g^{(2)}_{tt}\,=\,\delta_0-\frac{\delta_1}{f_1r}+\frac{\delta_2(t)}{3m}\frac{e^{-m
r}}{mr}+\frac{\delta_3(t)}{6m^2}\frac{e^{m
r}}{mr}
\\\\
g^{(2)}_{rr}\,=\,-\frac{\delta_1}{f_1r}-\frac{\delta_2(t)}{3m}\frac{mr+1}{mr}e^{-mr}+\frac{\delta_3(t)}{6m^2}\frac{mr-1}{mr}e^{mr}
\\\\
R^{(2)}\,=\,\delta_2(t)\frac{e^{-mr}}{r}+\frac{\delta_3(t)}{2m}\frac{e^{mr}}{r}
\end{array}
\right.
\end{eqnarray}
where

\begin{eqnarray}\label{yukawa-length}
m\,\doteq\,\sqrt{-\frac{f_1}{6f_2}}
\end{eqnarray}
with the dimension of \emph{length}$^{-1}$. We note that the definition of mass (\ref{yukawa-length}) is compatible
with the definition of (\ref{mass_definition_TS}). Let us notice that the
integration constant $\delta_0$ has to be dimensionless,
$\delta_1$ has the dimension of \emph{length}, while the time -
dependent functions $\delta_2$ and $\delta_3$, respectively, have
the dimensions of \emph{length}$^{-1}$ and \emph{length}$^{-2}$.
The functions $\delta_i(t)$ ($i\,=\,2,3$) are completely arbitrary
since the differential equation system (\ref{eq23}) contains only
spatial derivatives. Besides, the integration constant $\delta_0$
can be set to zero, as in the theory of the potential,  since it
represents an unessential additive quantity. When we consider the
limit $f(R)\,\rightarrow\,R$, in the case of a point-like source (\ref{point_like}),
we recover the perturbed version of standard Schwarzschild
solution (\ref{schwarz-solution-stand-coord}) at $\mathcal{O}$(2)-order with
$\delta_1\,=\,r_g$. In order to match at infinity the Minkowskian
prescription of the metric, we discard the Yukawa growing mode
present in (\ref{yukawa-solution-O(2)-order}). Then we have, in standard coordinates,

\begin{eqnarray}\label{yukawa-solution-O(2)-order-1}
\left\{\begin{array}{ll}
ds^2\,=\,\biggl[1-\frac{r_g}{f_1r}+\frac{\delta_2(t)}{3m}\frac{e^{-mr}}{mr}\biggr]dt^2-\biggl[1+\frac{r_g}{f_1r}+\frac{\delta_2(t)}{3m}
\frac{mr+1}{mr}e^{-mr}\biggr]dr^2-r^2d\Omega
\\\\
R\,=\,\frac{\delta_2(t)e^{-mr}}{r}\end{array}\right.
\end{eqnarray}
At this point one can provide the solution in term of the
gravitational potential. In such a case, we have an explicit
Newtonian-like term into the definition, according to previous
results obtained with less rigorous methods \cite{stel, qua-sch}.
The first of (\ref{yukawa-solution-O(2)-order}) provides the
second order solution in term of the metric expansion (see the
definition (\ref{definexpans})), but, this term coincides with the
gravitational potential at the Newtonian order
(\ref{gravitational-potential-metric}). In particular the gravitational potential of a FOG-model,
analytic in the Ricci scalar $R$, is

\begin{eqnarray}\label{gravpot}
\Phi\,=\,-\frac{GM}{f_1r}+\frac{\delta_2(t)}{6m}\frac{e^{-m
r}}{m r}
\end{eqnarray}

As first remark, one has to notice that the structure of the
potential (\ref{gravpot}), for a given $f(R)$-gravity, is determined only
by the parameter $m$, (\ref{yukawa-length}), which depends
on the first and the second derivative of the $f(R)$-function, once developed
around a vanishing value of the  Ricci scalar.

As second remark, the solution for the gravitational potential has
a Yukawa-like behavior depending on a characteristic length on which it evolves.

\emph{In other words, the correction to the Newtonian gravitational
potential is always characterized by a Yukawa-like correction and
only the first two terms of the Taylor expansion of a generic $f(R)$-function
turn out to be relevant. This is indeed a general
result.}

Let us now consider   system (\ref{sys1}) at third order
approximation. The first important issue is that, at this order,
one has to consider even the off-diagonal equation

\begin{eqnarray}\label{off-d}
f_1g^{(2)}_{rr,t}+2f_2rR^{(2)}_{,tr}\,=\,0
\end{eqnarray}
which relates the time derivative of the Ricci scalar to the time
derivative of $g^{(2)}_{rr}$. From this relation, it is possible
to draw a  relevant conclusion. One can deduce that, if the
Ricci scalar depends on time, so it is the same for the metric
components and even the gravitational potential turns out to be
time-dependent. This result agrees with the analysis provided in
\cite{spher_symm_fR} where a complete description of the weak field
limit of FOG has been provided in term of the dynamical evolution
of the Ricci scalar. Moreover it has been demonstrated that
supposing the time independent Ricci scalar, a static
spherically symmetric solution is found.

Eq. (\ref{off-d}) confirms this result and provides the formal
theoretical explanation of such a behavior. In particular,
together with Eqs.(\ref{yukawa-solution-O(2)-order-1}), it
suggests that if one considers the problem at a lower level of
approximation (\emph{i.e.} the second order) the background spacetime
metric can be only factorized with a function space-depending
and an arbitrary function time-depending. Then the Birkhoff
theorem at Newtonian level is modified. The static solutions
according to the Birkhoff theorem in GR are not directly obtained.
Obviously this is still no more verified when the problem is faced with
approximations of higher order. In other words, the debated issue
to prove the validity of the Birkhoff theorem in FOG, finds here
its physical answer. In \cite{spher_symm_fR}, the validity
of this theorem is demonstrated for FOG only when the
Ricci scalar is time-independent or, in addition, when the
solutions of (\ref{fieldequationfR2}) are investigated up to the second
order of approximation in the metric coefficients
(\ref{definexpans}). \emph{Therefore, the Birkhoff theorem is not
a general result for FOG but, on the other hand, in the limit of
small velocities and weak fields (which is enough to deal with the
Solar System gravitational experiments), one can assume that the
gravitational potential is effectively time independent according
to (\ref{yukawa-solution-O(2)-order-1}) and (\ref{gravpot})}.

The above results fix a fundamental difference between GR and FOG.
While, in GR, a spherically symmetric solution
represents a stationary and static configuration difficult to be
related to a cosmological background evolution, this is no more
true in the case of generic FOGs. In the latter case, a spherically
symmetric background can show time-dependent evolution together
with the radial dependence. In this sense, a relation between a
spherical solution and the cosmological Hubble flow could be, in principle,
achieved.

\section{The Newtonian and Post-Newtonian Limit of $f(R)$-gravity  in isotropic coordinates}\label{postnewtlimisitropic}

Let us consider now a form of metric tensor generalizing the metric
(\ref{definexpans}). It is interesting, using the isotropic coordinates
(\ref{me4}) and a more general approach, to solve the field equations as shown in  \cite{postnewtonian_limit_fR}.
The metric which we take into account is the following

\begin{eqnarray}\label{metric_tensor_PPN}
  g_{\mu\nu}\,\sim\,\begin{pmatrix}
  1+g^{(2)}_{tt}(t,\mathbf{x})+g^{(4)}_{tt}(t,\mathbf{x})+\dots & g^{(3)}_{ti}(t,\mathbf{x})+\dots \\
  g^{(3)}_{ti}(t,\mathbf{x})+\dots & -\delta_{ij}+g^{(2)}_{ij}(t,\mathbf{x})+\dots\end{pmatrix}
\end{eqnarray}
and the set of coordinates adopted is $x^\mu\,=\,(t,\mathbf{x})\,=\,(t,x^1,x^2,x^3)$.
The $n$-th derivative of function $f$ can be developed as in Eq. (\ref{sertay}) with $R_0\,=\,0$

\begin{eqnarray}
f^{n}(R)\,\sim\,f^{n}(R^{(2)}+R^{(4)}+\dots)\,\sim\,f^{n}(0)+f^{n+1}(0)R^{(2)}+f^{n+1}(0)R^{(4)}+\frac{1}{2}f^{n+2}(0)
{R^{(2)}}^2+\dots
\end{eqnarray}
From lowest order of field Eqs. (\ref{fieldequationfR}), we find the same condition  (\ref{eq0}) ($f(0)\,=\,0$).
At $\mathcal{O}$(2)-order (Newtonian level) Eqs. (\ref{fieldequationfR}) becomes

\begin{eqnarray}\label{PPN-field-equation-general-theory-fR-O2}
\left\{\begin{array}{ll}
H^{(2)}_{tt}=f'(0)R^{(2)}_{tt}-\frac{f'(0)}{2}R^{(2)}-f''(0)\triangle
R^{(2)}=\mathcal{X}\,T^{(0)}_{tt}\\\\
H^{(2)}_{ij}=f'(0)R^{(2)}_{ij}+\biggl[\frac{f'(0)}{2}R^{(2)}+f''(0)\triangle
R^{(2)}\biggr]\delta_{ij}-f''(0)R^{(2)}_{,ij}=0\\\\
H^{(2)}=-3f''(0)\triangle
R^{(2)}-f'(0)R^{(2)}=\mathcal{X}\,T^{(0)}
\end{array}\right.
\end{eqnarray}
while at $\mathcal{O}$(3)-order becomes

\begin{eqnarray}\label{PPN-field-equation-general-theory-fR-O3}
H^{(3)}_{ti}\,=\,f'(0)R^{(3)}_{ti}-f''(0){R^{(2)}}_{,ti}\,=\,\mathcal{X}\,T^{(1)}_{ti}
\end{eqnarray}
Remembering the expressions of Christoffel symbols and using the
following approximation for the determinant of metric tensor
$\ln\sqrt{-g}\,\sim\,\frac{1}{2}[g^{(2)}_{tt}-g^{(2)}_{mm}]+\dots$, at
$\mathcal{O}$(4)-order,  we have

\begin{eqnarray}\label{PPN-field-equation-general-theory-fR-O4}
\left\{\begin{array}{ll}
H^{(4)}_{tt}\,=\,f'(0)R^{(4)}_{tt}+f''(0)R^{(2)}R^{(2)}_{tt}-\frac{f'(0)}{2}R^{(4)}-\frac{f'(0)}{2}g^{(2)}_{tt}R^{(2)}-
\frac{f''(0)}{4}
{R^{(2)}}^2\\\\\,\,\,\,\,\,\,\,\,\,\,\,\,\,\,\,\,\,\,\,\,-f''(0)\biggl[g^{(2)}_{mn,m}{R^{(2)}}_{,n}+\triangle
R^{(4)}+g^{(2)}_{tt}\triangle R^{(2)}
+g^{(2)}_{mn}{R^{(2)}}_{,mn}-\frac{1}{2}\nabla
g^{(2)}_{mm}\cdot\nabla R^{(2)}\biggr]\\\\\,\,\,\,\,\,\,\,\,\,\,\,
\,\,\,\,\,\,\,\,\,-f'''(0)\biggl[|\nabla
R^{(2)}|^2+R^{(2)}\triangle
R^{(2)}\biggr]=\mathcal{X}\,T^{(2)}_{tt}\\\\\\
H^{(4)}=-3f''(0)\triangle
R^{(4)}-f'(0)R^{(4)}-3f'''(0)\biggl[|\nabla
R^{(2)}|^2+R^{(2)}\triangle
R^{(2)}\biggr]\\\\\,\,\,\,\,\,\,\,\,\,\,\,\,\,\,\,\,\,\,\,\,+3f''(0)\biggl[R^{(2)}_{,tt}-g^{(2)}_{mn}R^{(2)}_{,mn}-
\frac{1}{2}\nabla(g^{(2)}_{tt}-g^{(2)}_{mm})\cdot\nabla
R^{(2)}-g^{(2)}_{mn,m}R^{(2)}_{,n}\biggr]=\mathcal{X}\,T^{(2)}
\end{array}\right.
\end{eqnarray}
Note that the propagation of Ricci scalar $R^{(4)}$ has the same dynamics of
previous one (third line of Eqs. (\ref{PPN-field-equation-general-theory-fR-O2})). The complete knowledge of correction at fourth order
for the $tt$-component of Ricci tensor fix the third derivative of
$f(R)$ in $R\,=\,0$. Also at this level, there is a degeneracy of
$f(R)$-theory: different theories,  considering only  the first three
derivatives, admit the same gravitational field without
radiation emission.

We want to rewrite and generalize the outcome of Eqs. (\ref{yukawa-solution-O(2)-order-1})
by introducing the Green function method (we remember that the Newtonian limit corresponds also
to the linearization of field equations). Let us start
from the trace equation. The solution for the Ricci scalar $R^{(2)}$ in the third
line of Eqs.(\ref{PPN-field-equation-general-theory-fR-O2}) is

\begin{eqnarray}\label{scalar_ricci_sol_gen}
R^{(2)}(t,\textbf{x})\,=\,\frac{m^2\mathcal{X}}{f'(0)}\int
d^3\mathbf{x}'\mathcal{G}(\mathbf{x},\mathbf{x}')T^{(0)}(t,\mathbf{x}')
\end{eqnarray}
where we defined

\begin{eqnarray}\label{mass_definition_0}
m^2\,\doteq\,-\frac{f'(0)}{3f''(0)}
\end{eqnarray}
and $\mathcal{G}(\mathbf{x},\mathbf{x}')$ is the Green function of
field operator $\triangle-m^2$. We note that the definition of mass
(\ref{mass_definition_0}) is compatible with the definitions of
(\ref{mass_definition_TS}) and (\ref{yukawa-length}).

The solution for $g^{(2)}_{tt}$, from the first line of
(\ref{PPN-field-equation-general-theory-fR-O2}) by considering
that $R^{(2)}_{tt}\,=\,\frac{1}{2}\triangle g^{(2)}_{tt}$ (first line of (\ref{PPN-ricci-tensor}) or (\ref{PPN-ricci-tensor-HG})), is

\begin{eqnarray}\label{new_sol}
g^{(2)}_{tt}(t,\mathbf{x})\,=\,-\frac{\mathcal{X}}{2\pi f'(0)}\int
d^3\textbf{x}'\frac{T^{(0)}_{tt}(t,\textbf{x}')}{|\textbf{x}-
\textbf{x}'|}-\frac{1}{4\pi}\int
d^3\textbf{x}'\frac{R^{(2)}(t,\textbf{x}')}{|\textbf{x}-
\textbf{x}'|}-\frac{2}{3m^2}R^{(2)}(t,\textbf{x})\end{eqnarray}
We can check immediately that when $f(R)\rightarrow R$ we find
$g^{(2)}_{tt}(t,\textbf{x})\rightarrow-2G\int
d^3\textbf{x}'\frac{\rho(\textbf{x}')}{|\textbf{x}-
\textbf{x}'|}$. The expression (\ref{new_sol}) is the "modified"
gravitational potential (here we have a factor 2) for
$f(R)$-gravity and generalize, for any matter distribution,
the outcome (\ref{gravpot}). \emph{This solution, which is
the Newtonian limit of $f(R)$-gravity, is also gauge-free}.

Since we have a linearized version of field equations, this limit
corresponds to one of the Einstein equation and the linear
superposition is satisfied. So the $tt$-component of
energy-momentum tensor is, in this limit, the sum of mass- energy-volume density of sources, that is:
$T^{(0)}_{tt}\,=\,\Sigma_aM_a\delta(\mathbf{x}-\mathbf{x}_a)$
where $\delta(\mathbf{x})$ is the delta function.

As it is evident, the Gauss theorem is not valid since the force
law is not $\propto|\mathbf{x}|^{-2}$. The equivalence between a
spherically symmetric distribution and a point-like distribution is
not valid and how the matter is distributed in the space becomes extremely
important in this situation. However, we have to say that the Bianchi identities hold in any case so the consistency of the theory is guaranteed.

From  field Eq. (\ref{PPN-field-equation-general-theory-fR-O3}), by using the gauge harmonic condition (\ref{gauge-harmonic}), we find the
general solution for $g^{(3)}_{ti}$

\begin{eqnarray}\label{postnew_sol_ti}
g^{(3)}_{ti}(t,\textbf{x})\,=\,-\frac{\mathcal{X}}{2\pi f'(0)}\int
d^3\textbf{x}'\frac{T^{(1)}_{ti}(t,\textbf{x}')}{|\textbf{x}-
\textbf{x}'|}+\frac{1}{6\pi m^2}\frac{\partial}{\partial t}\int
d^3\textbf{x}'\frac{\nabla_{i'}R^{(2)}(t,\textbf{x}')}{|\textbf{x}-
\textbf{x}'|}
\end{eqnarray}
The choice of harmonic gauge enable us to solve
Eq. (\ref{PPN-field-equation-general-theory-fR-O3}) but we
lose  information on time evolution of
$g^{(2)}_{tt}(t,\textbf{x})$. This is  important
to obtain, at least in perturbative approach, some information about
the Birkhoff theorem. By hypothesizing a perturbative approach
(Newtonian-like), we confine   time-evolution
only to the timel-variation of matter source. In fact
in this working  hypothesis, the motion of bodies embedded in
gravitational fields evolves  very slowly with respect to the internal motions of
matter. Then we have, in any case,  an instantaneous readjustment of
spacetime. In other words, the motion of bodies is adiabatic and it
enables us to factorize the solutions that, by a time
transformation,  become  static solutions.

Still more, also the corrections to the gravito-magnetic effects
(\ref{PPN-field-equation-general-theory-fR-O3}) are depending on
the only first two derivatives of $f(R)$ in $R\,=\,0$. This means that different
theories, from the third derivative on, admit the same Newtonian
solution.

From second line of
(\ref{PPN-field-equation-general-theory-fR-O2}), by using the gauge harmonic condition (\ref{gauge-harmonic}), the solution for
$g^{(2)}_{ij}$ follows

\begin{eqnarray}\label{post_new_ij}
g^{(2)}_{ij}(t,\textbf{x})\,=&\biggl[&\frac{1}{4\pi}\int
d^3\textbf{x}'\frac{R^{(2)}(t,\textbf{x}')}{|\textbf{x}-
\textbf{x}'|}+\frac{2}{3m^2}R^{(2)}(t,\textbf{x})-\frac{1}{6\pi
m^2}\frac{1}{|\mathbf{x}|^3}\int_{\Omega_{|\mathbf{x}|}}d^3\mathbf{x}'R^{(2)}
(t,\mathbf{x}')\biggr]\delta_{ij}\nonumber\\\nonumber\\&+&\biggl[\frac{1}{2\pi
m^2|\mathbf{x}|^3}\int_{\Omega_{|\mathbf{x}|}}d^3\mathbf{x}'R^{(2)}
(t,\mathbf{x}')-\frac{2}{3m^2}R^{(2)}(t,\mathbf{x})\biggr]\frac{x_i x_j}{|\mathbf{x}|^2}
\end{eqnarray}
where $\Omega_{|\mathbf{x}|}$ represents the integration volume
with radius $|\mathbf{x}|$ (for the details see \cite{weinberg}).
By the solutions (\ref{new_sol}), (\ref{postnew_sol_ti}),
(\ref{post_new_ij}) we can affirm that it is possible to have
solution non-Ricci-flat in vacuum. This means that: \emph{Higher Order Gravity
mimics a matter source}. It is evident, from (\ref{new_sol}), that the
Ricci scalar can be considered a "matter source" which  curves the spacetime
also in absence of ordinary matter. Then it is clear also that the
knowledge of behavior of Ricci scalar inside mass distribution is
fundamental to obtain the behavior of metric tensor outside the
matter.

From the fourth order of field equations, we note also that the Ricci
scalar $R^{(4)}$ propagates with the same $m$ (the second line
of (\ref{PPN-field-equation-general-theory-fR-O4})) and the
solution at second order originates a supplementary matter source
in {\it r.h.s.} of (\ref{fieldequationfR}). The solution is

\begin{eqnarray}\label{Ricci_quarto}
R^{(4)}(t,\textbf{x})=&&\int
d^3\mathbf{x}'\mathcal{G}(\mathbf{x},\mathbf{x}')\biggl\{\frac{m^2\mathcal{X}}{f'(0)}T^{(2)}(t,\mathbf{x}')-g^{(2)}_
{mn,m}(t,\textbf{x}')R^{(2)}_{,n}(t,\textbf{x}')-g^{(2)}_{mn}(t,\textbf{x}')R^{(2)}_{,mn}(t,\textbf{x}')
\nonumber\\\nonumber\\
&&+R^{(2)}_{,tt}(t,\textbf{x}')-\frac{m^2}{\mu^4}\biggl[|\nabla_{\textbf{x}'}R^{(2)}(t,\textbf{x}')|^2+R^{(2)}
(t,\textbf{x}')\triangle_{\textbf{x}'}R^{(2)}(t,\textbf{x}')\biggr]
\nonumber\\\nonumber\\&&-\frac{1}{2}\nabla_{\textbf{x}'}\biggl[g^{(2)}_{tt}(t,\textbf{x}')-g^{(2)}_{mm}(t,\textbf{x}')
\biggr]\cdot
\nabla_{\textbf{x}'}R^{(2)}(t,\textbf{x}') \biggr\}
\end{eqnarray}
where we introduced a new parameter

\begin{eqnarray}\label{mass_definition_fR3}
\mu^4\,\doteq\,-\frac{f'(0)}{3f'''(0)}
\end{eqnarray}

Also in this case we
can have a non-vanishing curvature in absence of matter. The
solution for $g^{(4)}_{tt}$, from the first line of
(\ref{PPN-field-equation-general-theory-fR-O4}), is

\begin{eqnarray}\label{temp_temp_quarto}
g_{tt}^{(4)}(t,\textbf{x})=&&\int
d^3\mathbf{x}'\frac{1}{|\mathbf{x}-\mathbf{x}'|}\biggl\{-\frac{\mathcal{X}T^{(2)}_{tt}(t,\textbf{x}')}{2\pi
f'(0)}+\frac{1}{6\pi\mu^4}\biggl[|\nabla
R^{(2)}(t,\textbf{x}')|^2+R^{(2)}(t,\textbf{x}')\triangle
R^{(2)}(t,\textbf{x}')\biggr]
\nonumber\\\nonumber\\
&&+\frac{1}{4\pi}\biggl[g^{(2)}_{mn}(t,\textbf{x}')g^{(2)}_{tt,mn}(t,\textbf{x}')-g^{(2)}_{tt,tt}(t,\textbf{x}')
-|\nabla_{\textbf{x}'}g^{(2)}_{tt}(t,\textbf{x}')|^2-R^{(4)}(t,\textbf{x}')-g^{(2)}_{tt}(t,\textbf{x}')R^{(2)}
(t,\textbf{x}')\biggr]
\nonumber\\\nonumber\\
&& +\frac{1}{6\pi
m^2}\biggl[\frac{{R^{(2)}}^2(t,\textbf{x}')}{4}-\frac{R^{(2)}(t,\textbf{x}')\triangle
g^{(2)}_{tt}(t,\textbf{x}')}{2}+g^{(2)}_{mn,m}(t,\textbf{x}'){R^{(2)}}_{,n}
(t,\textbf{x}')+\triangle R^{(4)}(t,\textbf{x}')
\nonumber\\\nonumber\\
&&+g^{(2)}_{tt}(t,\textbf{x}')\triangle R^{(2)}(t,\textbf{x}')
+g^{(2)}_{mn}(t,\textbf{x}'){R^{(2)}}_{,mn}(t,\textbf{x}')-\frac{1}{2}\nabla
g^{(2)}_{mm}(t,\textbf{x}')\cdot\nabla
R^{(2)}(t,\textbf{x}')\biggr]\biggr\}
\end{eqnarray}

In summary, we  have  shown the more general solutions of field equations of $f(R)$-gravity in the
Newtonian and post-Newtonian limits assuming a coordinates transformation where the gauge harmonic
condition is verified. Now we shall apply  such an approach to obtain the explicit form of the metric
tensor for a static and spherically symmetric matter source.

\subsection{Solutions generated  by an extended spherically symmetric source with harmonic gauge conditions}

Let us consider a spherical  source with mass $M$ and radius $\xi$.
Since the metric is given by Eqs.(\ref{metric_tensor_PPN}), the energy-momentum tensor (\ref{point_like})  becomes

\begin{eqnarray}\label{emtensorPPN}
\left\{\begin{array}{ll}
T_{tt}(t,\mathbf{x})\,\sim\,\rho(\mathbf{x})+\rho(\mathbf{x})g^{(2)}_{tt}(t,\mathbf{x})\,
=\,T^{(0)}_{tt}(t,\mathbf{x})+T^{(2)}_
{tt}(t,\mathbf{x})
\\\\
T\,=\,\rho(\mathbf{x})\,=\,T^{(0)}(t,\mathbf{x})
\end{array}\right.
\end{eqnarray}
where the density is

\begin{eqnarray}\label{ball_like}
\rho(\mathbf{x})\,=\,\frac{3M}{4\pi\xi^3}\Theta(\xi-|\mathbf{x}|)
\end{eqnarray}
$\Theta(x)$ is the Heaviside function. We are not interested here to the internal structure of the source.
The possible choices of the  Green functions of the field operator $\triangle-m^2$, for spherically symmetric systems (\emph{i.e.} $\mathcal{G(\mathbf{x},
\mathbf{x}')}\,=\,\mathcal{G}(|\mathbf{x}-\mathbf{x}'|)$), are the following

\begin{eqnarray}\label{green_function}
\mathcal{G}(\mathbf{x},\mathbf{x}')\,=\,\left\{\begin{array}{ll}-\frac{1}{4\pi}\frac{e^{-m|\mathbf{x}-\mathbf{x}'|}}
{|\mathbf{x}-\mathbf{x}'|}\,\,\,\,\,\,\,\,\,\,\,\,\,\,\,\,\,\,\,\,\,\,\,\,\,\,\,\,\,\,\,\,\,\,\,\,\,\,\,\,\,\,\,\,\,
\,\,\,\text{if}\,\,\,\,\,\,\,\,\,\,\,\,\,\,m^2\,\,>\,0
\\\\
C_1\frac{e^{-im|\mathbf{x}-\mathbf{x}'|}}{|\mathbf{x}-\mathbf{x}'|}+C_2\frac{e^{im|\mathbf{x}-\mathbf{x}'|}}
{|\mathbf{x}-\mathbf{x}'|}
\,\,\,\,\,\,\,\,\,\,\,\,\,\,\,\text{if}\,\,\,\,\,\,\,\,\,\,\,\,\,\,m^2\,<\,0\end{array}\right.
\end{eqnarray}
with $C_1+C_2\,=\,-\frac{1}{4\pi}$.

In the Newtonian limit of GR, the equation for the
gravitational potential, generated by
a point-like source

\begin{eqnarray}
\triangle_\mathbf{x}\mathcal{G}_{New. mech.}(\mathbf{x}-\mathbf{x}')\,=\,-4\pi\delta(\mathbf{x}-\mathbf{x}')
\end{eqnarray}
is not satisfied by the Green functions (\ref{green_function}). If we
consider the flux of gravitational field $\mathbf{g}_{New. mech.}$
defined as

\begin{eqnarray}
\mathbf{g}_{New. mech.}\,=\,-\frac{GM(\mathbf{x}-\mathbf{x}')}{|\mathbf{x}-\mathbf{x}'|^3}\,=\,-GM
\mathbf{\nabla}_{\mathbf{x}}\mathcal{G}_{New.
mech.}(\mathbf{x}-\mathbf{x}')
\end{eqnarray}
we obtain, as standard, the Gauss theorem

\begin{eqnarray}
\int_\Sigma d\Sigma\,\,\,\,\mathbf{g}_{New. mech.}\cdot\hat{n}\propto M
\end{eqnarray}
where $\Sigma$ is a generic two-dimensional surface and $\hat{n}$
its surface normal vector. The flux of field $\mathbf{g}_{New.
mech.}$ on the surface $\Sigma$ is proportional to the matter
content $M$, inside the surface independently of the particular
shape of surface (Gauss theorem, or Newton theorem for the
gravitational field \cite{bin-tre}). On the other hand, if we
consider the flux defined by the new Green function, its value is
not proportional to the enclosed mass but depends on the
particular choice of the surface

\begin{eqnarray}
\int_\Sigma d\Sigma\,\,\,\,\mathbf{g}_{New. mech.}\cdot\hat{n}\propto M_\Sigma
\end{eqnarray}
Hence $M_\Sigma$ is a mass-function depending on the surface
$\Sigma$. Then we have to find the solution inside/outside the
matter distribution by evaluating the integral quantities and
imposing the boundary condition on the separation surface.

We have to note that, for any function of modulus $h(|\mathbf{x}|)$,  it is

\begin{eqnarray}\label{int1}I\,=\,\int
d^3\mathbf{x}'\mathcal{G}(\mathbf{x},\mathbf{x}')h(\textbf{x}')\,=\,-\frac{1}{4\pi}\int
d|\mathbf{x}'||\mathbf{x}'|^2h(|\textbf{x}'|)\int_0^{2\pi}
d\phi'\int_0^\pi
d\theta'\frac{\sin\theta'e^{-m\sqrt{|\mathbf{x}|^2+|\mathbf{x}'|^2-2|\mathbf{x}||\mathbf{x}'|\cos\alpha}}}
{\sqrt{|\mathbf{x}|^2+|\mathbf{x}'|^2-2|\mathbf{x}||\mathbf{x}'|\cos\alpha}}
\end{eqnarray}
where
$\cos\alpha\,=\,\cos\theta\cos\theta'+\sin\theta\sin\theta'\cos(\phi-\phi')$
and $\alpha$ is the angle between two vectors $\mathbf{x}$,
$\mathbf{x}'$. In the spherically symmetric case we can choose
$\theta\,=\,0$ without losing generality (the symmetry of system is independent by the angle). By making the angular
integration we get

\begin{eqnarray}\label{int2}
I\,=\,-\frac{1}{2m|\textbf{x}|}\int
d|\mathbf{x}'||\mathbf{x}'|h(|\mathbf{x}'|)\biggl[e^{-m||\textbf{x}|-|\textbf{x}'||}-e^{-m(|\textbf{x}|+|\textbf{x}'|)}
\biggr]
\end{eqnarray}
An analogous relation is useful also for the Green function of Newtonian mechanics
$|\mathbf{x}-\mathbf{x}'|^{-1}$

\begin{eqnarray}\label{int3}
\int d^3\mathbf{x}'\frac{h(|\textbf{x}'|)}{|\mathbf{x}-\mathbf{x}'|}\,=\,-\frac{2\pi}{|\mathbf{x}|}\int
d|\mathbf{x}'||\mathbf{x}'|\biggl[||\mathbf{x}|-|\mathbf{x}'||-|\mathbf{x}|-|\mathbf{x}'|\biggr]h(|\textbf{x}'|)
\end{eqnarray}

\subsubsection{Solutions at $\mathcal{O}(2)$- and $\mathcal{O}(3)$-order}

By supposing that $m^2\,>\,0$ (\emph{i.e.}
$\text{sign}[f'(0)]\,=\,-\,\text{sign}[f''(0)]$ (an analogous condition
used in (\ref{yukawa-solution-O(2)-order})), the Ricci scalar (\ref{scalar_ricci_sol_gen}) is\footnote{We have set for
simplicity $f'(0)\,=\,1$, otherwise we have to renormalize the coupling constant
$\mathcal{X}$ in the action (\ref{FOGaction}).}

\begin{eqnarray}\label{sol_ric}
R^{(2)}(t,\textbf{x})\,=\,-\frac{3r_g}{\xi^3}\,
\biggl[1-e^{-m\xi}(1+m\,\xi)\frac{\sinh m|\mathbf{x}|}{m|\mathbf{x}|}\biggr]\Theta(\xi-|\mathbf{x}|)
-r_g\,m^2F(\xi)\frac{e^{-m|\mathbf{x}|}}{|\mathbf{x}|}\,\Theta(|\mathbf{x}|-\xi)
\end{eqnarray}
where we introduced a \emph{shape function}

\begin{eqnarray}\label{shapefunction}
F(x)\,\doteq\,3\frac{mx\cosh m x-\sinh
m x}{m^3x^3}
\end{eqnarray}
The solutions of (\ref{new_sol}),
(\ref{postnew_sol_ti}) and (\ref{post_new_ij}), given the relations (\ref{int2}) and (\ref{int3}), respectively are

\begin{eqnarray}\label{sol_pot}
g^{(2)}_{tt}(t,\mathbf{x})\,=\,&-&r_g\,
\biggl[\frac{3}{2\xi}+\frac{1}{m^2\xi^3}-\frac{|\textbf{x}|^2}{2\xi^3}-\frac{e^{-m\xi}(1+m\,\xi)}{m^2\xi^3}\frac{\sinh m|\mathbf{x}|}{m|\mathbf{x}|}\biggr]\,\Theta(\xi-|\mathbf{x}|)\nonumber\\\nonumber\\&-&r_g\,\biggl[\frac{1}{|\textbf{x}|}+\frac{F(\xi)}{3}
\frac{e^{-m|\mathbf{x}|}}{|\mathbf{x}|}\biggr]\Theta(|\mathbf{x}|-\xi)
\end{eqnarray}

\begin{eqnarray}
g^{(3)}_{ti}(t,\textbf{x})\,=\,0
\end{eqnarray}

\begin{eqnarray}\label{gsol2}
g^{(2)}_{ij}(t,\textbf{x})=-&r_g&\biggl\{\biggl[\frac{3}{2\xi}-\frac{5}{3m^2\xi^3}-\frac{|\textbf{x}|^2}{2\xi^3}+
\frac{(1+m\xi)e^{-m\xi}}{3m^2\xi^3}\biggl(2F(\textbf{x})+3\frac{\sinh m|\textbf{x}|}{m|\textbf{x}|}\biggr)\biggr]\Theta(\xi-|\mathbf{x}|)\nonumber\\\nonumber\\&&\,\,\,\,
+\biggl[\frac{1}{|\textbf{x}|}-\frac{2}
{3m^2|\textbf{x}|^3}-\frac{F(\xi)}{3}\biggl(\frac{1}{|\mathbf{x}|}-\frac{2}{m|\textbf{x}|^2}-\frac{2}
{m^2|\textbf{x}|^3}\biggr)e^{-m|\textbf{x}|}\biggr]\Theta(|\mathbf{x}|-\xi)\biggr\}
\delta_{ij}\nonumber\\\nonumber\\
-&r_g&\biggl\{\biggl[\frac{2(1+m\xi)e^{-m\xi}}{m^2\xi^3}\biggl(\frac{\sinh m|\textbf{x}|}{m|\textbf{x}|}-F(\textbf{x})\biggr)\biggr]\Theta(\xi-|\textbf{x}|)\nonumber\\\nonumber\\
&&\,\,\,\,+\biggl[
\frac{2}{m^2|\textbf{x}|^3}-\frac{2F(\xi)}{3}\biggl(
\frac{1}{|\textbf{x}|}
+\frac{3}{m|\textbf{x}|^2}
+\frac{3}{m^2|\textbf{x}|^3}\biggr)e^{-m|\textbf{x}|}
\biggr]\Theta(|\textbf{x}|-\xi)\biggr\}\frac{x_ix_j}{|\textbf{x}|^2}
\end{eqnarray}

We note that the corrections to GR behavior are ruled by
$\mathcal{G}(\mathbf{x},\mathbf{x}')$. If we perform a Taylor
expansion for $m|\mathbf{x}|\ll 1$, we have

\begin{eqnarray}
\frac{\sinh m|\mathbf{x}|}{m|\mathbf{x}|}\simeq 1+\frac{m^2|\mathbf{x}|^2}{6}+\dots
\end{eqnarray}
For fixed values of the distance $|\mathbf{x}|$, the solutions
$g^{(2)}_{tt}$ and $g^{(2)}_{ij}$ depend on the value of the
radius $\xi$, then the Gauss theorem does not hold also if the
Bianchi identities hold, as already said above \cite{stel}.
In other words, since the Green function does not scale as the inverse of distance but
has also an exponential behavior, the Gauss theorem is not
verified. We can affirm: \emph{the potential does not depend only
on the total mass but also on the mass-distribution in the
space}. We can write

\begin{eqnarray}
\lim_{\xi\rightarrow\infty}F(m\xi)\,=\,\infty
\end{eqnarray}
obviously the limit of $\xi$ has to be interpreted up to the
maximal value  where the generic position $|\mathbf{x}|$ in the
space is fixed. If we consider the limit $\xi\rightarrow 0$ (the
point-like source limit), we obtain

\begin{eqnarray}
\lim_{\xi\rightarrow
0}F(m\xi)\,=\,1
\end{eqnarray}

\begin{figure}[htbp]
  \centering
  \includegraphics[scale=1]{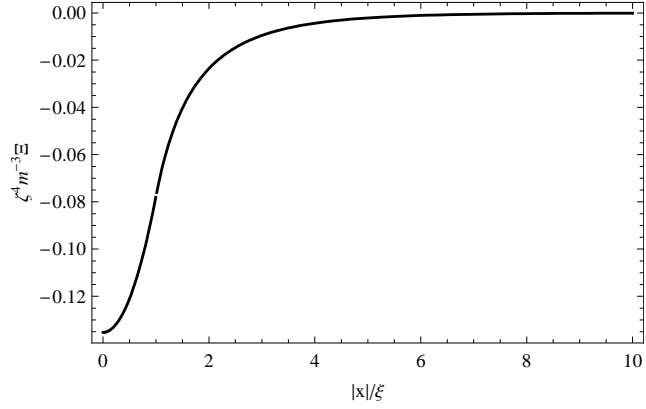}\\
  \caption{Plot of dimensionless function $\zeta^4m^{-3}\Xi$ for $\zeta\,\doteq\,m\xi\,=\,0.5$ representing the
  spatial behavior of Ricci scalar at second order. In GR we would have $\Xi(\textbf{x})\,=\,\frac{3}{\xi^3}\Theta(\xi-|\mathbf{x}|)$.}
  \label{plotricciscalar}
\end{figure}

\begin{figure}[htbp]
  \centering
  \includegraphics[scale=1]{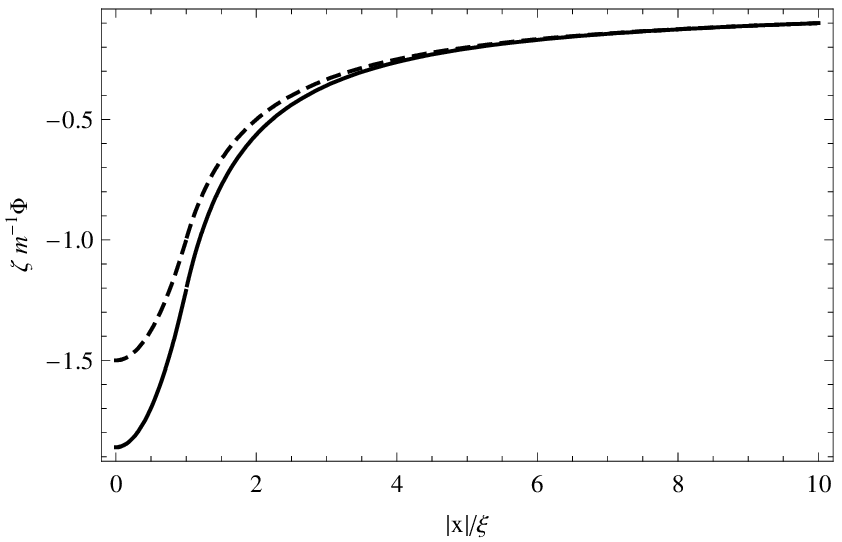}\\
  \caption{Plot of metric potential $\zeta m^{-1}\Phi$ vs distance from central mass with $\zeta\,\doteq\,m\xi\,=\,0.5$.
  The dashed line is the GR behavior: $\Phi\,=\,-\biggl[\frac{3}{2\xi}-\frac{|\textbf{x}|^2}{2\xi^3}\biggr]
  \Theta(\xi-\textbf{x})-\frac{\Theta(\textbf{x}-\xi)}{|\textbf{x}|}$.}
\label{plotpontential00}
\end{figure}

\begin{figure}[htbp]
  \centering
  \includegraphics[scale=1]{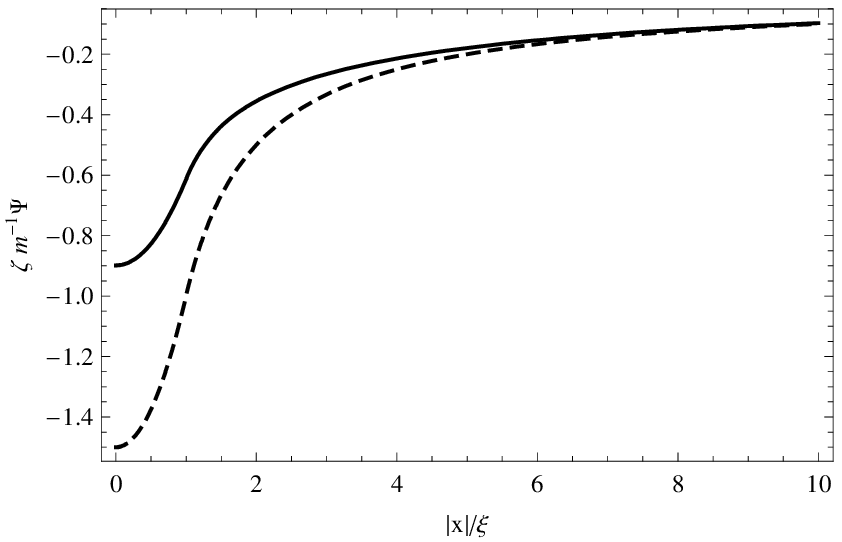}\\
  \caption{Plot of metric potential $\zeta m^{-1}\Psi$ vs distance from central mass with $\zeta\,\doteq\,m\xi\,=\,0.5$.
  The dashed line is the GR behavior (similar to metric potential $\Phi$).}
\label{plotpontentialij1}
\end{figure}

\begin{figure}[htbp]
  \centering
  \includegraphics[scale=1]{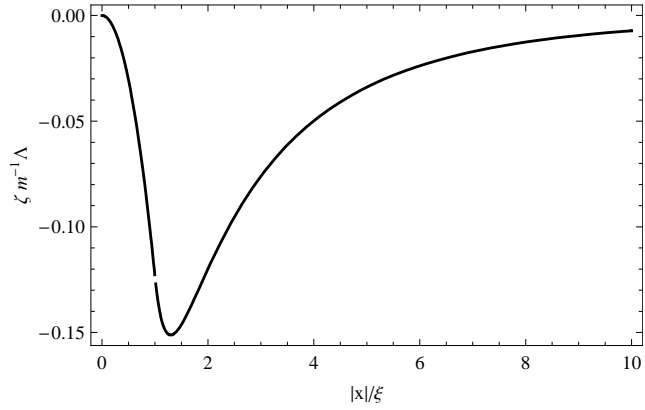}\\
  \caption{Plot of metric potential $\zeta m^{-1}\Lambda$ vs distance from central mass with $\zeta\,\doteq\,m\xi\,=\,0.5$.
  In GR  such a behavior is missing.}
\label{plotpontentialij2}
\end{figure}

By introducing three metric potentials $\Phi(\textbf{x})$,
$\Psi(\textbf{x})$ and $\Lambda(\textbf{x})$ (the dimension is the
inverse of length) we can rewrite (\ref{sol_pot}) and
(\ref{gsol2}) as follows

\begin{eqnarray}\label{PAS}
\left\{\begin{array}{ll}
g^{(2)}_{tt}(t,\textbf{x})\,=\,r_g\Phi(\textbf{x})
\\\\
g^{(2)}_{ij}(t,\textbf{x})\,=\,r_g\Psi(\textbf{x})\delta_{ij}+r_g\Lambda(\textbf{x})\frac{x_ix_j}{|\mathbf{x}|^2}
\end{array}\right.
\end{eqnarray}
and with a fourth function, $\Xi(\textbf{x})$, (the dimension is the cubic inverse of length) the Ricci scalar
(\ref{sol_ric}) is

\begin{eqnarray}\label{PAS_1}
R^{(2)}(t,\textbf{x})\,=\,r_g\Xi(\textbf{x})
\end{eqnarray}
The spatial behavior (\ref{sol_ric}) is shown in FIG.
\ref{plotricciscalar}. The metric potentials are shown in FIGs.
\ref{plotpontential00}, \ref{plotpontentialij1} and
\ref{plotpontentialij2}.
It is interesting to note as the function $\Phi$ assumes
smaller value of its equivalent in GR, then in terms of
gravitational attraction we have a potential well more deep. A
such scheme can be interpretable or assuming a variation of the
gravitational constant $G$ or requiring that there is a central
greater mass. These two affirmations are compatible on the one
hand with the tensor-scalar theories (in the which we have a
scaling of gravitational constant) and on the other hand with the
theory of GR plus the hypothesis of the existence of the dark
matter. In particular, if the mass distribution takes a bigger
volume, the potential increases and vice versa.


In the limit of point-like source, \emph{i.e.}
$\lim_{\xi\rightarrow 0}\frac{3M}{3\pi
\xi^3}\Theta(\xi-|\mathbf{x}|)\,=\,M\delta(\mathbf{x})$, we get

\begin{eqnarray}\label{sol_new_pfR}
\left\{\begin{array}{ll}
R^{(2)}(t,\textbf{x})\,=\,-r_gm^2\frac{e^{-m|\mathbf{x}|}}{|\mathbf{x}|}
\\\\
g^{(2)}_{tt}(t,\mathbf{x})\,=\,-r_g\biggl(\frac{1}{|\textbf{x}|}+\frac{1}{3}\frac{e^{-m|\mathbf{x}|}}{|\mathbf{x}|}\biggr)
\\\\
g^{(2)}_{ij}(t,\textbf{x})\,=\,-r_g\biggl\{\frac{1}{|\mathbf{x}|}-\frac{2}{3m^2|\textbf{x}|^3}
-\frac{1}{3}\biggl(\frac{1}{|\mathbf{x}|}-\frac{2}{m|\textbf{x}|^2}-\frac{2}
{m^2|\textbf{x}|^3}\biggr)e^{-m
|\mathbf{x}|}\biggr\}\delta_{ij}
\\\\\,\,\,\,\,\,\,\,\,\,\,\,\,\,\,\,\,\,\,\,\,\,\,\,\,\,\,\,\,\,\,-r_g\biggl[\frac{2}{m^2|\textbf{x}|^3}-\frac{2}{3}
\biggl(\frac{1}{|\textbf{x}|}
+\frac{3}{m|\textbf{x}|^2}
+\frac{3}{m^2|\textbf{x}|^3}\biggr)e^{-m|\textbf{x}|}\biggr]\frac{x_ix_j}{|\textbf{x}|^2}
\end{array}\right.
\end{eqnarray}

An important remark has to be done at this point. Now we can check the
compatibility of $f(R)$-gravity with respect to GR in the limit ($f(R)\,\rightarrow R)$.
Since the Gauss theorem is not verified for $f(R)$-gravity, while the
relations (\ref{schwarz-isotropic-PPN}) satisfy it, we have to consider the
relations (\ref{sol_new_pfR}) and not (\ref{sol_pot}),
(\ref{gsol2}). After making the limit $f(R)\,\rightarrow\,R$ we have

\begin{eqnarray}\label{sol_GR_vacuum}
\left\{\begin{array}{ll} R^{(2)}(t,\textbf{x})\,=\,0
\\\\
g^{(2)}_{tt}(t,\mathbf{x})\,=\,-\frac{r_g}{|\textbf{x}|}
\\\\
g^{(2)}_{ij}(t,\textbf{x})\,=\,-\frac{r_g}{|\mathbf{x}|}\delta_{ij}
\end{array}\right.
\end{eqnarray}
which suggest that the $f(R)$-gravity is compatible with respect
to GR. It is interesting to note that also in the case of
extended spherically symmetric distribution of matter, when we
perform the limit $f(R)\rightarrow R$, the solutions (\ref{sol_pot})
and (\ref{gsol2}) directly converge (in the vacuum) to solutions
(\ref{sol_GR_vacuum}), showing the validity of Gauss theorem
in GR.

Another important consideration is about the asymptotic behavior of
$f(R)$-gravity with respect to GR. In fact, increasing the distance
from the central mass, the gravitational field should converge to that
of GR. Such a convergence is the standard   consequence of the
spherically symmetry of the source with asymptotically flat boundary conditions.
In FIG. \ref{plotricciscalar}, we report the spatial
behavior of Ricci scalar (\ref{sol_ric}) approximating
asymptotically the given value in GR. In fact supposing
$f(R)$- gravity,  the Ricci scalar acquires dynamics, and in the
Newtonian limit, we find a characteristic scale length ($m^{-1}$)
related to  the scalar massive mode. Only for
distances larger than $m^{-1}$,  we recover the outcome of GR, that is
$R\,=\,0$.  The metric potentials are shown  in FIGs. 2, 3 and 4.

\begin{figure}[htbp] \centering
  \includegraphics[scale=1]{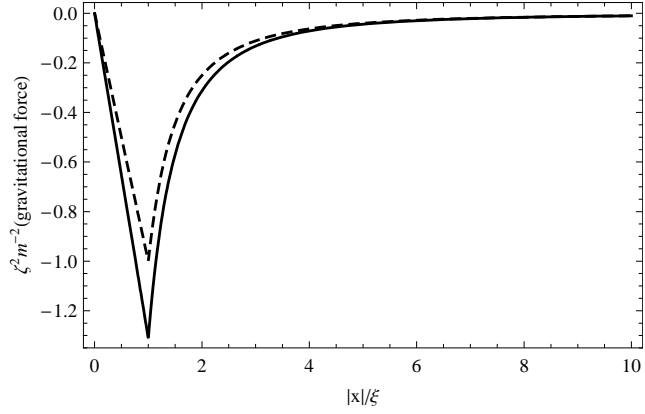}\\
  \caption{Comparison between gravitational forces induced by GR and $f(R)$-gravity with $\zeta\,\doteq\,m\xi\,=\,0.5$.
  The dashed line is the GR behavior.}
  \label{plotforce}
\end{figure}

To conclude this section, we show in FIG. \ref{plotforce} the
comparison between gravitational forces induced in GR and in
$f(R)$-gravity considered in the Newtonian limit. Obviously also considering
forces,  we could obtained an intensity different  than that in GR.

\subsubsection{The Newtonian Limit of $f(R)$-gravity in oscillating regime}

If we consider $m^2\,<\,0$ (i.e.
$\text{sign}[f'(0)]\,=\,\text{sign}[f''(0)]$) from
(\ref{green_function}) we can choose the "oscillating" Green
function

\begin{eqnarray}\label{green_function_2}
\mathcal{G}(\mathbf{x},\mathbf{x}')\,=\,-\frac{1}{4\pi}\frac{\cos
m|\mathbf{x}-\mathbf{x}'|+ \sin
m|\mathbf{x}-\mathbf{x}'|}{|\mathbf{x}-\mathbf{x}'|}
\end{eqnarray}
The Ricci scalar (\ref{scalar_ricci_sol_gen}) and the
$tt$-component of $g_{\mu\nu}$ at $\mathcal{O}(2)$ order
(\ref{new_sol}) become

\begin{eqnarray}\label{sol_ric_oscil}
R^{(2)}(t,\textbf{x})=\,-\frac{6r_g}{\xi^3}\,\biggl[1-H(\xi)\frac{\sin
m|\mathbf{x}|}{m|\mathbf{x}|}\biggr]\Theta(\xi-|\mathbf{x}|)
-2\,r_g\,m^2G(\xi)\frac{\cos m|\textbf{x}|+\sin
m|\textbf{x}|}{|\mathbf{x}|}\,\Theta(|\mathbf{x}|-\xi)
\end{eqnarray}

\begin{eqnarray}\label{sol_pot_oscil}
g^{(2)}_{tt}(t,\mathbf{x})\,=\,&-&r_g\,
\biggl[\frac{3}{2\xi}-\frac{2}{m^2\xi^3}-\frac{|\textbf{x}|^2}{2\xi^3}+\frac{2H(\xi)}{m^2\xi^3}\frac{\sin
m|\mathbf{x}|}{m|\mathbf{x}|}\biggr]\,\Theta(\xi-|\mathbf{x}|)\nonumber\\&-&r_g\,\biggl[\frac{1}{|\textbf{x}|}-\frac{2G(\xi)}{3}
\frac{\cos m|\mathbf{x}|+\sin
m|\mathbf{x}|}{|\mathbf{x}|}\biggr]\Theta(|\mathbf{x}|-\xi)
\end{eqnarray}
where we introduced  two new \emph{shape functions}

\begin{eqnarray}\label{shapefunction_2}
G(x)\,\doteq\,3\frac{mx\cos m x-\sin
m x}{m^3x^3}\,,\,\,\,\,\,\,\,\,\,\,\,\,\,\,\,\,H(x)\,\doteq\,(1-mx)\cos mx+(1+mx)\sin mx
\end{eqnarray}
with the properties $\lim_{\xi\rightarrow 0}G(\xi)\,=\,-1$ and
$\lim_{\xi\rightarrow 0}H(\xi)\,=\,1$. Since we have an
oscillating Green function which is not asymptotically zero, the
"gravitational potentials" (\ref{sol_pot_oscil}) at infinity are
zero a part a possible  constant value
($\lim_{a\rightarrow\infty}2r_g\,m\,G(\xi)(\sin ma-\cos ma)$).

\begin{figure}[htbp]
  \centering
  \includegraphics[scale=1]{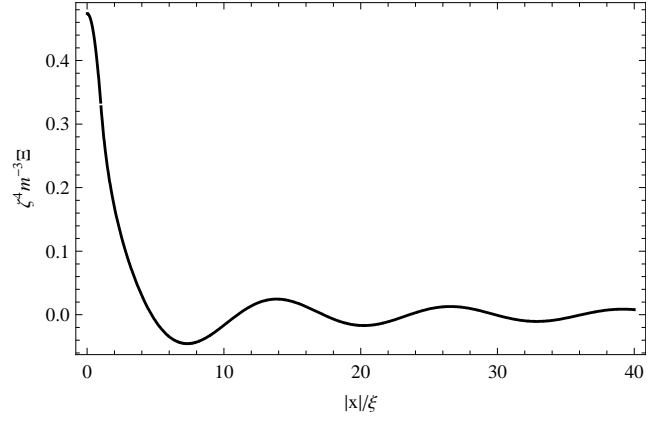}\\
  \caption{Plot of dimensionless function $\zeta^4m^{-3}\Xi$ with $\zeta\,\doteq\,m\xi\,=\,.5$ representing the
  spatial behavior of Ricci scalar at second order in the oscillating case.}
  \label{plotricciscalar_oscil}
\end{figure}

\begin{figure}[htbp]
  \centering
  \includegraphics[scale=1]{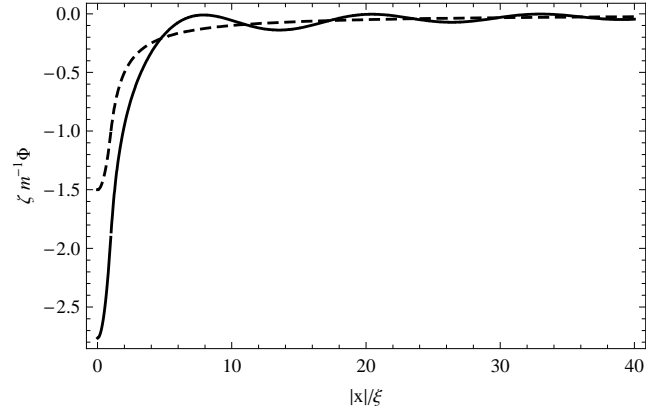}\\
  \caption{Plot of metric potential $\zeta m^{-1}\Phi$ vs distance from central mass with the choice
  $\zeta\,\doteq\,m\xi\,=\,0.5$ in the oscillating case. The dashed line is the GR behavior.}
\label{plotpontential00_oscil}
\end{figure}

The spatial behavior of Ricci scalar (\ref{sol_ric_oscil}) and
metric component (\ref{sol_pot_oscil}) are shown in FIGs.
\ref{plotricciscalar_oscil} and \ref{plotpontential00_oscil}.
The previous considerations hold also for the
solutions (\ref{sol_ric_oscil}) - (\ref{sol_pot_oscil}). The only
difference is that now we have oscillating behaviors instead of
exponential behaviors.

The correction term to the Newtonian
potential in the external solution can be interpreted as the
Fourier transform of the matter density $\rho(\mathbf{x})$. In
fact, we have

\begin{eqnarray}
\int\frac{d^3\mathbf{x}'}{(2\pi)^3}\rho(\mathbf{x}')e^{-i\mathbf{k}\cdot\mathbf{x}'}\,=\,-\frac{MG(|\mathbf{k}|\xi)}{(2\pi)^3}
\end{eqnarray}
and in the limit of point-like source

\begin{eqnarray}
\lim_{\xi\rightarrow 0}\int\frac{d^3\mathbf{x}'}{(2\pi)^3}\rho(\mathbf{x}')e^{-i\mathbf{k}\cdot
\mathbf{x}'}=\frac{M}{(2\pi)^3}
\end{eqnarray}

\begin{figure}[htbp] \centering
  \includegraphics[scale=1]{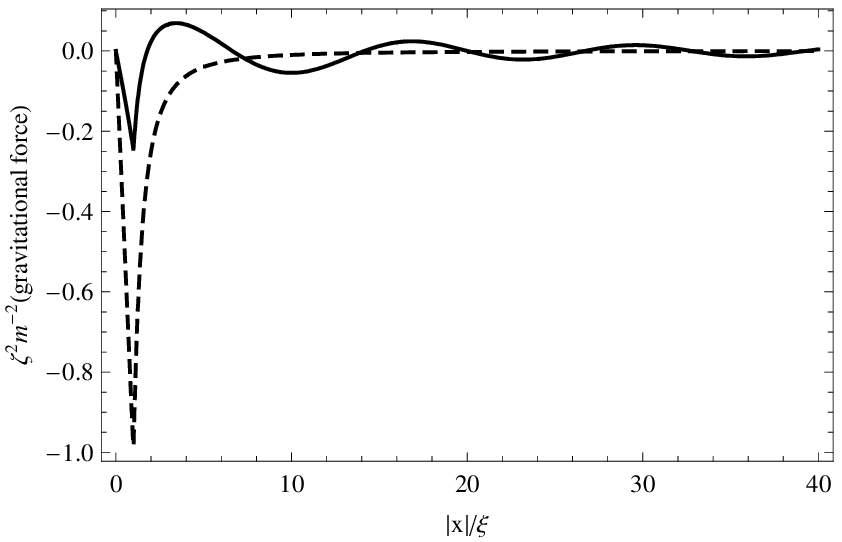}\\
  \caption{Comparison between gravitational forces induced by GR and $f(R)$-gravity with $\zeta\,\doteq\,m\xi\,=\,0.5$ in
  the oscillating case. The dashed line is the GR behavior.}
  \label{plotforce_oscil}
\end{figure}

Also in this case we show in FIG. \ref{plotforce_oscil} the comparison between gravitational forces
induced in GR and in $f(R)$-gravity in the Newtonian limit.
Obviously also in this last case we obtained a different force with respect to GR.

\subsubsection{Solutions at $\mathcal{O}$(4)-order}

The metric potentials and the function $\Xi(\mathbf{x})$, respectively defined in (\ref{PAS}) and (\ref{PAS_1}), satisfy the
following properties with respect to derivative of coordinate $l$-th\footnote{We remember that $|\textbf{x}|_{,l}\,=\,|\textbf{x}|^{-1}\,x_l$.} in the matter

\begin{eqnarray}\label{relations_matter}
\left\{\begin{array}{ll}
\Xi_{,l}(\textbf{x})\,=\,\frac{m^2(1+m\xi)}{\xi^3}\,e^{-m\xi}\,F(\mathbf{x})\,x_l\,=\,
\Xi_0(\textbf{x})\,x_l
\\\\
\Phi_{,l}(\textbf{x})\,=\,\biggl[\frac{1}{\xi^3}+\frac{(1+m\xi)}{3\xi^3}\,e^{-m\xi}\,F(\mathbf{x})\biggr]\,x_l\,=\,\Phi_0(\textbf{x})\,x_l
\\\\
\Psi_{,l}(\textbf{x})\,=\,\biggl[\frac{1}{\xi^3}-\frac{(1+m\xi)}{\xi^3}\,e^{-m\xi}\,\frac{(m^2|\textbf{x}|^2+6)\sinh
m|\mathbf{x}|+m|\textbf{x}|(m^2|\textbf{x}|^2-6)\cosh
m|\textbf{x}|}{m^5|\textbf{x}|^5}\biggr]\,x_l\,=\,\Psi_0
(\textbf{x})\,x_l
\\\\
\Lambda_{,l}(\textbf{x})\,=\,\frac{2(1+m\xi)}{\xi^3}\,e^{-m\xi}\,\frac{(4m^2|\textbf{x}|^2+9)\sinh
m|\mathbf{x}|-m|\textbf{x}|(m^2|\textbf{x}|^2+9)\cosh
m|\textbf{x}|}{m^5|\textbf{x}|^5}\,x_l\,=\,\Lambda_0(\textbf{x})
\,x_l\\\\
\Xi_{,ln}(\textbf{x})\,=\,\Xi_0(\textbf{x})\delta_{ln}+\frac{3(1+m\xi)}{\xi^3}\,e^{-m\xi}\,\frac{(m^2|\mathbf{x}|^2+3)\sinh
m |\mathbf{x}|-3m|\mathbf{x}|\cosh
m|\mathbf{x}|}{m|\mathbf{x}|^5}\,x_lx_n\,=\,\Xi_0(\textbf{x})\delta_{ln}+\Xi_1(\textbf{x})x_lx_n
\\\\
\Phi_{,ln}(\textbf{x})\,=\,\Phi_0(\textbf{x})\delta_{ln}+\frac{(1+m\xi)}{\xi^3}\,e^{-m\xi}\,\frac{(m^2|\mathbf{x}|^2+3)\sinh
m |\mathbf{x}|-3m|\mathbf{x}|\cosh
m|\mathbf{x}|}{m^3|\mathbf{x}|^5}
\,x_lx_n\,=\,\Phi_0(\textbf{x})\delta_{ln}+\Phi_1(\textbf{x})x_lx_n
\end{array}\right.
\end{eqnarray}
and in the vacuum

\begin{eqnarray}\label{relations}
\left\{\begin{array}{ll}
\Xi_{,l}(\textbf{x})\,=\,\frac{m^2(m|\textbf{x}|+1)}{|\textbf{x}|^3}\,F(\xi)\,e^{-m|\textbf{x}|}\,x_l\,=\,
\Xi_0(\textbf{x})\,x_l
\\\\
\Phi_{,l}(\textbf{x})\,=\,\biggl[\frac{1}{|\textbf{x}|^3}+\frac{m|
\textbf{x}|+1}{|\textbf{x}|^3}\frac{F(\xi)e^{-m|\textbf{x}|}}{3}\biggr]\,x_l\,=\,\Phi_0(\textbf{x})\,x_l
\\\\
\Psi_{,l}(\textbf{x})\,=\,\biggl[\frac{m^2|\textbf{x}|^2-2}{m^2|\textbf{x}|^5}-\frac{m^3
|\textbf{x}|^3-m^2|\textbf{x}|^2-6m|\mathbf{x}|-6}{m^2|\textbf{x}|^5}\frac{F(\xi)e^{-m|\textbf{x}|}}{3}\biggr]\,x_l\,=\,\Psi_0
(\textbf{x})\,x_l
\\\\
\Lambda_{,l}(\textbf{x})\,=\,\biggl[\frac{6}{m^2|\textbf{x}|^5}-\frac{m^3|\textbf{x}|
^3+4m^2|\textbf{x}|^2+9m|\textbf{x}|+9}{m^2|\textbf{x}|^5}\frac{2F(\xi)e^{-m|\textbf{x}|}}{3}\biggr]\,x_l\,=\,\Lambda_0(\textbf{x})
\,x_l\\\\
\Xi_{,ln}(\textbf{x})\,=\,\Xi_0(\textbf{x})\delta_{ln}-\frac{m^2(m^2|\textbf{x}|^2+3m|\textbf{x}|+3)}{
|\textbf{x}|^5}F(\xi)e^{-m|\textbf{x}|}x_lx_n\,=\,\Xi_0(\textbf{x})\delta_{ln}+\Xi_1(\textbf{x})x_lx_n
\\\\
\Phi_{,ln}(\textbf{x})\,=\,\Phi_0(\textbf{x})\delta_{ln}-\biggl[\frac{3}{|
\textbf{x}|^5}+\frac{m^2|\textbf{x}|^2+3m|\textbf{x}|+3}{|\textbf{x}|^5}\frac{F(\xi)e^{-m|\textbf{x}|}}{3}\biggr]x_lx_n\,=\,\Phi_0(\textbf{x})\delta_{ln}+\Phi_1(\textbf{x})x_lx_n
\end{array}\right.
\end{eqnarray}
Obviously when we consider the physics in the matter or in the
vacuum we have to choose the "right" quantities $\Xi_0(\textbf{x})$,
$\Xi_1(\textbf{x})$, $\Phi_0(\textbf{x})$, $\Phi_1(\textbf{x})$,
$\Psi_0(\textbf{x})$, $\Lambda_0(\textbf{x})$.

The expression of Ricci scalar at fourth order
(\ref{Ricci_quarto}) is

\begin{eqnarray}\label{Ricci_quarto_solution}
R^{(4)}(t,\textbf{x})\,=\,&&\frac{{r_g}^2}{2m|\textbf{x}|}\int_0^\infty
d|\mathbf{x}'||\mathbf{x}'|\biggl\{e^{-m||\textbf{x}|-|\textbf{x}'||}-e^{-m(|\textbf{x}|+|\textbf{x}'|)}\biggr\}\biggl
\{\frac{m^4}{\mu^4}\biggl[\Xi(\textbf{x}')^2+\frac{|\textbf{x}'|^2}{m^2}\Xi_0(\textbf{x}')^2\biggr]
\nonumber\\\nonumber\\
&&+\biggl[3\Lambda(\mathbf{x}')+\frac{\Phi_0(\mathbf{x}')-\Psi_0(\mathbf{x}')+\Lambda_0(\textbf{x}')}{2}|\textbf{x}'|^2
\biggr]\Xi_0(\textbf{x}')
+m^2\Psi(\textbf{x}')\Xi(\textbf{x}')+\Xi_1(\textbf{x}')\Lambda(\textbf{x}')|\textbf{x}'|^2\biggr\}
\end{eqnarray}
where we note two contributions. The first one still
depends on the quadratic term ($\propto R^2$) in the action
(\ref{FOGaction}), while the second one is related to the cubic term
($\propto R^3$). By introducing the two functions $\Xi_I(\textbf{x})$
and $\Xi_{II}(\textbf{x})$, Eq. (\ref{Ricci_quarto_solution}) can be rewritten
as follows

\begin{eqnarray}\label{PAS_Ricci_quarto_solution}
R^{(4)}(t,\textbf{x})\,=\,{r_g}^2\biggl[\Xi_I(\textbf{x})+\frac{m^4}{\mu^4}\Xi_{II}(\textbf{x})\biggr]
\end{eqnarray}

An analogous situation is found for the $tt$-component of metric
tensor at fourth order. In fact Eq. (\ref{temp_temp_quarto})
becomes

\begin{eqnarray}\label{g00_quarto_solution}
g_{tt}^{(4)}(t,\textbf{x})=&&\frac{r_g\mathcal{X}}{|\mathbf{x}|}\int_0^\xi
d|\mathbf{x}'||\textbf{x}'|\biggl\{||\mathbf{x}|-|\mathbf{x}'||-|\mathbf{x}|-|\mathbf{x}'|\biggr\}
\rho(\mathbf{x}')\Phi(\mathbf{x}')\nonumber\\\nonumber\\&&
+\frac{{r_g}^2}{|\mathbf{x}|}\int_0^\infty
d|\mathbf{x}'||\textbf{x}'|\biggl\{||\mathbf{x}|-|\mathbf{x}'||-|\mathbf{x}|-|\mathbf{x}'|\biggr\}\biggl\{
\frac{1}{2}\biggl[\Xi_I(\mathbf{x}')+\Phi(\mathbf{x}')\Xi(\mathbf{x}')+\Phi_0(\mathbf{x}')^2|\mathbf{x}'|^2
\biggr]\nonumber\\\nonumber\\&&
\,\,\,\,\,\,\,\,\,\,\,\,\,\,\,\,\,\,\,\,\,\,\,\,\,\,\,\,\,\,\,\,\,\,\,\,\,\,\,\,\,\,\,\,\,\,\,\,\,\,\,\,\,\,\,\,
\,\,\,\,\,\,\,\,\,\,\,\,\,\,\,
-\frac{1}{2}\biggl[\biggl(3\Psi(\mathbf{x}')+\Lambda(\mathbf{x}')\biggr)\Phi_0(\mathbf{x}')+
\biggl(\Psi(\mathbf{x}')+\Lambda(\mathbf{x}')\biggr)\Phi_1(\mathbf{x}')|\mathbf{x}'|^2\biggr]\biggr\}
\nonumber\\\nonumber\\&&
-\frac{{r_g}^2}{3m^2|\mathbf{x}|}\int_0^\infty
d|\mathbf{x}'||\textbf{x}'|\biggl\{||\mathbf{x}|-|\mathbf{x}'||-|\mathbf{x}|-|\mathbf{x}'|\biggr\}
\biggl\{\frac{\Xi(\textbf{x}')^2}{4}+\triangle_{\mathbf{x}'}\Xi_I(\mathbf{x}')+\Xi_1(\mathbf{x}')\Lambda
(\mathbf{x}')|\mathbf{x}'|^2 \nonumber\\\nonumber\\&&
\,\,\,\,\,\,\,\,\,\,\,\,\,\,\,\,\,\,\,\,\,\,\,\,\,\,\,\,\,\,\,\,\,\,\,\,\,\,\,\,\,\,\,\,\,\,\,\,\,\,\,\,\,\,\,\,
\,\,\,\,\,\,\,\,\,\,\,\,\,\,\,\,\,\,\,\,\,\,\,\,\,\,\,\,\,\,\,\,\,\,\,\,\,\,\,\,\,\,\,\,\,\,\,\,\,\,\,\,\,\,\,\,
+\frac{\Xi_0(\mathbf{x}')}{2}\biggr[6\Lambda(\mathbf{x}')+\biggl(5\Psi_0(\mathbf{x}')
+3\Lambda_0(\mathbf{x}')\biggr)|\mathbf{x}'|^2\biggr]\nonumber\\\nonumber\\&&
\,\,\,\,\,\,\,\,\,\,\,\,\,\,\,\,\,\,\,\,\,\,\,\,\,\,\,\,\,\,\,\,\,\,\,\,\,\,\,\,\,\,\,\,\,\,\,\,\,\,\,\,\,\,\,\,
\,\,\,\,\,\,\,\,\,\,\,\,\,\,\,\,\,\,\,\,\,\,\,\,\,\,\,\,\,\,\,\,\,\,\,\,\,\,\,\,\,\,
+\frac{\Xi(\mathbf{x}')}{2}\biggr[2m^2\biggl(\Phi(\mathbf{x}')+\Psi(\mathbf{x}')\biggr)-3\Phi_0(\mathbf{x}')
-\Phi_1|\mathbf{x}'|^2\biggr]\biggr\} \nonumber\\\nonumber\\&&
+\frac{m^4}{\mu^4}\frac{{r_g}^2}{|\mathbf{x}|}\int_0^\infty
d|\mathbf{x}'||\textbf{x}'|\biggl\{||\mathbf{x}|-|\mathbf{x}'||-|\mathbf{x}|-|\mathbf{x}'|\biggr\}
\biggl\{\frac{\Xi_{II}(\textbf{x}')}{2}-\frac{m^2\Xi(\textbf{x}')^2+|\textbf{x}'|^2\Xi_0(\textbf{x}')^2}{3m^4}
\nonumber\\\nonumber\\&&
\,\,\,\,\,\,\,\,\,\,\,\,\,\,\,\,\,\,\,\,\,\,\,\,\,\,\,\,\,\,\,\,\,\,\,\,\,\,\,\,\,\,\,\,\,\,\,\,\,\,\,\,\,\,\,\,
\,\,\,\,\,\,\,\,\,\,\,\,\,\,\,\,\,\,\,\,\,\,\,\,\,\,\,\,\,\,\,\,\,\,\,\,\,\,\,\,\,\,\,\,\,\,\,\,\,\,\,\,\,\,\,\,
\,\,\,\,\,\,\,\,\,\,\,\,\,\,\,\,\,\,\,\,\,\,\,\,\,\,\,\,\,\,\,\,\,\,\,\,\,\,\,\,\,\,\,\,\,\,\,\,\,\,\,\,\,\,\,\,
\,\,\,\,\,\,\,\,\,\,\,\,\,\,\,\,\,\,\,\,\,\,\,\,\,\,\,
\,\,\,-\frac{\triangle_{\mathbf{x}'}\Xi_{II}(\mathbf{x}')}{3m^2}\biggr\}
\end{eqnarray}
and by introducing other new functions $\Phi_I(\textbf{x})$, $\Phi_{II}(\textbf{x})$, finally we have

\begin{eqnarray}\label{PAS_g00_quarto_solution}
g^{(4)}_{tt}(t,\textbf{x})={r_g}^2\biggl[\Phi_I(\textbf{x})+\frac{m^4}{\mu^4}\Phi_{II}(\textbf{x})\biggr]
\end{eqnarray}

It is useful to note that we have generally four contributions to
$g_{tt}^{(4)}$ in (\ref{g00_quarto_solution}). The first one is
induced by the non-linearity of the metric tensor even in
static spherically symmetric case. The product
$\rho(\mathbf{x})\Phi(\mathbf{x})$ is non-zero only in the matter
but contributes to determination of $tt$-component in any point of
the space. The second one takes into account  the  contribution induced
by the solution of previous order for the determination of the
$tt$-component of the Ricci tensor at fourth order. These first
two terms are present also in GR. While the second two terms are
derived from the modification of  the theory. In fact the third
contribution depends on the addition of the quadratic term
($\propto R^2$) in the action and finally the fourth one from the
addition of the cubic term ($\propto R^3$).

\begin{figure}[htbp] \centering
  \includegraphics[scale=1]{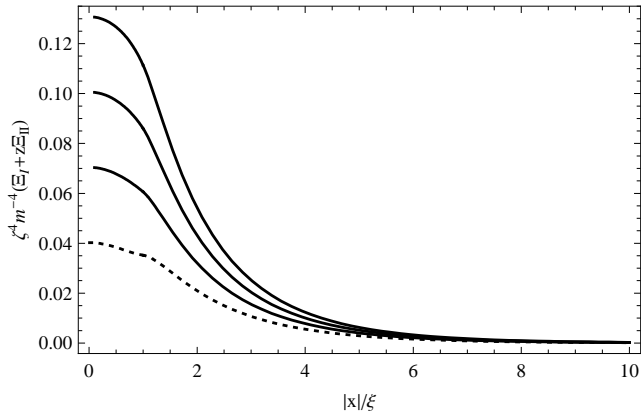}\\
  \caption{Plot of dimensionless function $\zeta^4m^{-4}(\Xi_I+z\,\Xi_{II})$, representing the Ricci scalar at fourth
  order, where $z\,=\,m^4/\mu^4$ and $\zeta\,\doteq\,m\xi\,=\,0.5$. The spatial behavior is shown for $0.3\,\leqslant\,z\,\leqslant\,0.9$
  (solid lines) while the dotted line corresponds to $R-\frac{1}{6m^2}R^2$-theory.}
  \label{plotricciquartoordine}
\end{figure}

The choice of free parameter $\mu$, which is linked to the third
derivative of $f(R)$,  is a crucial point in both the expressions
(\ref{PAS_Ricci_quarto_solution}) and
(\ref{PAS_g00_quarto_solution}) to obtain the right behavior. From the
mathematical interpretation of Newtonian limit one has
$|f'''(0)|\,<\,|f''(0)|$ and if $\mu^4\,>\,0$ (\emph{i.e.}
$\text{sign}[f'(0)]\,=\,-\,\text{sign}[f'''(0)]$, otherwise
$\mu^4$ is not a length) we have
$m^4/\mu^4\,=\,|f'''(0)|/3{f''(0)}^2$, so we find the constraint
$0\,<\,m^4/\mu^4\,<\,1$. In FIG. \ref{plotricciquartoordine}, we
report the spatial behavior of (\ref{PAS_Ricci_quarto_solution})
in the matter and in the vacuum,
($0.3\,\leqslant\,m^4/\mu^4\,\leqslant\,0.9$), showing that far from
the source we obtain a spacetime with a vanishing scalar curvature. At
Newtonian level, the Ricci scalar $R^{(2)}$ (\ref{sol_ric}) is negative defined
 while at post-Newtonian limit it is positive defined.

\begin{figure}[htbp] \centering
  \includegraphics[scale=1]{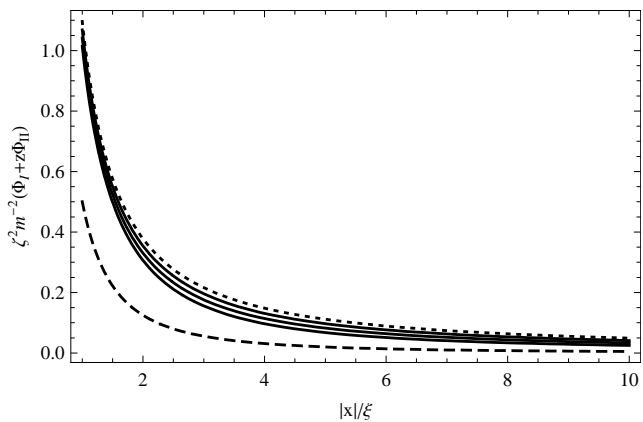}\\
  \caption{Plot of dimensionless function $\zeta^2m^{-2}(\Phi_I+z\,\Phi_{II})$ (solid lines) where $z\,=\,m^4/\mu^4$ and
  of function $1/2|\mathbf{x|^2}$ (dashed line). For $z\,=\,0$ (dotted line) we have the behavior of $R-\frac{1}{6m^2}R^2$-theory.
  The solid lines are obtained for $0.3\,\leqslant\,z\,\leqslant\,0.9$ and for $\zeta\,=\,m\,\xi\,=\,0.5$.}
  \label{plotg00quartoordine}
\end{figure}

In FIG. (\ref{plotg00quartoordine}), we report the time-time
component of metric tensor $g^{(4)}_{tt}$ on the same interval
of values of $m^4/\mu^4$, although the behavior is quite
insensitive to changes induced by the contributions of the cubic
term in the Lagrangian. Besides we can observe an important
analogy with respect the results of GR. In both cases, we have a
potential barrier, but for $f(R)$-gravity it is higher (as in the
Newtonian limit we found a deeper potential well).


\subsubsection{Solutions from isotropic to standard coordinates}

The metric solutions that we have found are expressed in isotropic
coordinates and often, for spherically symmetric problems, they are
conveniently rewritten in standard coordinates (the standard form in
which we write the Schwarzschild solution). Here the relativistic
invariant of metric (\ref{metric_tensor_PPN}) is

\begin{eqnarray}\label{metric0}
{ds}^2\,=\,\biggl[1+r_g\Phi(\mathbf{x})+{r_g}^2\biggl(\Phi_I(\mathbf{x})+\frac{m^4}{\mu^4}\Phi_{II}(\mathbf{x})
\biggr)\biggr]dt^2-\biggl[1-r_g\Psi(\mathbf{x})\biggr]|d\mathbf{x}|^2+r_g\Lambda(\mathbf{x})\frac{(\mathbf{x}\cdot
d\mathbf{x})^2}{|\mathbf{x}|^2}
\end{eqnarray}
From spherically symmetric form of (\ref{metric0}), it is convenient to
replace it with Eq. (\ref{me5}) by transformation (\ref{polar}).
The proper time interval (\ref{metric0}) then becomes

\begin{eqnarray}\label{metric1}
{ds}^2\,=\,\biggl[1+r_g\Phi(r)+{r_g}^2\biggl(\Phi_I(r)+\frac{m^4}{\mu^4}\Phi_{II}(r)\biggr)
\biggr]dt^2-\biggl[1-r_g\biggl(\Psi(r)+\Lambda(r)\biggr)\biggr]dr^2-\biggl[1-r_g\Psi(r)\biggr]r^2d\Omega
\end{eqnarray}
To get the metric in the standard form, we need
to impose a radial coordinate transformation

\begin{eqnarray}\label{condit_transf}
\biggl[1-r_g\Psi(r)\biggr]r^2\,=\,\tilde{r}^2
\end{eqnarray}
and we have a new set of coordinates $\{\tilde{r},\theta,\phi\}$. The metric (\ref{metric1}) becomes

\begin{eqnarray}\label{metricsolvedstandar}
{ds}^2\,=\,\biggl[1+r_g\tilde{\Phi}(\tilde{r})+{r_g}^2\biggl(\tilde{\Phi}_I(\tilde{r})+\frac{m^4}{\mu^4}
\tilde{\Phi}_{II}(\tilde{r})\biggr)
\biggr]dt^2-
\biggl[1-r_g\biggl(\tilde{\Psi}(\tilde{r})+\tilde{\Lambda}(\tilde{r})\biggr)\biggr]\biggr(\frac{dr}{d\tilde{r}}
\biggr)^2d\tilde{r}^2-\tilde{r}^2d\Omega
\end{eqnarray}
 The explicit expression of (\ref{metricsolvedstandar}) is not
displayed because the equation (\ref{condit_transf}) cannot be
solved algebraically. However, this technical problem is overcome
with the help of numerical methods when we are interesting to test
experimentally the theory.

\section{The Newtonian limit of Quadratic Gravity: $f(X,Y)\,=\,a_1R+a_2R^2+a_3R_{\alpha\beta}R^{\alpha\beta}$}\label{newtlimitquadr}

Since terms resulting from $R^n$ with $n \geq 3$ \emph{do not
contribute} in the Newtonian limit, as we have seen previously, we
provide explicit solutions for  different types of Lagrangians
generated by a primitive Lagrangian, the so-called \emph{Quadratic
Lagrangian} of the form \cite{newtonian_limit_R_Ric}

\begin{eqnarray}\label{quadratic-theory}
f(X,Y)\,=\,a_1R+a_2R^2+a_3R_{\alpha\beta}R^{\alpha\beta}
\end{eqnarray}
where $a_1$, $a_2$ and $a_3$ are arbitrary constants\footnote{Note
that the physical dimensions of the constants are
$[a_2]\,=\,[a_3]\,=\,$\emph{length}$^{2}$ and
$[a_1]\,=\,$\emph{length}$^{0}$.}. Such a Lagrangian belongs to
the general class of FOG (\ref{FOGaction}). The field equations
coming from (\ref{quadratic-theory}) are

\begin{eqnarray}\label{eq2}
\left\{\begin{array}{ll}
H_{\mu\nu}\,=\,(a_1+2a_2R)R_{\mu\nu}-\frac{a_1R+a_2R^2+a_3R_{\alpha\beta}R^{\alpha\beta}}{2}g_{\mu\nu}-2a_2R_{;\mu\nu}+2a_2g_{\mu\nu}\Box
R+2a_3{R_\mu}^\alpha R_{\alpha\nu}-2a_3{R^\alpha}_{(\mu;\nu)\alpha}\\\\\,\,\,\,\,\,\,\,\,\,\,\,\,\,\,\,\,\,\,\,\,+a_3\Box R_{\mu\nu}+a_3R_{\alpha\beta}^{\,\,\,\,\,\,\,\,;\alpha\beta}g_{\mu\nu}\,=\,\mathcal{X}\,T_{\mu\nu}
\\\\
H\,=\,-a_1R+2(3a_2+a_3)\Box R\,=\,\mathcal{X}\,T
\end{array}\right.
\end{eqnarray}
Till now the solutions of field equations (in general Eqs.
(\ref{fieldequationFOG})) are found by considering the trace Eq.
(\ref{tracefieldequationFOG}) as the dynamical equation for the
Ricci scalar. This approach allows  to study a set of second order
differential equations while the FOG equations are intrinsically
fourth order differential equations in metric formalism. By using
the trace equation (second line of (\ref{eq2})) or the definition
of the Ricci scalar (\ref{ricciscalar}) and substituting them into
the field equations (first line of (\ref{eq2})), we  have a set of
fourth order partial differential equations. In this section, we
will study  FOG field equations.

If we introduce the gravitational potentials in the isotropic metric (\ref{me4}) by
the quantities $\Phi$ and $\Psi$ linked to $g^{(2)}_{tt}$ and
$g^{(2)}_{ij}$

\begin{eqnarray}\label{metric-newtonian-order}
ds^2\,=\,\biggl[1+2\Phi\biggr]dt^2-\biggl[1-2\Psi\biggr]\delta_{ij}dx^idx^j
\end{eqnarray}
and by using the paradigm of Newtonian and post-Newtonian
developments, we can investigate, without assuming the harmonic
gauge condition, field Eqs.(\ref{eq2}) for the \emph{Quadratic
Lagrangian} (\ref{quadratic-theory}) at Newtonian order, that is

\begin{eqnarray}\label{PPN-fe-quadratic-theory}
\left\{\begin{array}{ll}
2a_1\triangle\Psi-2(4a_2+a_3)\triangle^2\Psi
+2(2a_2+a_3)\triangle^2\Phi\,=\,\mathcal{X}\rho
\\\\
\triangle\biggl[a_1(\Psi-\Phi)+(4a_2+a_3)\triangle\Phi-(8a_2+3a_3)\triangle\Psi\biggr]\delta_{ij}
-\biggl[a_1(\Psi-\Phi)+(4a_2+a_3)\triangle\Phi-(8a_2+3a_3)\triangle\Psi\biggr]_{,ij}\,=\,0
\end{array} \right.
\end{eqnarray}
By introducing two new auxiliary functions ($\tilde{\Phi}$ and
$\tilde{\Psi}$), Eqs.(\ref{PPN-fe-quadratic-theory}) become

\begin{eqnarray}\label{PPN-fe-quadratic-theory-1}
\left\{\begin{array}{ll}\frac{2a_2}{2a_2+a_3}\triangle^2\tilde{\Psi}-\frac{2a_3(3a_2+a_3)}{a_1(2a_2+a_3)}\triangle^2\tilde
{\Phi}-\frac{4a_2+a_3}{2a_2+a_3}\triangle\tilde{\Phi}-\frac{a_1}{2a_2+a_3}\triangle\tilde{\Psi}\,=\,\mathcal{X}
\rho\,\\\\
\triangle\biggl[\tilde{\Phi}+\triangle\tilde{\Psi}\biggr]\delta_{ij}-\biggl[\tilde{\Phi}+\triangle\tilde{\Psi}\biggr]_{,ij}
\,=\,0
\end{array} \right.
\end{eqnarray}
where $\tilde{\Phi}$ and $\tilde{\Psi}$ are linked to $\Phi$ and
$\Psi$ via the transformations

\begin{eqnarray}\label{trans}
\left\{\begin{array}{ll}
\Phi\,=\,-\frac{(8a_2+3a_3)\tilde{\Phi}+a_1\tilde{\Psi}}{2a_1(2a_2+a_3)}\,\\\\
\Psi\,=\,-\frac{(4a_2+a_3)\tilde{\Phi}+a_1\tilde{\Psi}}{2a_1(2a_2+a_3)}
\end{array} \right.
\end{eqnarray}
Obviously we must require $a_1(2a_2+a_3)\neq 0$, which is the
determinant of the transformations (\ref{trans}). Let us introduce
the new function $\gimel$ defined as follows

\begin{eqnarray}\label{definition}
\gimel\doteq\tilde{\Phi}+\triangle\tilde{\Psi}
\end{eqnarray}
At this point, we can use the new function $\Xi$ to uncouple the
system (\ref{PPN-fe-quadratic-theory}). With the choice
$\tilde{\Phi}\,=\,\gimel-\triangle\tilde{\Psi}$, it is possible to
rewrite Eqs. (\ref{PPN-fe-quadratic-theory}) as follows

\begin{eqnarray}
\left\{\begin{array}{ll}
\frac{2a_3(3a_2+a_3)}
{a_1(2a_2+a_3)}\triangle^3\tilde{\Psi}+\frac{6a_2+a_3}{2a_2+a_3}\triangle^2\tilde{\Psi}-\frac{a_1}{2a_2+a_3}\triangle
\tilde{\Psi}\,=\,\mathcal{X}\rho+\tau\\\\
\triangle\,\gimel\,\delta_{ij}-\gimel_{,ij}\,=\,0
\end{array} \right.
\end{eqnarray}
where
$\tau\,\doteq\,\frac{4a_2+a_3}{2a_2+a_3}\triangle\gimel+\frac{2a_3(3a_2+a_3)}
{a_1(2a_2+a_3)}\triangle^2\gimel$. We are interested in the
solution of (\ref{PPN-fe-quadratic-theory-1}) in terms of the
Green function $\daleth(\mathbf{x},\mathbf{x}')$  of field
operator
$\frac{2a_3(3a_2+a_3)}{a_1(2a_2+a_3)}\triangle^3+\frac{6a_2+a_3}{2a_2+a_3}\triangle^2-\frac{a_1}{2a_2+a_3}\triangle$.
Then  Eqs.(\ref{PPN-fe-quadratic-theory}) are equivalent to

\begin{eqnarray}\label{green-equation}
\left\{\begin{array}{ll}
\frac{2a_3(3a_2+a_3)}{a_1(2a_2+a_3)}\triangle^3\daleth(\mathbf{x},\mathbf{x}')+\frac{6a_2+a_3}{2a_2+a_3}\triangle^2\daleth(\mathbf{x},
\mathbf{x}')-\frac{a_1}{2a_2+a_3}\triangle
\daleth(\mathbf{x},\mathbf{x}')\,=\,\delta(\mathbf{x}-\mathbf{x}')\\\\
\triangle\,\gimel(\textbf{x})\,\delta_{ij}-\gimel(\textbf{x})_{,ij}\,=\,0
\end{array} \right.
\end{eqnarray}

The general solutions of Eqs.(\ref{PPN-fe-quadratic-theory-1}) for
$\Phi(\mathbf{x})$ and $\Psi(\mathbf{x}$), in terms of the Green
function $\daleth(\mathbf{x}, \mathbf{x}')$ and the function
$\gimel(\mathbf{x})$, are

\begin{eqnarray}\label{general-solution-newtonian-limit}
\left\{\begin{array}{ll}
\Phi(\mathbf{x})\,=\,\frac{(8a_2+3a_3)\triangle_\mathbf{x}-a_1}{2a_1(2a_2+a_3)}\int
d^3\mathbf{x}'\daleth(\mathbf{x},\mathbf{x}')\biggl[\mathcal{X}\rho(\mathbf{x}')
+\frac{4a_2+a_3}{2a_2+a_3}\triangle_{\mathbf{x}'}\gimel(\mathbf{x}')+\frac{2a_3(3a_2+a_3)}{a_1(2a_2+a_3)}\triangle^2_
{\mathbf{x}'}\gimel(\mathbf{x}')\biggr]-\frac{8a_2+3a_3}{2a_1(2a_2+a_3)}\gimel(\mathbf{x})\\\\
\Psi(\mathbf{x})\,=\,\frac{(4a_2+a_3)\triangle_\mathbf{x}-a_1}{2a_1(2a_2+a_3)}\int
d^3\mathbf{x}'\daleth(\mathbf{x},\mathbf{x}')\biggl[\mathcal{X}\rho(\mathbf{x}')+\frac{4a_2+a_3}{2a_2+a_3}\triangle_{\mathbf{x}'}
\gimel(\mathbf{x}')+\frac{2a_3(3a_2+a_3)}{a_1(2a_2+a_3)}\triangle^2_{\mathbf{x}'}\gimel(\mathbf{x}')\biggr]-\frac{4a_2+a_3}{2a_1(2a_2+a_3)}\gimel
(\mathbf{x})
\end{array} \right.
\end{eqnarray}

Eqs. (\ref{PPN-fe-quadratic-theory-1}) represent a coupled set of
fourth order differential equations. The total number of
integration constants is eight. With the substitution
(\ref{definition}), it has been possible to decouple the set of
equations, but now the differential order of the single equations
is changed. The total differential order is the same, indeed we
have an equation of sixth order and another equation of second
order, while, previously, we had two equations of fourth order.
The number of integration constants is preserved. We can conclude
that, with our approach, also the introduction of the new
quantities $\tilde{\Phi}$, $\tilde{\Psi}$ does not contradict the
paradigm of  metric theories of FOG. The price is that now the
{\it r.h.s.} of $tt$\,-\,component of the field equation has been
modified: there is an additional matter term $\tau$ coming from
the $ij$\,-\,component. In Table \ref{tablefieldequation}, we show
particular cases of (\ref{PPN-fe-quadratic-theory-1}) for
different choices of coupling constants of the theory with the
vanishing  determinant of transformations (\ref{trans}).

\begin{table}[htbp]
\centering
\begin{tabular}{c|c|c}
\hline\hline\hline
 Case & Choices of constants & Corresponding field equations \\
 \hline
 & & \\
 A & $\begin{array}{ll}a_2=0\\a_3=0\end{array}$ & $\begin{array}{ll}
 \triangle\Psi=\frac{\mathcal{X}}{2a_1}\rho\\\triangle\biggl[\Phi(\mathbf{x})+\frac{G}{a_1}\int d^3\mathbf{x}'\frac{\rho
 (\mathbf{x}')}{|\mathbf{x}-\mathbf{x}'|} \biggr]\delta_{ij}-\biggl[\Phi(\mathbf{x})+\frac{G}{a_1}\int d^3x'
 \frac{\rho(x')}{|\mathbf{x}-\mathbf{x}'|}\biggr]_{,ij}=0\end{array}$ \\
 \hline
 & & \\
 B & $\begin{array}{ll}a_1=0\\a_3=0\end{array}$ & $\begin{array}{ll}\triangle^2(2\Psi-\Phi)=-\frac{\mathcal{X}}{4a_2}\rho
 \\\triangle\biggl[\triangle(2\Psi-\Phi)\biggr]\delta_{ij}-\biggl[\triangle(2\Psi-\Phi)\biggr]_{,ij}=0
 \end{array}$ \\
 \hline
 & & \\
 C & $\begin{array}{ll}a_1=0\\a_2=0\end{array}$ &
 $\begin{array}{ll}\triangle^2(\Phi-\Psi)=\frac{\mathcal{X}}{2b_1}\rho\\\triangle\biggl[\triangle
 (\Phi-3\Psi)\biggr]\delta_{ij}-\biggl[\triangle(\Phi-3\Psi)\biggr]_{,ij}=0\end{array}$ \\
 \hline
 & & \\
 D & $a_3=-2a_2$ &
 $\begin{array}{ll}2a_2\triangle^2\Psi-a_1\triangle\Psi=-\frac{\mathcal{X}}{2}\rho\\\nabla^2\biggl[a_1\Phi(\mathbf{x})-2a_2
 \triangle\Phi(\mathbf{x})+G\int d^3\mathbf{x}'\frac{\rho(\mathbf{x}')}{|\mathbf{x}-\mathbf{x}'|}\biggr]\delta_{ij}-\biggl[
 a_1\Phi(\mathbf{x})-2a_2\triangle\Phi(\mathbf{x})+G\int d^3\mathbf{x}'\frac{\rho(\mathbf{x}')}{|\mathbf{x}-\mathbf{x}'|}
 \biggr]_{,ij}=0\end{array}$ \\
 \hline
 & & \\
 E & $\begin{array}{ll}a_1= 0\\a_3=-4a_2\end{array}$ & $\begin{array}{ll}\triangle^2\Phi=-\frac{\mathcal{X}}{4a_2}\rho
 \\\triangle\biggl[\triangle\Psi\biggr]\delta_{ij}- \biggl[\triangle\Psi\biggr]_{,ij}=0\end{array}$ \\
 \hline
 & & \\
 F & $\begin{array}{ll}a_1=0\\a_3=-2a_2\end{array}$ & $\begin{array}{ll}\triangle^2\Psi=-\frac{\mathcal{X}}{4a_2}\rho\\\
 \triangle\biggl[\triangle(\Psi-\Phi)\biggr]\delta_{ij}-\biggl[\triangle(\Psi-\Phi)\biggr]_{,ij}=0\end{array}$ \\
 \hline
 & & \\
 G & $\begin{array}{ll}a_1=0\\a_3=-\frac{8a_2}{3}\end{array}$ & $\begin{array}{ll}\triangle^2(2\Psi+\Phi)=-\frac{3\mathcal{X}}
 {4a_2} \rho\\\triangle\biggl[\triangle\Phi\biggr]\delta_{ij}-\biggl[\triangle\Phi\biggr]_{,ij}=0\end{array}$ \\
 \hline\hline\hline
 \end{tabular}
\caption{\label{tablefieldequation}Explicit form of the field
equations for different choices of the coupling constants for
which the determinant of  transformations (\ref{trans}) vanishes.}
\end{table}

\subsection{Green functions for systems with spherical symmetry}

As above,  we are interested in the solutions of
(\ref{PPN-fe-quadratic-theory-1}) at Newtonian order by using the
method of Green functions with spherical symmetry. Let us
introduce the radial coordinate $r\doteq|\mathbf{x}-\mathbf{x}'|$;
with this choice, the first equation of (\ref{green-equation}) for
$r\neq 0$ becomes

\begin{eqnarray}\label{eq16}
2a_3(3a_2+a_3)\triangle_r^3\daleth(r)+(6a_2+a_3)\triangle_r^2\daleth(r)-a^2_1\triangle_r
\daleth(r)\,=\,0
\end{eqnarray}
where $\triangle_r\,=\,r^{-2}\partial_r(r^{-2}\partial_r)$ is the radial
component of the Laplacian in polar coordinates. The solution of
(\ref{eq16}) is

\begin{eqnarray}\label{sol9}
\daleth(r)\,=\,K_1-\frac{1}{r}\biggr[K_2+\frac{a_3}{a_1}\biggl(K_3e^{-\sqrt{-\frac{a_1}{a_3}}r}
+K_4e^{\sqrt{-\frac{a_1}{a_3}}r}\biggr)-\frac{2(3a_2+a_3)}{a_1}\biggl(K_5e^{-\sqrt{\frac{a_1}{2(3a_2+a_3)}}r}
+K_6e^{\sqrt{\frac{a_1}{2(3a_2+a_3)}}r}\biggr)\biggr]
\end{eqnarray}
where $K_i$ are constants. We note that, if $a_2\,=\,a_3\,=\,0$,
the Green function of the Newtonian Mechanics is recovered. It is
the same situation of the Electromagnetism. The integration
constants $K_i$ have to be fixed by imposing the boundary
conditions at infinity and at  the origin. In fact Eq.
(\ref{sol9}) is a solution of (\ref{eq16}) and not of the first
equation in (\ref{green-equation}). A physically consistent
solution has to satisfy the condition
$\daleth(\mathbf{x},\mathbf{x}')\,\rightarrow\,0$, if
$|\mathbf{x}-\mathbf{x}'|\rightarrow \infty$, then the constants
$K_1$, $K_4$, $K_6$ in Eq.(\ref{sol9}) have to vanish. To obtain
the conditions on the constants $K_2$, $K_3$, $K_5$, we consider
the Fourier transformation of $\daleth(\mathbf{x},\mathbf{x}')$,
that is

\begin{eqnarray}\label{fourier}
\daleth(\mathbf{x},\mathbf{x}')\,=\,\int\frac{d^3\mathbf{k}}
{(2\pi)^3}\,\,\tilde{\daleth}(\mathbf{k})\,\,e^{i\mathbf{k}\cdot(\mathbf{x}-\mathbf{x}')}
\end{eqnarray}
By putting (\ref{fourier}) in the first equation of
(\ref{green-equation}), we obtain

\begin{eqnarray}
\daleth(\mathbf{x},\mathbf{x}')\,=\,-\int\frac{d^3\mathbf{k}}{(2\pi)^3}\frac{
e^{i\mathbf{k}\cdot(\mathbf{x}-\mathbf{x}')}}
{\frac{2a_3(3a_2+a_3)}{a_1(2a_2+a_3)}\mathbf{k}^6-\frac{6a_2+a_3}{2a_2+a_3}\mathbf{k}^4-\frac{a_1}{2a_2+a_3}
\mathbf{k}^2}
\end{eqnarray}
which, in the case of spherical symmetry, becomes\footnote{we introduced the polar
coordinates in the $\mathbf{k}$-space.}

\begin{eqnarray}\label{intfou}
\daleth(\mathbf{x},\mathbf{x}')\,=\,-\frac{1}{4\pi^2}\frac{a_1(2a_2+a_3)}{a_3(3a_2+a_3)}\frac{1}{|\mathbf{x}-\mathbf{x}'|}\int_0^\infty\frac{d
|\mathbf{k}|\sin|\mathbf{k}||\mathbf{x}-\mathbf{x}'|}{|\mathbf{k}|\biggl[\mathbf{k}^2-\frac{a_1}{a_3}\biggr]\biggl
[\mathbf{k}^2+\frac{a_1}{2(3a_2+a_3)}\biggr]}
\end{eqnarray}
The analytic expression of
$\daleth(\mathbf{x},\mathbf{x}')$ depends on the nature of the
poles of $|\mathbf{k}|$ and on the values of the arbitrary
constants $a_1$, $a_2$, $a_3$. If we define two "masses" $m_1$ and $m_2$

\begin{eqnarray}\label{scale2}
m_1^2\,\doteq\,-\frac{a_1}{2(3a_2+a_3)}\,,\,\,\,\,\,\,\,\,\,\,\,\,\,\,\,\,\,m_2^2\,\doteq\,\frac{a_1}{a_3}
\end{eqnarray}
we obtain, in Table \ref{tablegrennfunction}, the three particular
expressions of (\ref{intfou}). We note that the definition of mass
$m_1$ is a generalization of (\ref{mass_definition_0}) (and
obviously of (\ref{mass_definition_TS}) and
(\ref{yukawa-length})).

\begin{table}[htbp]
 \centering
 \begin{tabular}{c|c|c}
 \hline\hline\hline
 Case & Choices of constants & Green function \\
 \hline
 & & \\
 A & $\begin{array}{ll}a_3>0\\\\3a_2+a_3<0\end{array}$ & $
 \daleth^A(\mathbf{x},\mathbf{x}')\,=\,\frac{1}{12\pi}\frac{1}{|\mathbf{x}-\mathbf{x}'|}\biggl[\frac{m_1
 ^2-m_2^2}{m_1^2m_2^2}+\frac{e^{-m_1|\mathbf{x}-\mathbf{x}'|}}{m_1^2}-\frac{e^{-m_2|\mathbf{x}-\mathbf{x}'|}}{m_2^2}\biggr]$ \\
 \hline
 & & \\
 B & $\begin{array}{ll}a_3<0\\\\3a_2+a_3>0\end{array}$ & $ \daleth^B(\mathbf{x},\mathbf{x}')\,=\,\frac{1}{12\pi}
 \frac{1}{|\mathbf{x}-\mathbf {x}'|}\biggl[-\frac{m_1^2-m_2^2}{m_1^2m_2^2}-\frac{\cos(m_1|\mathbf{x}-\mathbf{x}'|)}{m_1^2}+\frac{\cos(m_2|\mathbf{x}-\mathbf{x}'|)}{m_2^2}\biggr]$ \\
 \hline
 & & \\
 C & $\begin{array}{ll}a_3<0\\\\3a_2+a_3<0\end{array}$ & $ \daleth^C(\mathbf{x},\mathbf{x}')\,=\,\frac{1}{12\pi}\frac
 {1}{|\mathbf{x}-\mathbf{x}'|}\biggl[-\frac{m_1^2+m_2^2}{m_1^2m_2^2}+\frac{e^{-m_1|\mathbf{x}-\mathbf{x}'|}}{m_1^2}+\frac{\cos(m_2|\mathbf{x}-
 \mathbf{x}'|)}{m_2^2}\biggr]$ \\
 \hline\hline\hline
 \end{tabular}
 \caption{\label{tablegrennfunction}The complete set of Green
 functions for (\ref{intfou}). It is possible to have a further choice for the scale lengths
 depending on the other two length scales. If we perform the substitution
 $m_1^2\rightleftharpoons-m_2^2$, we obtain a fourth choice. In addition, for a correct
 Newtonian component, we assumed $a_1\,>\,0$. In fact when $a_2\,=\,a_3=0$
 the field Eqs. (\ref{PPN-fe-quadratic-theory}) give  the Newtonian limit for $a_1\,=\,1$.}
\end{table}

Besides, we have obtained  another Yukawa-like correction to the
Newtonian potential related to the squared Ricci tensor correction
in the Lagrangian (\ref{quadratic-theory}). This behavior is
strictly linked to the sixth order of (\ref{green-equation}),
which depends on the coupled form of the system of equations
(\ref{PPN-fe-quadratic-theory}). In fact, if we consider the
Fourier transform of the potentials $\Phi$ and $\Psi$

\begin{eqnarray}
\Phi(\mathbf{x})\,=\,\int\frac{d^3\mathbf{k}}{(2\pi)^{3/2}}\,\,\hat{\Phi}(\mathbf{k})\,\,e^{i\mathbf{k}\cdot
\mathbf{x}}\,,\,\,\,\,\,\,\,\,\Psi(\mathbf{x})\,=\,\int\frac{d^3\mathbf{k}}{(2\pi)^{3/2}}\,\,\hat{\Psi}(\mathbf{k})\,\,
e^{i\mathbf{k}\cdot\mathbf{x}}
\end{eqnarray}
the solutions are

\begin{eqnarray}\label{general-solution-newtonian-limit-point-like}
\left\{\begin{array}{ll}
\Phi(\mathbf{x})\,=\,-\frac{\mathcal{X}}{2}\int\frac{d^3\mathbf{k}}{(2\pi)^{3/2}}\frac{[a_1+
(8a_2+3a_3)\mathbf{k}^2]\tilde{\rho}(\mathbf{k})e^{i\mathbf{k}\cdot\mathbf{x}}}{\mathbf{k}^2(a_1-a_3\mathbf{k}^2)[a_1+2(3a_2+
a_3)\mathbf{k}^2]}\\\\
\Psi(\mathbf{x})\,=\,-\frac{\mathcal{X}}{2}\int\frac{d^3\mathbf{k}}{(2\pi)^{3/2}}\frac{[a_1+
(4a_2+a_3)\mathbf{k}^2]\tilde{\rho}(\mathbf{k})e^{i\mathbf{k}\cdot\mathbf{x}}}{\mathbf{k}^2(a_1-a_3\mathbf{k}^2)[a_1+2(3a_2+
a_3)\mathbf{k}^2]}
\end{array} \right.
\end{eqnarray}
where $\tilde{\rho}(\mathbf{k})$ is the Fourier transform of the
matter density. We can see that the solutions have the same poles
as Eq.(\ref{intfou}). Finally, considering the Fourier transform
of the point-like source (\ref{point_like}), that is
$\tilde{\rho}(\mathbf{k})\,=\,\frac{M}{(2\pi)^3}$,  the solutions
(\ref{general-solution-newtonian-limit-point-like}) are

\begin{eqnarray}\label{solutionsfuorier}
\left\{\begin{array}{ll}
\Phi(\mathbf{x})\,=\,-\frac{GM}{a_1}\biggl(\frac{1}{|\mathbf{x}|}+\frac{1}{3}\frac{e^{-m_1|\mathbf{x}|}}{|\mathbf{x}|}
-\frac{4}{3}\frac{e^{-m_2|\mathbf{x}|}}{|\mathbf{x}|}\biggr)\\\\
\Psi(\mathbf{x})\,=\,-\frac{GM}{a_1}\biggl(\frac{1}{|\mathbf{x}|}-\frac{1}{3}
\frac{e^{-m_1|\mathbf{x}|}}{|\mathbf{x}|}-\frac{2}{3}\frac{e^{-m_2|\mathbf{x}|}}{|\mathbf{x}|}\biggr)
\end{array} \right.
\end{eqnarray}
where $a_3\,\neq\,0$ and $3a_2+a_3\,\neq\,0$ and
$m_1^2,\,\,m_2^2\,>0$. The solution for $\Phi$ generalizes  the
third line of (\ref{sol_new_pfR}) in the case of point-like
source. In fact, it is interesting to note that, if $a_3\,=\,0$
($m_2\,\rightarrow\,\infty$), we have the missing of a scale
length (a pole is missed) with only a Yukawa-like term analogously
to Electrodynamics. The Green function, in this case, is

\begin{eqnarray}\label{greenfunctionfR}
\tilde{\daleth}(\mathbf{k})_{a_3=0}\,=\,\frac{2a_2}{6a_2\mathbf{k}^4+a_1\mathbf{k}^2}
\end{eqnarray}
and  Lagrangian (\ref{quadratic-theory}) becomes
$f\,=\,a_1R+a_2R^2$. Finally the presence of the pole is achieved
considering a particular choice of the constants in the theory:
$a_3\,=\,-2a_2$. In Table \ref{tablefieldequation} (Case D), we
provide the field equations for this choice and the relative Green
function is

\begin{eqnarray}\label{greenfunctionD}
\tilde{\daleth}_{(2a_2\nabla^4-a_1\nabla^2)}(\mathbf{k})\propto\frac{1}{2a_2
\mathbf{k}^4+a_1\mathbf{k}^2}
\end{eqnarray}
The spatial behavior of (\ref{greenfunctionfR}) -
(\ref{greenfunctionD}) is the same but the coefficients are
different since the theories are different. In conclusion, we need
the Green function for the differential operator $\triangle^2$.
The only possible physical choice for the squared Laplacian  is

\begin{eqnarray}\label{greenfunctionlaplsqu}
\tilde{\daleth}_{(\triangle^2)}(\mathbf{x}-\mathbf{x}')\propto\frac{1}
{|\mathbf{x}-\mathbf{x}'|}
\end{eqnarray}
since the other choice is proportional to
$|\mathbf{x}-\mathbf{x}'|$ and cannot to be accepted. Considering
the last possibility, we will end up with a force law increasing
with distance \cite{old_papers_fR_2}. In  summary, we have shown
the general approach to find out solutions of the field equations
by using the Green functions. In particular, the vacuum solutions
with point-like source have been used to find out directly the
potentials, however it remains the  important issue to find out
solutions  for systems with extended matter distributions.

\subsection{Solutions using the Green functions approach}

Before  investigating the general solution of (\ref{PPN-fe-quadratic-theory}),
we want discuss all cases shown in the Table
\ref{tablefieldequation}. Later, we will
derive solutions in presence of matter using the Green
functions  in Table \ref{tablegrennfunction}.

Specifically, in Table \ref{tablefieldequationsolution}, we provide solutions in
terms of  Green functions of the corresponding differential
operators coming from the field equations shown in Table
\ref{tablefieldequation}. Case A corresponds to the Newtonian
theory and the arbitrary constant $a_1$ can be absorbed in the
definition of matter Lagrangian. The solutions are

\begin{eqnarray}
_A\Phi(\mathbf{x})\,=\,_A\Psi(\mathbf{x})\,=\,-G\int d^3\mathbf{x}'\frac{\rho(\mathbf{x}')}{|\mathbf{x}-
\mathbf{x}'|}
\end{eqnarray}
For Case D, instead, we have the
field equations  of a sort of modified electrodynamic-like
representation. The solution can be expressed as follows

\begin{eqnarray}\label{potsolD}
_D\Phi(\mathbf{x})\,=\,_D\Psi(\mathbf{x})\,=\,-G\int d^3\mathbf{x}'\biggl[\frac{1-e^{-\sqrt
{\frac{a_1}{2a_2}}|\mathbf{x}-\mathbf{x}'|}}{|\mathbf{x}-\mathbf{x}'|}\biggr]\rho(\mathbf{x}')
\end{eqnarray}
The solutions make sense only if $a_1/a_2\,>\,0$, then we can
introduce a new scale-length. A particular expression of
(\ref{potsolD}), for a fixed matter density $\rho(\mathbf{x})$,
will be found in a more general context in the next section.
Nevertheless these two cases are the only ones which exhibit the
Newtonian limit (obviously the first one!), while for the
remaining cases there are serious problems with divergences and
incompatibilities. In fact, Case B presents an incompatibility
between the solution obtained from the $tt$-component and the
one from the $ij$-component. The incompatibility can be
removed if we consider, as the Green function for the differential
operator $\nabla^4$, the trivial solution:
$\mathcal{G}_{(\triangle^2)}|_B\,=\,\text{const.}$ With this
choice, the arbitrary integration constant $\Phi_0$ can be
interpreted as $-GM$. However another problem remains: namely the
divergence at the origin. The interpretation of the constant
$\Phi_0$ as a total mass and not as a generic integral $\int
d^3\mathbf{x}'\rho(\mathbf{x}')$ does not avoid the singularity.
Finally, the solution

\begin{eqnarray}
2_B\Psi(\mathbf{x})-_B\Phi(\mathbf{x})\,=\,-\frac{GM}{|\mathbf{x}-\mathbf{x}'|}
\end{eqnarray}
holds only in vacuum. Before continuing the analysis of the
various cases, the term $\int
d^3\mathbf{x}'\daleth_{(\triangle^2)}(\mathbf{x},\mathbf{x}')\rho(\mathbf{x}')$
has to be discussed for the  choice (\ref{greenfunctionlaplsqu}).
A generic field equation with $\triangle^2$ (from Table
\ref{tablefieldequation}) is

\begin{eqnarray}
\triangle_\mathbf{x}^2\Phi(\mathbf{x})\propto\triangle_\mathbf{x}^2\int d^3\mathbf{x}'\frac{\rho(\mathbf{x}')}
{|\mathbf{x}-\mathbf{x}'|}\,\propto\,\triangle_\mathbf{x}\rho(\mathbf{x})\,\neq\,4\pi\rho(\mathbf{x})
\end{eqnarray}
from  which the only consistent possibility is to set
$\rho(\mathbf{x})\,=\,0$. In the remaining cases, we can only consider
 vacuum solutions.

\begin{table}[htbp]
 \centering
 \begin{tabular}{c|c|c}
 \hline\hline\hline
 Case & Solutions & Newtonian behavior \\
 \hline
 & & \\
 A & $\begin{array}{ll}
 \Phi(\mathbf{x})=\Psi(\mathbf{x})=-\frac{G}{a_1}\int d^3\mathbf{x}'\frac{\rho(\mathbf{x}')}{|\mathbf{x}-\mathbf{x}'|}
 \end{array}$ & yes \\
 \hline
 & & \\
 B & $\begin{array}{ll}
 2\Psi(\mathbf{x})-\Phi(\mathbf{x})=\frac{\Phi_0}{|\mathbf{x}|}\\\\ 2\Psi(\mathbf{x})-\Phi(\mathbf{x})=-\frac{2\pi G}{a_2}\int
 d^3\mathbf{x}'\daleth_{(\nabla^4)}(\mathbf{x},\mathbf{x}')\rho
 (\mathbf{x}')\end{array}$ & no \\
 \hline
 & & \\
 C & $\begin{array}{ll}
 \Phi(\mathbf{x})=\frac{\Phi_0}{|\mathbf{x}|}+\frac{6\pi G}{a_3}\int d^3\mathbf{x}'\daleth_{(\nabla^4)}(\mathbf{x},
 \mathbf{x}')\rho(\mathbf{x}')\\\\\Psi(\mathbf{x})=\frac{\Phi_0}{|\mathbf{x}|}+\frac{2\pi G}{a_3}\int d^3\mathbf{x}'\daleth_{(\nabla^4)}(\mathbf{x},\mathbf{x}')\rho(\mathbf{x}')\end{array}$ & no \\
 \hline
 & & \\
 D & $\begin{array}{ll}
 \Phi(\mathbf{x})=-4\pi G\int d^3\mathbf{x}'\daleth_{(2a_2\nabla^4-a_1\nabla^2)}(\mathbf{x},\mathbf{x}')\rho
 (\mathbf{x}') \\\\\Psi(\mathbf{x})=-4\pi G\int d^3\mathbf{x}'\daleth_{(2a_2\nabla^4-a_1\nabla^2)}(\mathbf{x},\mathbf{x}')
 \rho(\mathbf{x}')\end{array}$ & yes \\
 \hline
 & & \\
 E & $\begin{array}{ll}
 \Phi(\mathbf{x})=-\frac{2\pi G}{a_2}\int d^3\mathbf{x}'\daleth_{(\nabla^4)}(\mathbf{x},\mathbf{x}')\rho(\mathbf{x}')\\\\
 \Psi(\mathbf{x})=\frac{\Phi_0}{|\mathbf{x}|}\end{array}$ & no\\
 \hline
 & & \\
 F & $\begin{array}{ll}
 \Phi(\mathbf{x})=\frac{\Phi_0}{|\mathbf{x}|}-\frac{2\pi G}{a_2}\int d^3\mathbf{x}'\daleth_{(\nabla^4)}(\mathbf{x},
 \mathbf{x}')\rho(\mathbf{x}')\\\\\Psi(\mathbf{x})=-\frac{2\pi G}{a_2}\int
 d^3\mathbf{x}'\daleth_{(\nabla^4)}(\mathbf{x},\mathbf{x}')\rho
 (\mathbf{x}')\end{array}$ & no\\
 \hline
 & & \\
 G & $\begin{array}{ll}
 \Phi(\mathbf{x})=\frac{\Phi_0}{|\mathbf{x}|}\\\\
 \Psi(\mathbf{x})=-\frac{1}{2}\frac{\Phi_0}{|\mathbf{x}|}-\frac{3\pi G}{a_2}\int d^3\mathbf{x}'\daleth_{(\nabla^4)}
 (\mathbf{x},\mathbf{x}')\rho(\mathbf{x}')\end{array}$ & no\\
 \hline\hline\hline
 \end{tabular}
 \caption{\label{tablefieldequationsolution}
 Solutions of the field equations in Table
 \ref{tablefieldequation}. Solutions are found by setting
 $\beth=0$ in the $ij$\,-\,component of the field Eqs. (\ref{green-equation}). Solutions are displayed in terms of the Green functions. $\Phi_0$ is a generic integration constant.}
\end{table}

By considering solutions (\ref{general-solution-newtonian-limit})
with the Green function $\daleth^A(\mathbf{x},\mathbf{x}')$
from Table \ref{tablegrennfunction} and by assuming
$\gimel(\mathbf{x})\,=\,0$\footnote{We have to note that our working hypothesis, $\gimel(\mathbf{x})\,=\,0$,
is not particular, since when we considered the Hilbert-Einstein
Lagrangian to give the Newtonian solution, we  imposed an analogous condition.}, we have

\begin{eqnarray}
\left\{\begin{array}{ll}
_A\Phi(\mathbf{x})\,=\,\mathcal{X}\,\frac{(8a_2+3a_3)\triangle_\mathbf{x}-a_1}{2a_1(2a_2+a_3)}\int
d^3\mathbf{x}'\daleth^A(\mathbf{x},\mathbf{x}')\rho(\mathbf{x}')\\\\
_A\Psi(\mathbf{x})\,=\,\mathcal{X}\,\frac{(4a_2+a_3)\triangle_\mathbf{x}-a_1}{2a_1(2a_2+a_3)}\int
d^3\mathbf{x}'\daleth^A(\mathbf{x},\mathbf{x}')\rho(\mathbf{x}')
\end{array} \right.
\end{eqnarray}
After some algebraic calculations we get, for a spherically symmetric matter source (\ref{ball_like}),
the internal and external solutions for $\Phi$ 

\begin{eqnarray}\label{potfinA}
\left\{\begin{array}{ll}
_A\Phi_{in}(\mathbf{x})\,=\,-\frac{GM}{a_1}\biggl[\frac{8m_1^2+m_2^2(2+3m_1^2\xi^2)
}{2m_1^2m_2^2\xi^3}-\frac{|\mathbf{x}|^2}{2\xi^3}-\frac{e^{-m_1\xi}(1+m_1\xi)}{m_1^2\xi^3}\frac{\sinh m_1|\mathbf{x}|}
{m_1|\mathbf{x}|}+4\frac{e^{-m_2\xi}(1+m_2\xi)}{m_2^2\xi^3}\frac{\sinh m_2|\mathbf{x}|}{m_2|\mathbf{x}|}\biggr]
\\\\\\
_A\Phi_{out}(\mathbf{x})\,=\,-\frac{GM}{a_1}\biggl[\frac{1}{|\mathbf{x}|}+\frac{F(m_1\xi)}{3}
\frac{e^{-m_1|\mathbf{x}|}}{|\mathbf{x}|}-\frac{4F(m_2\xi)}{3}\frac{e^{-m_2|\mathbf{x}|}}{|\mathbf{x}|}\biggr]
\end{array}\right.
\end{eqnarray}
representing a generalization of metric potential (\ref{sol_pot}). It is easy to check that when $a_3\,\rightarrow\,0$ (\emph{i.e.} $m_2\,\rightarrow\,\infty$), we get exactly Eq. (\ref{sol_pot}).
If we consider the limit of point-like source, we get the first line of (\ref{solutionsfuorier}).

For the sake of completeness, let us derive  solutions for the
other Green functions. Starting from Case B in Table
\ref{tablegrennfunction}, yet we have the
internal and external potentials

\begin{eqnarray}\label{potfinB}
\left\{\begin{array}{ll}
_B\Phi_{in}(\mathbf{x})\,=\,-\frac{GM}{a_1}\biggl[\frac{m_2^2(2+3m_1^2
\xi^2)-8m_1^2}{2m_1^2m_2^2\xi^3}-\frac{|\mathbf{x}|^2}{2\xi^3}-\frac{\cos m_1\xi+m_1\xi\sin m_1\xi}{m_1^2\xi^3}\frac{\sin m_1|\mathbf{x}|}
{m_1|\mathbf{x}|}
+4\frac{\cos m_2\xi+m_2\xi\sin m_2\xi}{m_2^2\xi^3}\frac{\sin m_2|\mathbf{x}|}{m_2|\mathbf{x}|}
\biggr]
\\\\
_B\Phi_{out}(\mathbf{x})\,=\,-\frac{GM}{a_1}\biggl[\frac{1}{|\mathbf{x}|}+\frac{G(m_1\xi)}{3}\frac{\cos m_1|\mathbf{x}|}
{|\mathbf{x}|}-
\frac{4G(m_2\xi)}{3}\frac{\cos m_2|\mathbf{x}|}{|\mathbf{x}|}\biggr]\end{array}\right.
\end{eqnarray}
and its limit of point-like source is

\begin{eqnarray}
\lim_{\xi\rightarrow 0}\,\,_B\Phi_{out}(|\mathbf{x}|)\,=\,-\frac{GM}{a_1}\biggl[\frac{1}
{|\mathbf{x}|}-\frac{1}{6}\frac{\cos m_1|\mathbf{x}|}{|\mathbf{x}|}+\frac{2}{3}\frac{\cos m_2|\mathbf{x}|}
{|\mathbf{x}|}\biggr]
\end{eqnarray}
Finally for Case C in Table \ref{tablegrennfunction}, we have

\begin{eqnarray}\label{potfinC}\left\{\begin{array}{ll}
_C\Phi_{in}(\mathbf{x})\,=\,-\frac{GM}{a_1}\biggl[\frac{m_2^2(2+3m_1^2\xi^2)+
8m_1^2}{2m_1^2m_2^2\xi^3}-\frac{|\mathbf{x}|^2}{2\xi^3}-\frac{e^{-m_1\xi}(1+m_1\xi)}{m_1^2\xi^3}\frac{\sinh m_1|\mathbf{x}|}{m_1|\mathbf{x}|}-4\frac{\cos m_2\xi+m_2\xi\sin m_2\xi}{m_2^2\xi^3}
\frac{\sin m_2|\mathbf{x}|}{m_2|\mathbf{x}|}\biggl]
\\\\
_C\Phi_{out}(\mathbf{x})\,=\,-\frac{GM}{a_1}\biggl[\frac{1}{|\mathbf{x}|}+
\frac{F(m_1\xi)}{3}\frac{e^{-m_1|\mathbf{x}|}}{|\mathbf{x}|}+\frac{4G(m_2\xi)}{3}\frac{\cos m_2|\mathbf{x}|}{|\mathbf{x}|}\biggr]
\end{array}\right.
\end{eqnarray}
which in the point-like limit becomes

\begin{eqnarray}
\lim_{\xi\rightarrow 0}\,\,_C\Phi_{out}(\mathbf{x})\,=\,-\frac{GM}{a_1}\biggl[\frac{1}{|\mathbf{x}|}+
\frac{1}{3}\frac{e^{-m_1|\mathbf{x}|}}{|\mathbf{x}|}-\frac{4}{3}\frac{\cos m_2|\mathbf{x}|}{|\mathbf{x}|}\biggr]
\end{eqnarray}
Similar relations are found for $\Psi$.

These results mean that, for suitable distance scales, the validity of Gauss theorem is restored and the theory agrees with the standard Newtonian limit of GR.

\section{The Newtonian limit of $f(X,Y,Z)$-Gravity}\label{fXYZgravity}

We conclude this review providing the Newtonian limit of a generic $f(X,Y,X)$-gravity \cite{fXYZ_limit}. By considering the paradigm of Newtonian limit (\ref{PPN-metric}),
 the curvature invariants $X$, $Y$, $Z$ become

\begin{eqnarray}
\left\{\begin{array}{ll}
X\,\sim\,X^{(2)}+X^{(4)}+\dots\\\\
Y\,\sim\,Y^{(4)}+Y^{(6)}+\dots\\\\
Z\,\sim\,Z^{(4)}+Z^{(6)}\dots
\end{array}\right.
\end{eqnarray}
and the function $f$ can be developed as

\begin{eqnarray}
f(X,Y,Z)\,&\sim&\,f(0)+f_X(0)X^{(2)}+\frac{1}{2}f_{XX}(0){X^{(2)}}^2+f_X(0)X^{(4)}+f_Y(0)Y^{(4)}+f_Z(0)Z^{(4)}+\dots
\end{eqnarray}
and analogous relations for partial derivatives of $f$ are
obtained. From the lowest order of field Eqs. (\ref{fieldequationFOG}),  we have

\begin{eqnarray}\label{PPN-field-equation-general-theory-fR-O0}
f(0)\,=\,0
\end{eqnarray}
this means that not only in $f(R)$-gravity \cite{newtonian_limit_fR,spher_symm_fR}
but also in $f(X,Y,Z)$-gravity, a missing cosmological component in
the action (\ref{FOGaction}) implies that the space-time is
asymptotically Minkowskian. Eqs. (\ref{fieldequationFOG})
and (\ref{tracefieldequationFOG}) at $\mathcal{O}$(2)-order
become\footnote{We used the properties
${R_{\alpha\beta}}^{;\alpha\beta}\,=\,\frac{1}{2}\Box R$ and
${{R_\mu}^{\alpha\beta}}_{\nu;\alpha\beta}\,=\,{{R_\mu}^\alpha}_{;\nu\alpha}-\Box
R_{\mu\nu}$.}

\begin{eqnarray}\label{NL-field-equation}
\left\{\begin{array}{ll}
H^{(2)}_{tt}\,=\,f_X(0)R^{(2)}_{tt}-[f_Y(0)+4f_Z(0)]\triangle
R^{(2)}_{tt}
-\frac{f_X(0)}{2}X^{(2)}-[f_{XX}(0)+\frac{f_Y(0)}{2}]\triangle
X^{(2)}\,=\,\mathcal{X}\,T^{(0)}_{tt}\\\\
H^{(2)}_{ij}\,=\,f_X(0)R^{(2)}_{ij}-[f_Y(0)+4f_Z(0)]\triangle
R^{(2)}_{ij}
+\frac{f_X(0)}{2}X^{(2)}\delta_{ij}+[f_{XX}(0)+\frac{f_Y(0)}{2}]\triangle
X^{(2)}\delta_{ij}-f_{XX}(0){X^{(2)}}_{,ij}+\\\\
\,\,\,\,\,\,\,\,\,\,\,\,\,\,\,\,\,\,\,\,\,\,+[f_Y(0)+4f_Z(0)]R^{(2)}_{mi,jm}+f_Y(0)R^{(2)}_{mj,im}\,=\,0\\\\
H^{(2)}\,=\,-f_X(0)X^{(2)}-[3f_{XX}(0)+2f_Y(0)+2f_Z(0)]\triangle
X^{(2)}\,=\,\mathcal{X}\,T^{(0)}
\end{array}\right.
\end{eqnarray}
By introducing the quantities

\begin{eqnarray}\label{mass_definition}
\left\{\begin{array}{ll}
{m_1}^2\,\doteq\,-\frac{f_X(0)}{3f_{XX}(0)+2f_Y(0)+2f_Z(0)}\\\\
{m_2}^2\,\doteq\,\frac{f_X(0)}{f_Y(0)+4f_Z(0)]}
\end{array}\right.
\end{eqnarray}
we get three differential equations for curvature invariant
$X^{(2)}$, $tt$- and $ij$-component of Ricci tensor
$R^{(2)}_{\mu\nu}$

\begin{eqnarray}\label{NL-field-equation_2}
\left\{\begin{array}{ll}
(\triangle-{m_2}^2)R^{(2)}_{tt}+\biggl[\frac{{m_2}^2}{2}-\frac{{m_1}^2+2{m_2}^2}{6{m_1}^2}\triangle\biggr]
X^{(2)}\,=\,-\frac{{m_2}^2\mathcal{X}}{f_X(0)}\,T^{(0)}_{tt}\\\\
(\triangle-{m_2}^2)R^{(2)}_{ij}+\biggl[\frac{{m_1}^2-{m_2}^2}{3{m_1}^2}\,
\partial^2_{ij}-\biggl(\frac{{m_2}^2}{2}-\frac{{m_1}^2+2{m_2}^2}{6{m_1}^2}\triangle\biggr)\delta_{ij}\biggr]
X^{(2)}\,=\,0\\\\
(\triangle-{m_1}^2)X^{(2)}\,=\,\frac{{m_1}^2\mathcal{X}}{f_X(0)}\,T^{(0)}
\end{array}\right.
\end{eqnarray}
We note that the definitions (\ref{mass_definition}) are a generalization of (\ref{scale2}). While the interpretation
of $m_1$ is clear from the trace field equation of $f(R)$-gravity, now also  $m_2$ is clear. 
In fact, from the two first lines of (\ref{NL-field-equation_2}) the Ricci tensor has a characteristic length (${m_2}^{-1}$).

The solution for curvature invariant $X^{(2)}\,=\,R^{(2)}$ at third
line of (\ref{NL-field-equation_2}) is

\begin{eqnarray}\label{scalar_invariant_sol_gen}
X^{(2)}(t,\textbf{x})\,=\,\frac{{m_1}^2\mathcal{X}}{f_X(0)}\int
d^3\mathbf{x}'\mathcal{G}_1(\mathbf{x},\mathbf{x}')T^{(0)}(t,\mathbf{x}')
\end{eqnarray}
where $\mathcal{G}_1(\mathbf{x},\mathbf{x}')$ is the Green
function of the field operator $\triangle-{m_1}^2$ (this solution is formally equal to (\ref{scalar_ricci_sol_gen})). The solution for
$g^{(2)}_{tt}$, by (\ref{PPN-ricci-tensor}), is the
following

\begin{eqnarray}\label{tt_component_sol_gen}
g^{(2)}_{tt}(t,\textbf{x})\,=\,\frac{1}{2\pi}\int
d^3\mathbf{x}'d^3\mathbf{x}''\frac{\mathcal{G}_2(\mathbf{x}',\mathbf{x}'')}{|\mathbf{x}-\mathbf{x}'|}\biggl[
\frac{{m_2}^2\mathcal{X}}{f_X(0)}T^{(0)}_{tt}(t,\mathbf{x}'')-\frac{({m_1}^2+2{m_2}^2)\mathcal{X}}{6f_X(0)}\,T^{(0)}
t,\mathbf{x}'')+\frac{{m_2}^2-{m_1}^2}{6}X^{(2)}(t,\mathbf{x}'')\biggr]
\end{eqnarray}
where $\mathcal{G}_2(\mathbf{x},\mathbf{x}')$ is the Green
function of field operator $\triangle-{m_2}^2$. The general solution for $g^{(2)}_{ij}$ from
Eqs. (\ref{NL-field-equation_2}), if we consider the expression (\ref{PPN-ricci-tensor-HG}) (gauge harmonic condition), is

\begin{eqnarray}
g^{(2)}_{ij}|_{HG}\,=\,\frac{1}{2\pi}\int
d^3\mathbf{x}'d^3\mathbf{x}''\frac{\mathcal{G}_2(\mathbf{x}',\mathbf{x}'')}{|\mathbf{x}-\mathbf{x}'|}
\biggl[\frac{{m_1}^2-{m_2}^2}{3{m_1}^2}\,
\partial^2_{i''j''}-\biggl(\frac{{m_2}^2}{2}-\frac{{m_1}^2+2{m_2}^2}{6{m_1}^2}\triangle_{\mathbf{x}''}\biggr)\delta_{ij}\biggr]
X^{(2)}(\mathbf{x}'')
\end{eqnarray}

If we choose the metric (\ref{metric-newtonian-order}), from  (\ref{PPN-ricci-tensor}) we have
$R^{(2)}_{ij}\,=\,\triangle\Psi\,\delta_{ij}+(\Psi-\Phi)_{,ij}$
and the second field equation of (\ref{NL-field-equation_2})
becomes

\begin{eqnarray}\label{NL-field-equationij}
\left\{\begin{array}{ll} \triangle\Psi\,=\,\int
d^3\mathbf{x}'\mathcal{G}_2(\mathbf{x},\mathbf{x}')\biggl(\frac{{m_2}^2}{2}-\frac{{m_1}^2+2{m_2}^2}{6{m_1}^2}
\triangle_{\mathbf{x}'}\biggr)X^{(2)}(\mathbf{x}')\\\\
(\Phi-\Psi)_{,ij}\,=\, \frac{{m_1}^2-{m_2}^2}{3{m_1}^2}\int
d^3\mathbf{x}'\mathcal{G}_2(\mathbf{x},\mathbf{x}')\,
X^{(2)}_{,i'j'}(\mathbf{x}')
\end{array}\right.
\end{eqnarray}
Then the general solution for $g^{(2)}_{ij}$ from
(\ref{NL-field-equation_2}), \emph{without gauge conditions} and by using
the first line of (\ref{NL-field-equationij}), is

\begin{eqnarray}\label{solpsi}
g^{(2)}_{ij}\,=\,2\,\Psi\,\delta_{ij}\,=\,-\frac{\delta_{ij}}{2\pi}\int
d^3\mathbf{x}'d^3\mathbf{x}''\frac{\mathcal{G}_2(\mathbf{x}',\mathbf{x}'')}{|\mathbf{x}-\mathbf{x}'|}
\biggl(\frac{{m_2}^2}{2}-\frac{{m_1}^2+2{m_2}^2}{6{m_1}^2}
\triangle_{\mathbf{x}''}\biggr)X^{(2)}(\mathbf{x}'')
\end{eqnarray}
and the second line of (\ref{NL-field-equationij}) is only a
constraint condition for metric potentials. In fact from its trace,
we have

\begin{eqnarray}\label{cond}
\triangle(\Phi-\Psi)\,=\,\frac{{m_1}^2-{m_2}^2}{3{m_1}^2}\int
d^3\mathbf{x}'\mathcal{G}_2(\mathbf{x},\mathbf{x}')\,
\triangle_{\mathbf{x}'}X^{(2)}(\mathbf{x}')
\end{eqnarray}
and we can affirm that only in GR the metric potentials $\Phi$ and
$\Psi$ are equals.

\begin{figure}[htbp]
  \centering
  \includegraphics[scale=1]{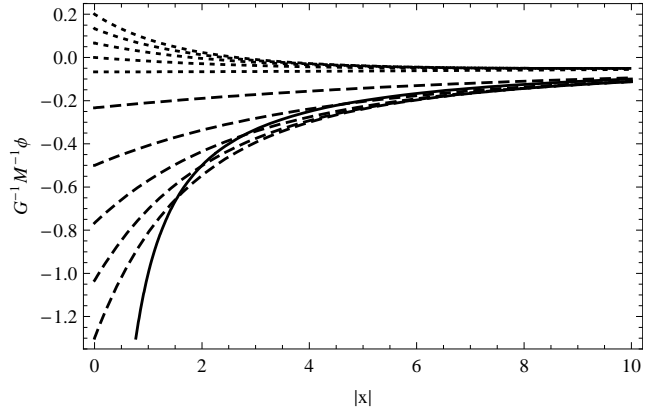}\\
  \caption{Plot of metric potential $\Phi$ in Eqs. (\ref{sol_pot_fXYZ}). $m_2\,=\,\zeta\,m_1$ and $m_1\,=\,0.1$ (dotted line),
  $m_1\,=\,\zeta\,m_2$ and $m_2\,=\,0.1$ (dashed line). The behavior of GR is shown by the solid line.
  The dimensionless quantity $\zeta$ runs between $0\div 10$. We set $f_X(0)\,=\,1$.}
  \label{plotpotential}
\end{figure}

\begin{figure}[htbp]
  \centering
  \includegraphics[scale=1]{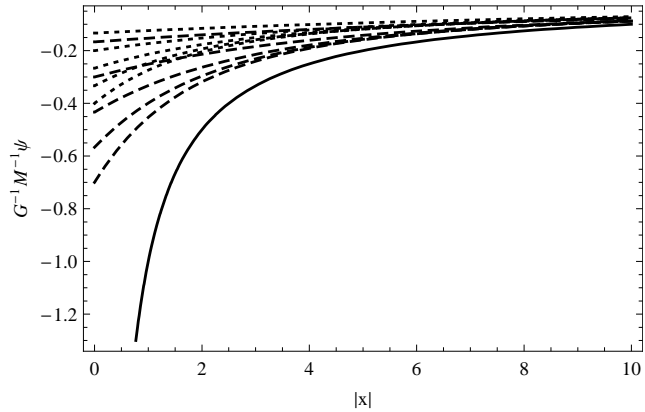}\\
  \caption{Plot of metric potential $\Psi$ in Eqs.(\ref{sol_pot_fXYZ}). $m_2\,=\,\zeta\,m_1$ and $m_1\,=\,0.1$ (dotted line),
  $m_1\,=\,\zeta\,m_2$ and $m_2\,=\,0.1$ (dashed line). The behavior of GR is shown by the solid line.
  The dimensionless quantity $\zeta$ runs between $0\div 10$. We set $f_X(0)\,=\,1$.}
  \label{plotpotential_2}
\end{figure}

Let us consider the point-like source (\ref{point_like}). If we choose
${m_1}^2\,>\,0$ and ${m_2}^2\,>\,0$, the curvature invariant $X^{(2)}$
(\ref{scalar_invariant_sol_gen}) and the metric potentials $\Phi$
(\ref{tt_component_sol_gen}) and $\Psi$ (\ref{solpsi}) are

\begin{eqnarray}\label{sol_pot_fXYZ}
\left\{\begin{array}{ll}
X^{(2)}\,=\,
-\frac{r_g\,{m_1}^2}{f_X(0)}\frac{e^{-m_1|\mathbf{x}|}}{|\mathbf{x}|}
\\\\
\Phi\,=\,-\frac{GM}{f_X(0)}\biggl[\frac{1}{|\textbf{x}|}
+\frac{1}{3}\frac{e^{-m_1|\mathbf{x}|}}{|\mathbf{x}|}
-\frac{4}{3}\frac{e^{-m_2|\mathbf{x}|}}{|\mathbf{x}|}\biggr]
\\\\
\Psi\,=\,-\frac{GM}{f_X(0)}\biggl[\frac{1}{|\textbf{x}|}
-\frac{1}{3}\frac{e^{-m_1|\mathbf{x}|}}{|\mathbf{x}|}
-\frac{2}{3}\frac{e^{-m_2|\mathbf{x}|}}{|\mathbf{x}|}\biggr]
\end{array}\right.
\end{eqnarray}
The modified gravitational potential of $f(R)$-gravity is further modified by
the presence of functions of $R_{\alpha\beta}R^{\alpha\beta}$ (as in the \emph{Quadratic Lagrangian}-theory) and
$R_{\alpha\beta\gamma\delta}R^{\alpha\beta\gamma\delta}$. The
curvature invariant $X^{(2)}$ presents a
massive propagation and when $f(X,Y,Z)\rightarrow f(R)$ we find
the mass definition $m^2\,=\,-f'(R\,=\,0)/3f''(R\,\,=0)$
given in (\ref{mass_definition_0}).  In this case, the propagation
mode with $m_2$ disappears. Obviously the two expressions for gravitational potentials
in (\ref{sol_pot_fXYZ}) satisfy the constraint
condition (\ref{cond}). In FIGs. \ref{plotpotential} and
\ref{plotpotential_2} we sketch  the spatial behavior of metric
potentials for some values  of parameters $m_1$ and $m_2$.


The same outcome has been found for \emph{Quadratic Lagrangian} if we identify $a_1\,=\,f_X(0)$.
We can affirm, then, the Newtonian limit of
any $f(X,Y,Z)$-Gravity can be reinterpreted by introducing the
\emph{Quadratic Lagrangian} and the coefficients have to satisfy
the following relations

\begin{eqnarray}\label{equivalence}
a_1\,=\,f_X(0),\,\,\,\,\,\,\,\,\,a_2\,=\,\frac{1}{2}f_{XX}(0)-f_Z(0),\,\,\,\,\,\,\,\,\,
a_3\,=\,f_Y(0)+4f_Z(0)
\end{eqnarray}
if we want the same definition of parameters in (\ref{scale2}) and (\ref{mass_definition}).

A first considerations about (\ref{equivalence}) concerns the
characteristic lengths induced by $f(X,Y,Z)$-theory. The second
length ${m_2}^{-1}$ is originated by the presence, in the
Lagrangian, of squared Ricci and Riemann tensors, but also a theory
containing only squared Ricci tensor  shows the same outcome.
Obviously the same is valid also with the only squared Riemann  tensor .
 Such terms give rise to  massive gravitational modes and then to the possibility of massive 
 gravitational waves  (see \cite{bogdanos} and references therein).

A second consideration concerns the Gauss - Bonnet
invariant defined by the relation $G_{GB}\,=\,X^2-4Y+Z$
\cite{dewitt_book}. In fact the induced field equations satisfy, in
four dimensions, the  condition

\begin{eqnarray}\label{fieldequationGB}
H^{GB}_{\mu\nu}\,=\,H^{X^2}_{\mu\nu}-4H^{Y}_{\mu\nu}+H^{Z}_{\mu\nu}\,=\,0
\end{eqnarray}
and, by substituting them at Newtonian level
($H^Z_{tt}\,\sim\,-4\triangle R^{(2)}_{tt}$) in Eqs. (\ref{fieldequationFOG}), we find the field equations (ever at
Newtonian Level) of \emph{Quadratic Lagrangian}.

A third and last consideration is about the solutions
(\ref{sol_pot_fXYZ}). When we perform the
limit in the origin $|\mathbf{x}|\,=\,0$ we have no 
divergence. In fact we find

\begin{eqnarray}\label{divergence}
\lim_{|\mathbf{x}|\rightarrow 0
}\Phi\,\propto\,\frac{m_1-4m_2}{3},\,\,\,\,\,\,\,\,\,\,\,\,\lim_{|\mathbf{x}|\rightarrow
0 }\Psi\,\propto\,-\frac{m_1+2m_2}{3}
\end{eqnarray}
and only if we remove, in the action (\ref{FOGaction}), the
dependence on the squared Ricci  or squared Riemann tensors, we get the
divergence of GR. For a physical interpretation of solution
(\ref{sol_pot_fXYZ}), we must impose the condition $m_1-4m_2\,<\,0$ to
have a potential well with a negative minimum in
$|\mathbf{x}|\,=\,0$ and $m_1\,<\,m_2$ to have a negative profile
of potential (see FIG. \ref{plotpotential}). Then, if we suppose
$f_X(0)\,>\,0$, we get a constraint on the derivatives of $f$ with
respect to curvature invariants, that is

\begin{eqnarray}\label{condition}
f_{XX}(0)+f_Y(0)+2f_Z(0)\,<\,0
\end{eqnarray}
In the case of $f(R)$-gravity ($f_Y(0)\,=\,f_Z(0)\,=\,0$) we
reobtain the same condition between the first and second derivatives
of $f(R)$.

\section{Conclusions}\label{conclus}
The weak field limit is a crucial issue  that  has to be addressed in any relativistic theory of gravity. It is also the test bed of such  theories in order
 to compare them with the well-founded experimental results of GR, at  least at Solar system level. 
 
 In this review paper, we have considered the problem of weak field limit of Fourth Order Gravity, that is of  gravitational theories where curvature invariants, a part the standard Ricci scalar, linear in the Hilbert-Einstein action, are taken into account.  In particular,  we have analyzed the Newtonian and the post-Newtonian limits of theories involving non-linear combinations of Ricci scalar, Ricci tensor and Riemann tensor. The calculations have been essentially developed in the so-called Jordan frame but we have also considered  the conformal  transformations and the possible shortcomings emerging in carrying the weak field limit results in
 the Einstein frame without an appropriate interpretation of post-Newtonian parameters. 
 
The general feature  that emerges from the weak field limit is that corrections to the Newtonian potential naturally come out. These corrections  are   Yukawa-like terms bringing characteristic masses and lengths. Conversely, the standard Newtonian potential is just a feature emerging in the particular case $f(R)=R$.  These characteristic masses (and lengths) come out as combinations of  the parameters of the theory and fix the scales where corrections become relevant. 

These results open new intriguing possibilities since accurate measurements of PPN parameters could confirm or rule out these theories in view of the  forthcoming space experiments as GAIA and  GAME (see \cite{vecchiato1,vecchiato2,vecchiato3} for a detailed review of the status of art).

On the other hand, it is well-known that the new features related to extended theories of gravity could have interesting applications in other fields of astrophysics as galactic dynamics \cite{cardone}, large scale structure \cite{salzano1}  and cosmology \cite{salzano2} in order to address dark  matter and dark energy issues. The fact that such "dark" structures have not been definitely discovered at fundamental quantum scales but operate at large astrophysical (infra-red scales) could be due to these corrections to the Newtonian potential which can be hardly detected at laboratory or Solar System scales.
Finally, the presence of unavoidable light massive modes could open new opportunities also for the gravitational waves detection of experiments like VIRGO, LIGO and the forthcoming  LISA \cite{bogdanos}.


\begin{thebibliography}{99}

\bibitem{cosmic_acceleration}
         Riess A.G. {\it et al.} Astron. J. {\bf 116}, 1009 (1998)\\
         Perlmutter S. {\it et al.} Astrophys. J. {\bf 517}, 565 (1999)\\
         Spergel D.N. {\it et al.} Astrophys. J. Suppl. {\bf 148}, 175 (2003)\\
         Riess A. G. {\it et al.} Astrophys. J. {\bf 607}, 665 (2004)\\
         Schmidt H. J., ArXiv: gr-qc/0407095 (2004)\\
         Cole S. {\it et al.}, Mon. Not. Roy. Astron. Soc. {\bf 362}, 505 (2005)\\
         Perlmutter S. {\it et al.} Astron. Astrophys. {\bf 447}, 31 (2006)\\
         Spergel D.N. {\it et al.} Astrophys. J. Suppl. {\bf 170}, 377
         (2007).

\bibitem{theo_aspect_1}
         Farhoudi M., Gen. Rel. Grav. \textbf{38}, 1261 (2006).
\bibitem{theo_aspect_2}
         Nojiri S., Odintsov S.D., Int. J. Meth. Mod. Phys. \textbf{4}, 115
         (2007).
\bibitem{theo_aspect_3}
         Capozziello S., Francaviglia M., Gen.Rel. Grav. \textbf{40}, 357
         (2008).
\bibitem{theo_aspect_4}
         Sotiriou T.P., Faraoni V.,  Rev. Mod. Phys. \textbf{82}, 451
         (2010).
\bibitem{theo_aspect_5}
         De Felice A., Tsujikawa S., Living Rev. Rel. \textbf{13}, 3
         (2010).
\bibitem{theo_aspect_6}
         Capozziello S., Faraoni V., {\it Beyond Einstein
         Gravity}, Fundamental Theories of Physics, Vol. \textbf{170},
         Ed. Springer, New York  (2010).

\bibitem{anderson}
         Anderson J.D. \emph{et al.} Phys. Rev. Lett. {\bf 81}, 2858 (1998)\\
         Anderson J.D. \emph{et al.} Phys. Rev. D {\bf 65}, 082004
         (2002.

\bibitem{bertolami}
         Bertolami O., B\"{o}hmer C.G., Harko T., Lobo F.S.N., Phys. Rev. D \textbf{75}, 104016 (2007)

\bibitem{old_papers_fR}
         Weyl H., Raum-Zeit-Materie, Springer Berlin (1921)\\
         Eddington A.S., \emph{The mathematical theory of relativity}, Cambridge University
         Press London (1924)\\
         Lanczos C., Z. Phys. \textbf{73}, 147 (1931)\\
         Buchdahl H.A., Nuovo Cimento \textbf{23}, 141 (1962)\\
         Bicknell G.V., Journ. phys. A \textbf{7}, 1061 (1974).

\bibitem{noethergae}
         Capozziello S.,  Lambiase G., Gen. Rel. Grav. \textbf{32}, 295 (2000).

\bibitem{stel}
         Stelle K., Gen. Rel. Grav. \textbf{9}, 353 (1978).

\bibitem{curvquin}
         Capozziello S.,  Int. J. Mod. Phys. D \textbf{11}, 483 (2002).

\bibitem{francaviglia}
         Magnano G., Ferraris M., Francaviglia M., Gen. Rel. Grav. \textbf{19}, 465 (1987)\\
         Allemandi G., Borowiec A., Francaviglia M., Phys. Rev. D \textbf{70}, 103503 (2004)\\
         Amarzguioui M., Elgaroy O., Mota D.F., Multamaki  T., Astron. Astrophys. \textbf{454}, 707
         (2006).

\bibitem{torsion}
         Capozziello S., Cianci R., Stornaiolo C., Vignolo S.,
         Class. Quant. Grav. \textbf{24}, 6417 (2007).

\bibitem{condition}
         Santos J., Alcaniz J.S., Reboucas M.J.,  Carvalho F.C., Phys. Rev. D {\bf 76}, 083513 (2007).

\bibitem{old_papers_fR_2}
         Havas P., Gen. Rel. Grav. \textbf{8}, 631 (1977).

\bibitem{olmo}
         Olmo G. J., Phys. Rev. D \textbf{72}, 083505 (2005)\\
         Olmo G. J., Phys. Rev. Lett. \textbf{95}, 261102 (2005)\\
         Olmo G. J., Phys. Rev. D \textbf{75}, 023511 (2007)

\bibitem{Damour:Esposito-Farese:1992}
         Damour T., Esposito-Far\`{e}se G., Class. Quant. Grav. \textbf{9}, 2093 (1992)

\bibitem{clifton}
         Clifton T., Phys. Rev. D \textbf{77}, 024041 (2008)

\bibitem{will} Will C. M., \emph{Theory and Experiments in Gravitational Physics} Cambridge Univ. Press 1993 Cambridge

\bibitem{einstein1} Einstein A., Ann. der Physik \textbf{49}, 769 (1916)

\bibitem{eisenhart} Eisenhart E., \emph{Riemannian Geometry}, Princeton Univ. Press 1955 Princeton

\bibitem{schroedinger} Schr\"odinger E., \emph{Space-Time Structure}, Cambridge Univ. Press 1960 Cambridge

\bibitem{landau} Landau L.D., Lif\v{s}its E.M, \emph{Theoretical physics} vol. II

\bibitem{weinberg} Weinberg S., \emph{Gravitation and Cosmology}, Wiley 1972, New York

\bibitem{levicivita} Levi Civita T., \emph{The Absolute Differential Calculus}, Blackie and Son 1929, London

\bibitem{kle} Klein O., New Theories in Physics \textbf{77}, Intern.Inst. of Intellectual Cooperation, League of Nations (1938)

\bibitem{cartan} Cartan E., Ann. Ec. Norm. \textbf{42}, 17 (1925)

\bibitem{palatini} Palatini A., Rend. Circ. Mat. Palermo \textbf{43}, 203 (1919)

\bibitem{arnold} Arnold V. I., \emph{Mathematical Methods of Classical Mechanics}, Springer Verlag 1978 Berlin

\bibitem{app-fre} Appelquist T., Chodos A., Freund P. G. O., \emph{Modern Kaluza-Klein Theories} 1978 Addison-Wesley Reading

\bibitem{kaku} Kaku M., Quantum Field Theory, 1933 Oxford Univ. Press, Oxford

\bibitem{mag-sok} Magnano G., Soko{\l}owski L. M., Phys. Rev. D \textbf{50}, 5039 (1994)

\bibitem{cqg} Capozziello S., de Ritis R., A.A. Marino, Class. Quant.
Grav. \textbf{14}, 3243 (1997).

\bibitem{spher_symm_fR} Capozziello S., Stabile A., Troisi A., Class. Quant. Grav. {\bf 25}, 085004 (2008)

\bibitem{trautman} Trautman A., Comp.rend. heb. sean. \textbf{257}, 617 (1963)

\bibitem{friedrichs} Friedrichs K.,  Math. Ann. \textbf{98}, 566 (1927)

\bibitem{kilmister} Kilmister C. W., J. Math. Phys. \textbf{12}, 1 (1963)

\bibitem{dautcourt} Dautcourt G., Acta Phys. Polon. \textbf{25}, 637 (1964)

\bibitem{kuenzle} Kuenzle H. P., Gen. Rel. Grav. \textbf{7}, 445 (1976)

\bibitem{ehlers} Ehlers J., Ann. N. Y. Acad. Scien. \textbf{336}, 279 (1980)

\bibitem{ehlers1} Ehlers J., Grundlagenprobleme der modernen Physik, Eds.\ J.\ Nitsch, J.\ Pfarr, E.W.\ Stachow, B. I. - Wissenschaftsverlag, Mannheim, 65 (1981)

\bibitem{dick} Dick R., Gen. Rel. Grav. \textbf{36}, 217 (2004).

\bibitem{newtonian_limit_fR} Capozziello S., Stabile A., Troisi A., Phys. Rev. D {\bf 76}, 104019 (2007)\\
                             Capozziello S., Stabile A., Troisi A., Modern physics letters A {\bf 24}, 659 (2009)

\bibitem{postnewtonian_limit_fR} Stabile A., arXiv: gr-qc/1004.1973v2

\bibitem{chietall} Olmo G. J., Phys. Rev. Lett. \textbf{95}, 261102 (2005)\\
                   Chiba T., Phys. Lett. B \textbf{575}, 1 (2005)\\
                   Erickcek A. L., Smith T. L., Kamionkowski M., Phys.Rev. D \textbf{74}, 121501 (2006)\\
                   Soussa M. E., Woodard R. P., Gen. Rel. Grav. \textbf{36}, 855
                   (2004).
\bibitem{wald}
Wald, R.M., {\it General Relativity\/}, The University of Chicago
Press, Chicago (1984).

\bibitem{vignolo} Capozziello S.,  Vignolo S.,  Class. Quantum Grav., {\bf 26}, 175013 (2009).

\bibitem{bransdicke} Brans C., Dicke R.H., Phys.\ Rev. {\bf 124}, 925 (1961)

\bibitem{cap-tro} Capozziello S., Troisi A., Phys. Rev. D \textbf{72}, 044022 (2005)

\bibitem{faraoni4} Faraoni V., Phys. Rev. D \textbf{74}, 023529 (2006)

\bibitem{nav-van1} Navarro I., Van Acoleyen K., Phys. Lett. B \textbf{622}, 1 (2005)

\bibitem{ohanlon} O'Hanlon J., Phys. Rev. Lett. \textbf{29}, 137 (1972)

\bibitem{TS-fR-analogy} Capozziello S., Stabile A., Troisi A., Phys. Lett. B \textbf{686}, 79 (2010)

\bibitem{qua-sch} Quant I., Schmidt H. J., Astron. Nachr. \textbf{312}, 97 (1991)

\bibitem{bin-tre} Binney J., Tremaine S., \emph{Galactic dynamics}, Princeton University Books, Princeton
(1987).

\bibitem{newtonian_limit_R_Ric} Capozziello S., Stabile A., Class. Quant. Grav. {\bf 26}, 085019 (2009)

\bibitem{fXYZ_limit} Stabile A., arXiv: gr-qc/1007.1917v1

\bibitem{dewitt_book} de Witt B. S., \emph{Dynamical theory of groups and fields}, Gordon and Breach, New York (1965).

\bibitem{bogdanos}
 Bogdanos C.,  Capozziello S.,   De Laurentis M.,  Nesseris S.,    Astrop. Phys.  {\bf 34},  236 ( 2010).
 
 
\bibitem{vecchiato1}Vecchiato, A., Lattanzi, M. G., Bucciarelli, B., Crosta, M., de Felice,
F., Gai, M.,  Astron. Astrophys., {\bf 399}, 337 (2003). 


\bibitem{vecchiato2} Vecchiato A., Gai M., Lattanzi M. G., Crosta M. T., Sozzetti A.,
 Adv. in Space  Res. {\bf  44}, 579 (2009).

\bibitem{vecchiato3} Gai M., Vecchiato A., Lattanzi M. G., Ligori S., Loreggia D., 2009.
 Adv.  in Space  Res.  {\bf  44},  588  (2009).
 
\bibitem{cardone} 
 Capozziello S., Cardone V.F.,  Troisi A,   Mon. Not. Roy. Astr. Soc. {\bf 375} , 1423  (2007). 
 
\bibitem{salzano1} 
 Capozziello S., De Filippis E,  Salzano V.,   Mon. Not. Roy. Astr. Soc.,  {\bf 394}, 947  (2009). 

\bibitem{salzano2} 
 Capozziello S., Cardone V.F.,  Salzano V.,   Phys.Rev. {\bf D 78}, 063504 (2008). 


\end{thebibliography}
\end{document}